\documentclass[dvips,rmp,twocolumn,groupedaddress,floatfix]{revtex4}
\makeatletter
\renewcommand\@biblabel[1]{#1.}
\makeatother
\bibpunct{(}{)}{,}{n}{,}{,}
\usepackage[utf8]{inputenc}
\usepackage[T1]{fontenc}

\usepackage{graphicx}
\usepackage[dvips,dvipsnames]{color}
\usepackage{amssymb}\usepackage{amsmath}
\usepackage{times}
\usepackage{helvet}
\definecolor{darkgrey}{rgb}{0.25,0.25,0.25}
\definecolor{darkred}{rgb}{0.5,0.0,0.0}
\definecolor{modulecolor}{rgb}{0.84,0.09,0.13}
\definecolor{nodecolor}{rgb}{0.16,0.51,0.71}
\usepackage{url}
\usepackage[dvips,breaklinks=true,colorlinks=true,citecolor=darkgrey,pagecolor=darkred,linkcolor=darkgrey,menucolor=darkred,urlcolor=darkred,pdfborder={1 0 0}]{hyperref}
\hypersetup{pdftitle={Maps of Information Flow Reveal Community Structure In Complex Networks},pdfauthor={Martin Rosvall and Carl Bergstrom}}
\usepackage{memhfixc}

\begin{document}

\title{Maps of random walks on complex networks reveal community structure}

\author{M. Rosvall}
\email{rosvall@u.washington.edu}
\affiliation{Department of Biology, University of Washington, Seattle, WA 98195-1800}
\email{rosvall@u.washington.edu} 
\author{C. T. Bergstrom}
\email{cbergst@u.washington.edu}
\affiliation{Department of Biology, University of Washington, Seattle, WA 98195-1800}
\affiliation{Santa Fe Institute, 1399 Hyde Park Rd., Santa Fe, NM 87501}
\homepage{http://octavia.zoology.washington.edu/}
\date{\today}
%\pacs{89.75.-k, 89.75.Fb, 89.70.+c}
\keywords{complex networks | clustering | information theory | compression}
\begin{abstract}
To comprehend the multipartite organization of large-scale biological and social systems,
we introduce a new information theoretic approach that reveals community structure in weighted and directed networks.
The method decomposes a network into modules by optimally compressing a description of information flows on the network.
The result is a map that both simplifies and highlights the regularities in the structure and their relationships. We illustrate the method by making a map of scientific communication as 
captured in the citation patterns of more than 6000 journals.
We discover a multicentric organization with fields that vary dramatically in size and degree of integration into the network of science. Along the backbone of the network --- including physics, chemistry, molecular biology, and medicine --- information flows bidirectionally, but the map reveals a directional pattern of citation from the applied fields to the basic sciences.
\end{abstract}

\maketitle

Biological and social systems are differentiated, multipartite, integrated, and dynamic.
Data about these systems, now available on unprecedented scales, are often schematized as networks.
Such abstractions are powerful \cite{newmanSIAM,newman-pap}, but even as abstractions they remain highly complex.
It is therefore helpful to decompose the myriad nodes and links into modules that represent the network \cite{girvan_newman,palla,guimerahierarchy}.
A cogent representation
will retain the important information about the network and reflect the fact
that interactions between the elements in complex systems are weighted, directional, interdependent, and conductive.
Good representations both simplify and highlight the underlying structures and the relationships which they depict; they are maps \cite{tufte,guimera-nature}.

To create a good map,
the cartographer must attain a fine balance between omitting important structures by oversimplification, and obscuring significant relationships in a barrage of superfluous detail. The best maps convey a great deal of information, but require minimal bandwidth:
the best maps are also good compressions.
By adopting an information-theoretic approach, we can measure how efficiently a map represents the underlying geography --- and we can measure how much detail is lost in the process of simplification. This allows us to quantify and resolve the cartographer's tradeoff.

\section*{Network maps and coding theory}
In this paper, we use maps to describe the dynamics across the links and nodes in directed, weighted networks that represent the local interactions among the subunits of a system. These local interactions induce a system-wide flow of information that characterizes the behavior of the full system \cite{ziv,donath,enright,girvan,kasper}.
Consequently, if we want to understand how network structure relates to system behavior, we need to understand the flow of information on the network. We therefore identify the modules which compose the network by finding an optimally compressed 
description of how information flows on the network. A group of nodes among which information flows quickly and easily can be aggregated and described as a single well-connected module; the links between modules capture the avenues of information flow between those modules. 

\begin{figure*}[thbp]
\centering
\includegraphics[width=\textwidth]{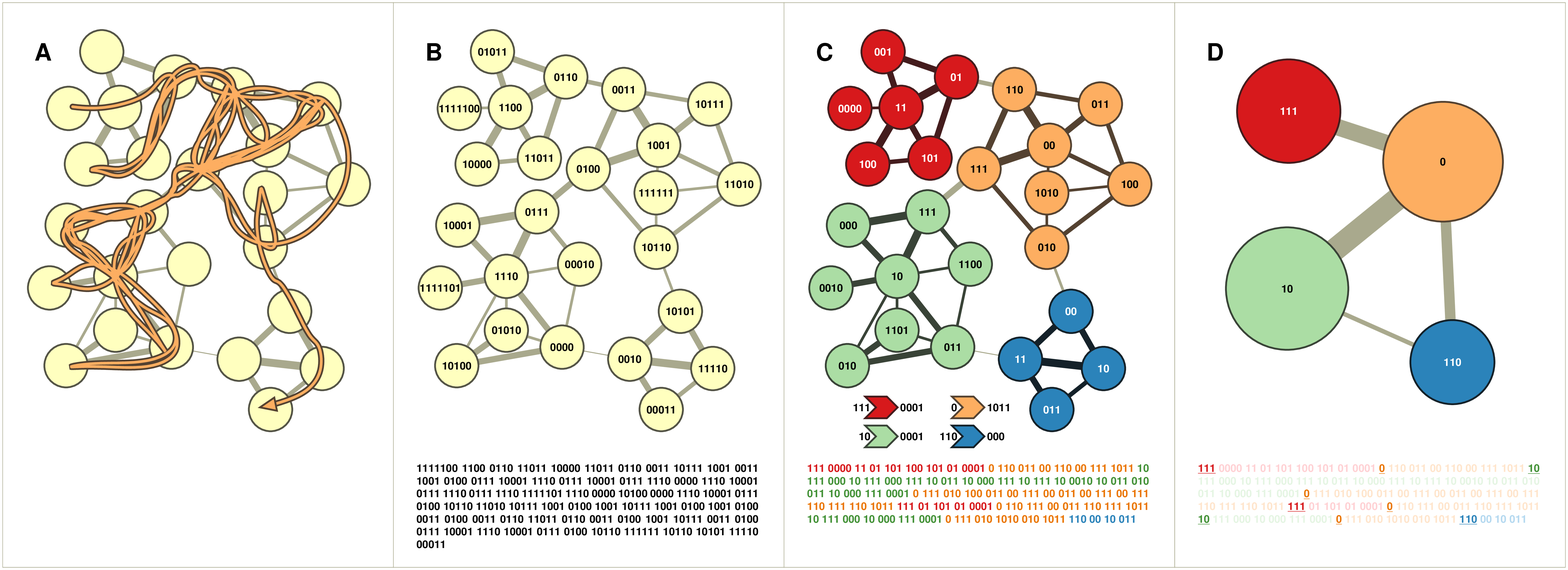}
\caption{\label{fig1}Detecting communities by compressing the description of information flows on networks. (A) 
We want to describe the trajectory of a random walk on the network, such that important structures have unique names. The orange line shows one sample trajectory. (B) A basic approach is to give a unique name to every node in the network. The Huffman code illustrated here is an efficient way to do so.
The 314 bits shown under the network describes the sample trajectory in A, starting with $1111100$ for the first node on the walk in the upper left corner, $1100$ for the second node etc., and ending with $00011$ for the last node on the walk in the lower right corner.
(C) A two-level description of the random walk, in which major clusters receive unique names but the names of nodes within clusters are reused, yields on average a 32\% shorter description for this network. The codes naming the modules and the codes used to indicate an exit from each module are shown to the left and the right of the arrows under the network, respectively. Using this code, we can describe the walk in A by the 243 bits shown under the the network in C. The first three bits $111$ indicate that the walk begins in the red module, the code $0000$ specifies the first node on the walk etc. (D) Reporting only the module names, and not the locations within the modules, provides an optimal coarse-graining of the network.}
\end{figure*}

Succinctly describing information flow is a coding or compression problem. The key idea in coding theory is that a data stream can be compressed by a code that exploits regularities in the process that generates the stream \cite{shannon}.  We use a random walk as a proxy for the information flow, because a random walk uses all of the information in the network representation and nothing more. Thus it provides a default mechanism for generating a dynamics from a network diagram alone \cite{ziv}. 

Taking this approach, we develop an efficient code to describe a random walk on a network. We thereby show that finding community structure in networks is equivalent to solving a coding problem \cite{RosvallAndBergstrom07,rissanen1978,grunwald}. We exemplify this by making a map of science, based on how information flows among scientific journals by means of citations.

\subsection*{Describing a path on a network}
To illustrate what coding has to do with map-making, consider the following communication game. Suppose that you and I both know the structure of a weighted directed network. We aim to choose a code that will allow us to efficiently describe paths on the network that arise from a random walk process, in a language that reflects the underlying structure of the network. How should we design our code? 

If maximal compression were our only objective, we could encode the path at or near the entropy rate of the corresponding Markov process. Shannon showed that one can achieve this rate by assigning to each node a unique dictionary over the outgoing transitions \cite{shannon48}. But compression is not our only objective; here want our language to reflect the network structure, we want the {\em words} we use to refer to {\em things} in the world. Shannon's approach does not do this for us, because every codeword would have a different meaning depending on where it is used. Compare maps: useful maps assign unique names to important structures. Thus we seek a way of describing or encoding the random walk in which important structures indeed retain unique names.

Let us look at a concrete example. Figure \ref{fig1}A shows a weighted network with $n=25$ nodes. The link thickness indicates the relative probability that a random walk will traverse any particular link. Overlaid upon the network is a specific 71-step realization of a random walk that we will use to illustrate our communication game. In panels \ref{fig1}B--D, we describe this walk with increasing levels of compression, exploiting more and more of the regularities in the network.

\subsection*{Huffman coding} 
A straightforward method of giving names to nodes is to use a Huffman code \cite{huffman}. Huffman codes save space by assigning short codewords to common events or objects, and long codewords to rare ones, much as common words are short in spoken languages \cite{zipf}. Figure \ref{fig1}B shows a prefix-free Huffman coding for our sample network. Each codeword specifies a particular node, and the codeword lengths are derived from the ergodic node visit frequencies of an infinitely long random walk. With the Huffman code pictured in Fig.~\ref{fig1}B, we are able to describe the specific 71-step walk in 314 bits. If we instead had chosen a uniform code, in which all codewords are of equal length, each codeword would be  $\lceil \log{25} \rceil = 5$ bits long and $71\cdot5=355$ bits would have been required to describe the walk. 

Though in this example we assign actual codewords to the nodes for illustrative purposes, in general we will not be interested in the codewords themselves, but rather in the theoretical limit of how concisely we can specify the path.  Here we invoke Shannon's source coding theorem \cite{shannon48} which implies that when you use $n$ codewords to describe the $n$ states of a random variable $X$ that occur with frequencies $p_i$, the average length of a codeword can be no less than the entropy of the random variable $X$ itself: $H(X) = -\sum_{1}^{n} p_i \log(p_i)$.  This theorem provides us with the necessary apparatus to see that in our Huffman illustration,  the average number of bits needed to describe a single step in the random walk is bounded below by the entropy $H(P)$, where $P$ is the distribution of visit frequencies to the nodes on the network. We define this lower bound on code length to be $L$. For example, $L=4.50$ bits/step  in Fig.~\ref{fig1}B.

\subsection*{Highlighting important objects}
Matching the length of codewords to the frequencies of their use gives us efficient codewords for the nodes, but no map. Merely assigning appropriate-length names to the nodes does little to simplify or highlight aspects of the underlying structure. To make a map, we need to separate the important structures from the insignificant details. We therefore divide the network into two levels of description. We retain unique names for large-scale objects, the clusters or modules to be identified within our network, but we reuse the names associated with fine-grain details, the individual nodes within each module. This is a familiar approach for assigning names to objects on maps: most US cities have unique names, but street names are reused from one city to the next, such that each city has a Main Street and a Broadway and a Washington Avenue and so forth. The reuse of street names rarely causes confusion, because most routes remain within the bounds of a single city. 

A two-level description allows us to describe the path in fewer bits than we could do with a one-level description. We capitalize on the network's structure --- and in particular, on the fact that a random walker is statistically likely to spend long periods of time within certain clusters of nodes. Figure \ref{fig1}C illustratess this approach. We give each cluster a unique name, but use a different Huffman code to name the nodes within each cluster. A special codeword, the exit code, is chosen as part of the within-cluster Huffman coding and indicates that the walk is leaving the current cluster. The exit code is always followed by the ``name'' or module code of the new module into which the walk is moving (see supporting online material for more details).  Thus we assign unique names to coarse-grain structures, the cities in the city metaphor, but reuse the names associated with fine-grain details, the streets in the city metaphor. The savings are considerable; in the two-level description of Fig~\ref{fig1}C the limit $L$ is $3.05$ bits/step compared to 4.50 for the one-level description.

Herein lies the duality between finding community structure in networks and the coding problem: to find an optimal code, we look for a module partition $\mathsf{M}$ of $n$ nodes into $m$ modules so as to minimize the expected description length of a random walk. Using the module partition $\mathsf{M}$, the average description length of a single step is given by 

\begin{equation}\label{map}
L(\mathsf{M}) = q_{\curvearrowright} H(\mathcal{Q}) + \sum_{i=1}^{m}p_{\circlearrowright}^iH(\mathcal{P}^i).
\end{equation}

This equation comprises two terms: first is the entropy of the movement between modules, and second is the entropy of movements within modules (where exiting the module is also considered a movement). Each is weighted by the frequency with which it occurs in the particular partitioning. Here $q_{\curvearrowright}$ is the probability that the random walk switches modules on any given step. $H(\mathcal{Q})$ is the entropy of the module names, i.e., the entropy of the underlined codewords in Fig.~\ref{fig1}D. $H(\mathcal{P}^i)$ is the entropy of the within-module movements --- including the exit code for module $i$. The weight $p_{\circlearrowright}^i$ is the fraction of within module movements that occur in module $i$, plus the probability of exiting module $i$ such that $\sum_{i=1}^mp_{\circlearrowright}^i=1+q_{\curvearrowright}$ (see supporting online material for more details).

For all but the smallest networks, it is infeasible to check all possible partitions to find the one that minimizes the description length in the map equation (Eq.~\ref{map}). Instead we use computational search. We first compute the  fraction of time each node is visited by a random walker using the power method, and using these visit frequencies we explore the space of possible partitions using a deterministic greedy search algorithm \cite{clauset-2004-70,wakita} . We refine the results with a simulated annealing approach \cite{guimera-nature} using the heat-bath algorithm (see supporting online material for more details).

Figure \ref{fig1}D shows the map of the network, with the within-module descriptors faded out; here the significant objects have been highlighted and the details have been filtered away.\bigskip

In the interest of visual simplicity, the illustrative network in Fig.~\ref{fig1} has weighted but undirected links. Our method is developed more generally, so that we can extract information from networks with links that are directed in addition to being weighted. The map equation remains the same, only the path that we aim to describe must be slightly modified to achieve ergodicity.  We introduce a small ``teleportation probability'' $\tau$ in the random walk: with probability $\tau$ the process jumps to a random node anywhere in the network. This converts our random walker into the sort of ``random surfer'' that drives Google's Page\-Rank algorithm \cite{google}. Our clustering results are highly robust to the particular choice of the small fraction $\tau$.  For example, so long as $\tau < 0.45$ the optimal partitioning of the network in Fig.~\ref{fig1} remains exactly the same. In general, the more significant the regularities, the higher $\tau$ can be before frequent teleportation swamps the network structure. We choose $\tau=0.15$ corresponding to the well known damping factor $d=0.85$ in the Page\-Rank algorithm \cite{google}.

\section*{Mapping flow compared to maximizing modularity}

\begin{figure}[tbp]
\centering
\includegraphics[width=\columnwidth]{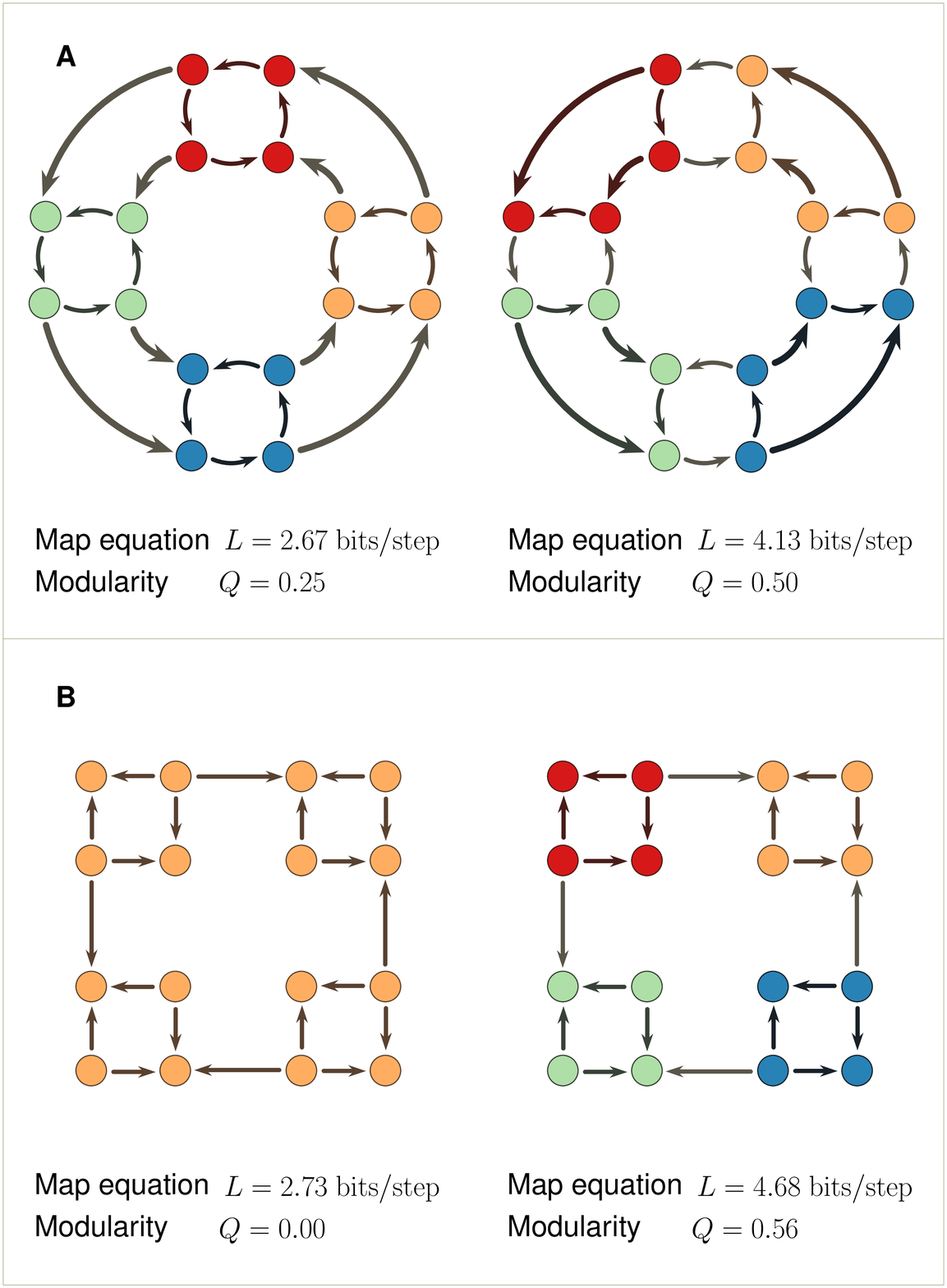}
\caption{Mapping flow compared to optimizing modularity in directed and weighted networks. 
The coloring of nodes illustrates alternative partitions of two sample networks. The left-hand partitions show the modular structure as optimized by the map equation (minimum $L$) and the right-hand partitions show the structure as optimized by modularity (maximum $Q$). In the network shown in panel A, the left-hand partition minimizes the map equation, because the persistence times in the modules are long; with the weight of the bold links set to twice the weight of other links, a random walker without teleportation takes on average 3 steps in a module before exiting. The right-hand clustering gives a longer description length, because a random walker takes on average only 12/5 steps in a module before exiting. The right-clustering maximizes the modularity, because modularity counts weights of links, the in-degree, and the out-degree in the modules; the right-hand partitioning places the heavily weighted links inside of the modules. In panel B, for the same reason, the right-hand partition again maximizes modularity. But not so the map equation. Because every node is either a sink or a source in this network, the links do not induce any long-range flow and the one step walks are best described as in the left-hand partition, with all nodes in the same cluster.\label{fig2}}
\end{figure}

The traditional way of identifing community structure in directed and weighted networks has been to simply disregard the directions and the weights of the links.  But such approaches discard valuable information about the network structure. By mapping the system-wide flow induced by local interactions between nodes, we retain the information about the directions and the weights of the links. We also acknowledge their interdependence in networks inherently characterized by flows. This makes it interesting to compare our flow-based approach with recent topological approaches based on modularity optimization that also makes use of information about weight and direction \cite{newman-fast,arenasdirectedweighted,guimeradirected,leichtdirected}. In its most general form, the modularity for a given partitioning of the network into $m$ modules is the sum of the total weight of all links in each module minus the expected weight
\begin{equation}\label{modularity}
 Q = \sum_{i=1}^{m}\frac{w_{ii}}{w} - \frac{w_{i}^{\mathrm{in}}w_{i}^{\mathrm{out}}}{w^2}.
\end{equation}
Here $w_{ii}$ is the total weight of links starting and ending in module $i$, $w_{i}^{\mathrm{in}}$ and $w_{i}^{\mathrm{out}}$  the total in- and out-weight of links in module $i$, and $w$ the total weight of all links in the network. To estimate the community structure in a network, Eq.~\ref{modularity} is maximized over all possible assignments of nodes into any number $m$ of modules. 

The two equations (\ref{map}) and (\ref{modularity}) reflect two different senses of what it means to have a network. The former, which we pursue here, finds the essence of a network in the patterns of flow that its structure induces. The latter effectively situates the essence of network in the combinatoric properties of its links (as we did in ref.~\cite{RosvallAndBergstrom07}). 

Does this conceptual distinction make any practical difference? Figure \ref{fig2} illustrates two simple networks for which the map equation and modularity give different partitionings. The weighted, directed links shown in the network in panel A induce a structured pattern of flow with long persistence times in, and limited flow between, the four clusters as highlighted on the left. The map equation picks up on these structural regularities and thus the description length is much shorter for the partitioning in the left-hand figure (2.67 bits/step) than for the right-hand one (4.13 bits/step). Modularity is blind to the interdependence in networks characterized by flows, and thus cannot pick up on this type of structural regularity. It only counts weights of links, in-degree, and out-degree in the modules, and thus prefers to partition the network as shown on the right with the heavily weighted links inside of the modules. 

In panel B, by contrast, there is no pattern of extended flow at all. Every node is either a source or a sink, and no movement along the links on the network can exceed more than one step in length. As a result, random teleportation will dominate (irrespective of teleportation rate) and any partition into multiple modules will lead to a high flow between the modules. For a network such as in panel B, where the links do not induce a pattern of flow, the map equation will always partition the network into one single module. Modularity, because it looks at pattern in the links and in- and out-degree, separates the network into the clusters shown at right. 

Which method should a researcher use? It depends on which of the two senses of network, described above, that one is studying. For analyzing network data where links represented patterns of movement among nodes, flow-based approaches such as the map equation are likely to identify the most important aspects of structure. For analyzing network data where links represent not flows but rather pairwise relationships, it may be useful to detect structure even where no flow exists. For these systems, combinatoric methods such as modularity \cite{girvan} or cluster-based compression \cite{RosvallAndBergstrom07} may be preferable.

\section*{Mapping scientific communication}
Science is a highly organized and parallel human endeavor to find patterns in nature;  the process of communicating research findings is as essential to progress as is the act of conducting the research in the first place. Thus science is not merely a set of ideas, but also the flow of these ideas through a multipartite and highly differentiated social system. Citation patterns among journals allow us to glimpse this flow, and provide the trace of communication between scientists \cite{price,small73,small,anegon,borner}. 
To highlight important fields and their relationships, to uncover differences and changes, to simplify and make the system comprehensible --- we need a good map of science.

\begin{figure*}[tbp]
\centering
\includegraphics[width=\textwidth]{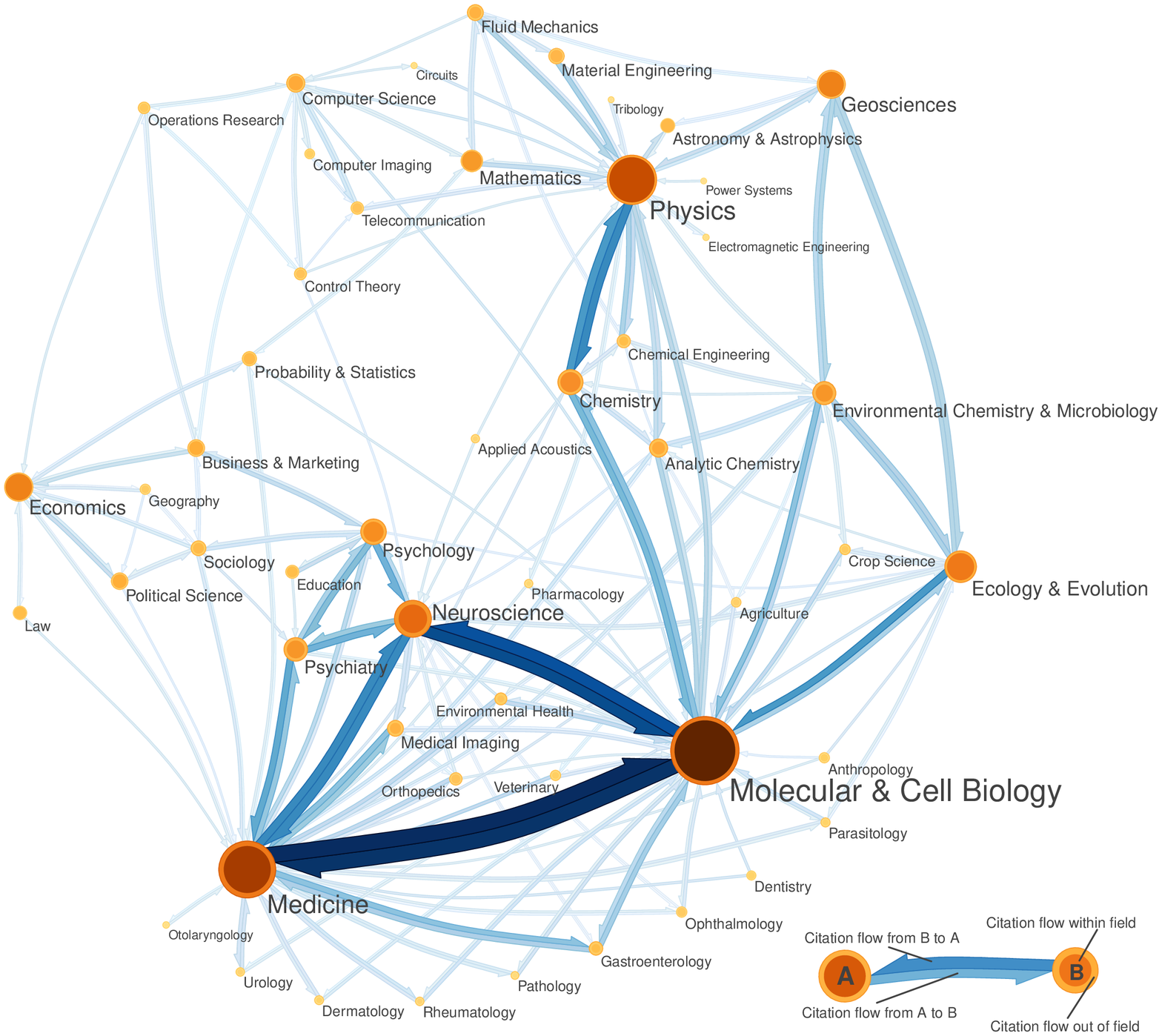}
\caption{A map of science based on citation patterns. We partitioned 6,128 journals connected by 6,434,916 citations into 88 modules and 3,024 directed and weighted links. For visual simplicity we show only the links that the random surfer traverses more than 1/5000'th of her time, and we only show the modules that are visited via these links (see supporting online material for the complete list). Because of the automatic ranking of nodes and links by the random surfer \cite{google}, we are assured of showing the most important links and nodes. For this particular level of detail we capture 98\% of the node weights and 94\% of all flow. 
\label{fig3}}
\end{figure*}

\begin{figure*}[tbp]
\centering
\includegraphics[width=1.0\textwidth]{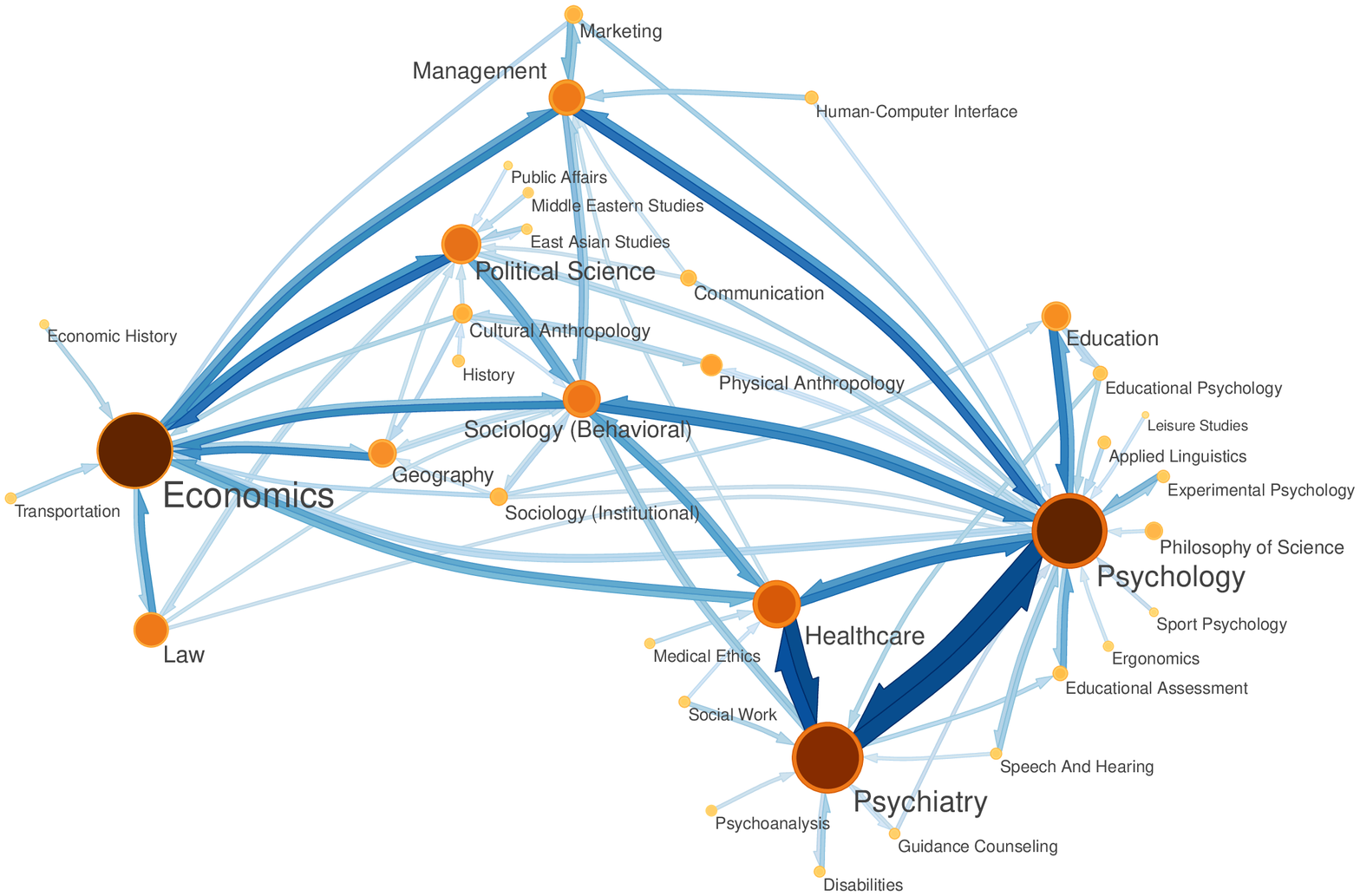}
\caption{
A map of the social sciences. The journals listed in the 2004 social science edition of Journal Citation Reports are a subset of those illustrated in Fig.~\ref{fig3}, totaling 1431 journals and 217,287 citations. When we map this subset on its own, we get a finer level of resolution. The 10 modules that correspond to the social sciences now are partitioned into 54 modules, but for simplicity we show only links which the random surfer visits at least 1/2000'th of her time together with the modules they connect (see supporting online material for the complete list). For this particular level of detail we capture 97\% of the node weights and 90\% of all flow.\label{fig4}}
\end{figure*}

Using the information theoretic approach presented above, we map the flow of citations among 6,128 journals in the sciences (Fig. \ref{fig3}) and social sciences (Fig. \ref{fig4}). The 6,434,916 citations in this cross-citation network represent a trace of the scientific activity during 2004 \cite{jsr}. Our data tally on a journal-by-journal basis the citations from articles published in 2004 to articles published in the previous five years. We exclude journals that that publish fewer than 12 articles per year, and those which do not cite other journals within the data set. We also exclude the only three major journals that span a broad range of scientific disciplines:  \emph{Science}, \emph{Nature}, and \emph{Proceedings of the National Academy of Sciences}; the broad scope of these journals otherwise creates an illusion of tighter connections among disciplines, when in fact few readers of the physics articles in \emph{Science} are also close readers of the biomedical articles therein. Because we are interested in relationships between journals, we also exclude journal self-citations. 

Through the operation of our algorithm, the fields and the boundaries between them emerge directly from the citation data, rather than from our preconceived notions of scientific taxonomy (see Figs.~\ref{fig3} and \ref{fig4}). Our only subjective contribution has been to suggest reasonable names for each cluster of journals that the algorithm identifies: economics, mathematics, geosciences, and so forth.

The physical size of each module or ``field'' on the map reflects the fraction of time that a random surfer spends
following citations within that module. Field sizes vary dramatically. Molecular biology includes 723 journals that span the areas of genetics, cell biology, biochemistry, immunology, and developmental biology; a random surfer spends 26\% of her time in this field, indicated by the size of the module. Tribology (the study of friction) includes only 7 journals, in which a random surfer spends 0.064\% of her time.

The weighted and directed links between fields represent citation flow, with the color and width of the arrows indicating flow volume. The heavy arrows between medicine and molecular biology indicate a massive traffic of citations between these disciplines. The arrows point in the direction of citation: $A \rightarrow B$ means ``$A$ cites $B$'' as shown in the legend. These directed links reveal the relationship between applied and basic sciences. We find that the former cite the latter extensively, but the reverse is not true, as we see e.g.\ with geotechnology citing geosciences, plastic surgery citing general medicine, and power systems citing general physics. The thickness of the module borders reflect the probability that a random surfer within the module will follow a citation to a journal outside of the module. These outflows show a large variation; for example the outflow is 30\% in general medicine but only 12\% in economics.

The map reveals a ring-like structure in which all major disciplines are connected to one another by chains of citations --- but these connections are not always direct, because fields on opposite sides of the ring are linked only through intermediate fields. For example, while psychology rarely cites general physics or visa versa, psychology and general physics are connected via the strong links to and between the intermediaries molecular biology and chemistry. Once we consider the weights of the links among fields, however, it becomes clear that the structure of science is more like the letter $\mathbf{U}$ than like a ring, with the social sciences at one terminal and engineering at other, joined mainly by a backbone of medicine, molecular biology, chemistry, and physics. Because our map shows the pattern of citations to research articles published within five years, it represents what de Sola Price called the ``research frontier,'' \cite{price} rather than the long-term interdependencies among fields.
For example, while mathematics are essential to all natural sciences,
the field of mathematics is not central in our map because only certain subfields
(e.g.\ areas of physics and statistics) rely heavily on the most recent
developments in pure mathematics and contribute in return to the
research agenda in that field.

When a cartographer designs a map, the scale or scope of the map influences the choice of which objects are represented. A regional map omits many of the details that appear on a city map.  Similarly, in the approach that we have developed here, the appropriate size or resolution of the modules depends on the universe of nodes that are included in the network. If we compare the map of a network to a map of a subset of the same network, we would expect to see the map of the subset to reveal finer divisions, with modules composed of fewer nodes. Figure \ref{fig4} illustrates this by partitioning a subset of the journals included in the map of science: the 1,431 journals in the the social sciences.  The basic structure of the fields and their relations remains unchanged, with psychiatry and psychology linked via sociology and management to economics and political science, but the map also reveals further details. Anthropology fractures along the physical / cultural divide. Sociology divides into behavioral and institutional clusters. Marketing secedes from management. Psychology and psychiatry reveal a set of applied subdisciplines. 

The additional level of detail in the more narrowly focused map would have been clutter on the full map of science.
When we design maps to help us comprehend the world, we must find that balance where we eliminate extraneous detail but highlight the relationships among important structures. Here we have shown how to formalize this cartographer's precept using the mathematical apparatus of information theory.

\begin{acknowledgments}
We are grateful to Jevin West for processing the data used to construct the maps in Figs.~3 and 4, and to Cynthia A. Brewer, \texttt{www.ColorBrewer.org},
for providing the color schemes we have used in Figs.~1--4. This work was supported by the National Institute of General Medical Sciences Models of Infectious Disease Agent Study program cooperative agreement 5U01GM07649.
\end{acknowledgments}

\clearpage
\renewcommand{\figurename}{Fig.\rule{-0.6em}{0cm}}
\renewcommand{\thefigure}{}
\twocolumngrid
\noindent \textbf{\large Supporting material}\\

\bigskip

\noindent \textbf{Here we present a detailed description of the map equation that serves as our objective function, followed by a step-by-step overview of the computational procedure that we use to minimize it. Thereafter we illustrate the procedure for the greedy search in 22 subsequent slides, which show how the method operates to find successively shorter encodings that highlight and exploit regularities in the network structure. Finally we present the maps from Figs.~3 and 4 in the paper, along with a listing of the complete set of fields and a listing of all journals within the fields.
}

\section*{The map equation}
\noindent Define a {\em module partition} $\mathsf{M}$ as a hard partition of a set of $n$ nodes into $m$ modules such that each node is assigned to one and only one module. The map equation $L(\mathsf{M})$ gives the average number of bits per step that it takes to describe an infinite random walk on a network partitioned according to  $\mathsf{M}$:
\begin{align}
L(\mathsf{M}) = {\color{modulecolor}q_{\curvearrowright} H(\mathcal{Q})} {\color{nodecolor} + \sum_{i=1}^{m}p_{\circlearrowright}^iH(\mathcal{P}^i)}.
 \end{align}
Below, we define and expand these terms, but first a note about the general approach. The map equation calculates the minimum description length of a random walk on the network for a two-level code that separates the important structures from the insignificant details based on the partition $\mathsf{M}$. As described in the main text, this two-level code uses unique codewords to name the modules specified by partition $\mathsf{M}$ but reuses the codewords used to name the individual nodes within each module. The first term of this equation ({\color{modulecolor}in red}) gives the average number of bits necessary to describe movement between modules, and the second term ({\color{nodecolor}in blue}) gives the average number of bits necessary to describe movement within modules. 

To efficiently describe a random walk using a two-level code of this sort, the choice of partition $\mathsf{M}$ must reflect the patterns of flow within the network, with each module corresponding to a cluster of nodes in which a random walker spends a long period of time before departing for another module. To find the best such partition, we therefore seek to minimize the map equation over all possible partitions $\mathsf{M}$. 
We begin by expanding the terms in the map equation. For clarity we here use $i,j$ to enumerate modules, $\alpha,\beta$ to enumerate nodes, {\color{modulecolor}red terms to describe movements between the modules}, and {\color{nodecolor}blue terms to describe movements within the modules}.
\bigskip

\noindent The per step probability that the random walker switches modules is 
\begin{align}\label{map1}
{\color{modulecolor}q_{\curvearrowright}=\sum_{i=1}^m q_{i \curvearrowright}},
\end{align}
where $q_{i \curvearrowright}$ is the per step probability that the random walker exits module $i$. This probability depends on the partitioning of the network and will be derived in Eq.~\ref{q_jump}.

The entropy of movements between modules is
\begin{align}\label{map2}
{\color{modulecolor}H(\mathcal{Q})= \sum_{i=1}^m\frac{q_{i \curvearrowright}}{\sum_{j=1}^m q_{j \curvearrowright}}\log \left( \frac{q_{i \curvearrowright}}{\sum_{j=1}^m q_{j \curvearrowright}} \right)},
\end{align}
which is the lower limit of the average length of a codeword used to name a module. Here we have used Shannon's source coding theorem \cite{sup-shannon48} and treated the modules as $m$ states of a random variable $X$ that occur with frequencies $q_{i\curvearrowright}/\sum_{j=1}^m q_{j \curvearrowright}$. Combining Eqs.~\ref{map1} and \ref{map2}, the first term in the map equation is the per step average description length of movements between modules within the random walk.
\bigskip

\noindent To weight the entropy of movements within module $i$, we compute
\begin{align}\label{map3}
{\color{nodecolor}p_{\circlearrowright}^i = q_{i \curvearrowright }+\sum_{\alpha \in i} p_\alpha},
\end{align}
where the notation $\alpha \in i$ means ``over all nodes $\alpha$ in module $i$'' and $p_\alpha$ is the ergodic node visit frequency at node $\alpha$ within the random walk. We use the power method, to be explained in detail on next page, to calculate this probability. Because the exit codewords are necessary to separate within-module movements from between-module movements, we include the probability of exiting module $i$, $q_{i\curvearrowright}$, in the weight of within-module movements in module $i$.
In this way we can guarantee efficient coding: by encoding the exit codewords together with the within-module codewords, we appropriately adjust the length of the exit codewords to the frequency of their use.

Finally, the entropy of movements within module $i$ is
\begin{subequations}
\label{map4}
\begin{align}
{\color{nodecolor}H(\mathcal{P}^i)} &= {\color{nodecolor}\frac{q_{i \curvearrowright} }{q_{i \curvearrowright }+\sum_{\beta \in i} p_\beta} \log \left( \frac{q_{i \curvearrowright} }{q_{i \curvearrowright }+\sum_{\beta \in i} p_\beta} \right)}\label{map4a}\\
&{\color{nodecolor}+ \sum_{\alpha \in i}\frac{p_\alpha}{q_{i \curvearrowright }+\sum_{\beta \in i} p_\beta}\log \left( \frac{p_\alpha}{q_{i \curvearrowright }+\sum_{\beta \in i} p_\beta} \right)}\label{map4b}
\end{align}
\end{subequations}
which is the lower limit of the average length of a codeword used to name a node (exit code included) in module $i$.
The single term in Eq.~\ref{map4a} is the contribution from the exit codeword and the sum in Eq.~\ref{map4b} is the contribution from the codewords naming the nodes. Combining Eqs.~\ref{map3} and \ref{map4} and summing over all modules makes it easy to identify the second term in the map equation as the per step average description length of movements within modules of the random walk.
\bigskip

\noindent By collecting the terms and simplifying, we get the final expression for the map equation
\begin{subequations}
\begin{align}
L(\mathsf{M}) &= 
{\color{modulecolor}\left(\sum_{i=1}^m q_{i \curvearrowright}\right) \log \left( \sum_{i=1}^m q_{i \curvearrowright} \right)}\\
&- ({\color{modulecolor}1} + {\color{nodecolor}1}) \sum_{i=1}^m q_{i \curvearrowright}\log \left( q_{i \curvearrowright} \right) 
{\color{nodecolor} - \sum_{\alpha=1}^{n} p_\alpha \log \left( p_\alpha \right)}\label{map5b}\\
&{\color{nodecolor}+ \sum_{i=1}^{m}\left(q_{i \curvearrowright }+\sum_{\alpha \in i} p_\alpha \right) \log \left( q_{i \curvearrowright }+\sum_{\alpha \in i} p_\alpha \right)}.
\end{align}
\end{subequations}
Note that the map equation is only a function of the ergodic node visit frequencies $p_\alpha$ and the exit probabilities $q_{i\curvearrowright}$, which both can be easily calculated. Moreover, because the term ${\color{nodecolor}\sum_{1}^{n} p_\alpha \log \left( p_\alpha \right)}$ is independent of partitioning and $p_\alpha$ otherwise only shows up summed over all nodes in a module, it is sufficient to keep track of changes in $q_{i\curvearrowright}$ and $\sum_{\alpha \in i} p_\alpha$ in the optimization algorithm. They 
can easily be derived for any partition of the network or quickly updated when they change in each step of the optimization procedure using the ergodic node visit frequencies.
%We therefore start our algorithm description below by explaining how we calculate $p_\alpha$.

\begin{itemize}
\item \emph{Ergodic node visit frequencies}.
We use the power method to calculate the steady state
visit frequency for each node.
To guarantee a unique steady state distribution for directed networks,
we introduce a small teleportation probability $\tau$ in the random walk
that links every node to every other node with positive probability
and thereby convert the random walker into a \emph{random surfer}.
The movement of the random surfer can now be
described by an irreducible and aperiodic Markov chain that has a unique
steady state by the Perron-Frobineous theorem.
To generate the ergodic node visit frequencies,
we start with a distribution of $p_\alpha=1/n$
for the random surfer to start at each node $\alpha$. The surfer moves as
follows: at each time step, with probability $1-\tau$ the random
surfer follows one of the outgoing links from the node $\alpha$ that it
currently occupies, with probability proportional to the 
weights of the outgoing links $w_{\alpha\beta}$ from $\alpha$ to $\beta$.
It is therefore convenient to set $\sum_\beta w_{\alpha\beta} = 1$.
With the remaining probability $\tau$, or with probability $1$ if the node does not have any outlinks,
the random surfer ``teleports''  with uniform probability to a random node anywhere in the system.
As in Google's Page\-Rank algorithm \cite{sup-google}, we use $\tau=0.15$,
but emphasize that the results are robust to this choice.
\item \emph{Exit probabilities}.
Given the ergodic node visit frequencies $p_\alpha,\, \alpha=1,\ldots,n$ and an initial partitioning of the network,
it is easy to calculate the ergodic module visit frequencies $\sum_{\alpha \in i} p_\alpha$ for module $i$.
The exit probability for module $i$, with teleportation taken into account is then
\begin{align}\label{q_jump}
q_{i \curvearrowright }=\tau\frac{n-n_i}{n-1}\sum_{\alpha \in i} p_\alpha + (1-\tau)\sum_{\alpha \in i}\sum_{\beta \notin i} p_\alpha w_{\alpha\beta},
\end{align}
where $n_i$ is the number of nodes in module $i$. This equation follows since every node teleports a fraction $\tau(n-n_i)/(n-1)$  and guides a fraction $(1-\tau)\sum_{\beta \notin i} w_{\alpha\beta}$ of its weight $p_\alpha$ to nodes outside of the module.
\end{itemize}

\section*{Implementation}

\noindent Finding the map that provides the minimal description length of the data, given the requirement that modules receive unique names, is now a standard computational optimization problem. Below we describe how we first use a deterministic greedy search algorithm \cite{sup-clauset-2004-70,sup-wakita} and then refine the results with a simulated annealing approach \cite{sup-kirkpatrick,sup-guimera-nature} with the heat-bath algorithm. 

\begin{enumerate}
\item \emph{Greedy search}.
We first calculate the ergodic node visit frequencies and then assign every node to a unique module and derive the
the exit probabilities as described above.
We use the map equation to calculate the description length and
repeatedly merge the two modules that give the largest
decrease in description length until further merging gives a longer description \cite{sup-clauset-2004-70,sup-wakita}.
With the improved version in ref.\ \cite{sup-wakita} of the greedy search algorithm in ref.\ \cite{sup-clauset-2004-70}, we have successfully partitioned networks with 2.6 million nodes and 29 million links.
Starting at the next page, we illustrate the greedy search for the example network in Fig.~1 of the paper.
% In the other extreme in terms of accuracy and speed, we describe a simulated annealing approach which we have used to refine the maps in the paper.
\item \emph{Simulated annealing}.
The result of the previous step can typically be refined by simulated annealing \cite{sup-kirkpatrick,sup-guimera-nature}.
We use the heat-bath algorithm \cite{sup-newmanheatbath} and start with the
module configuration achieved by the greedy search.
Starting the heat-bath algorithm at several different temperatures, we select the run
that gives the shortest description of the map, i.e., the minimal value of the map equation.
This step can improve the description length by up to several percent over that found by
the greedy search alone.
\item \emph{Visualization}. 
 We set the area of every module to be proportional to the fraction of time a random
surfer spends in the module, and the area of the bordering ring to be proportional
to the exit probability. Similarly, we vary the widths of
the links in accord with the transition probabilities between modules (excluding teleportation).
\end{enumerate}

\begin{figure*}
\centering
\includegraphics[width=\textwidth]{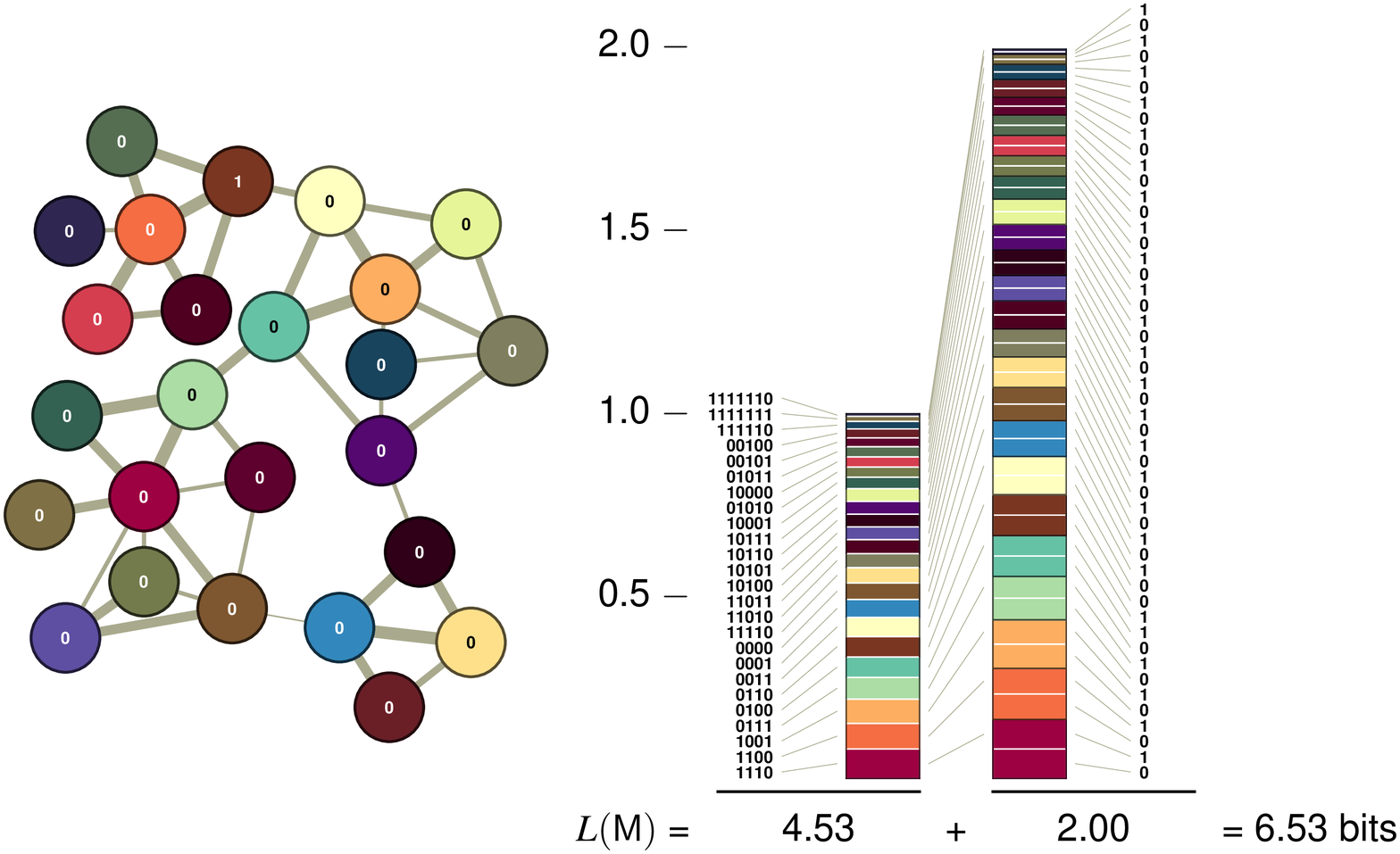}
\caption{S1 The greedy search algorithm begins with each node $\alpha=1,\ldots,n$ in its own module $i=1,\ldots,m$, as shown here.   The slides on the following 22 pages show how the greedy search operates, finding successively shorter encodings that highlight and exploit regularities in the network structure. Node colors indicate module identity. To illustrate the duality between module detection and coding, the code structure corresponding to the current partition is shown by the stacked boxes on the right. The height of each box corresponds to the per step rate of codeword use.\\
The left stack represents the codes associated with movements between modules. The height of each box is equal to the exit probability $q_{i \curvearrowright }$ of the corresponding module $i$. Boxes are ordered according to their heights. The codewords naming the modules are the Huffman codes calculated from the probabilities $q_{i \curvearrowright } / q_{\curvearrowright }$, where $q_{\curvearrowright }=\sum_1^m q_{i \curvearrowright }$ is the total height of the left stack. The length of the codeword naming module $i$ is approximately $-\log{q_{i \curvearrowright } / q_{\curvearrowright }}$, the Shannon limit in the map equation. In the figure, the per step description length of the random walker's movements between modules is the sum of the length of the codewords weighted by their use. This length is bounded below by the limit $-\sum_1^m q_{i \curvearrowright }\log{q_{i \curvearrowright } / q_{\curvearrowright }}$ that we use in the paper.\\
The right stack is associated with movements within modules. The height of each box in module $i$ is equal to the ergodic node visit probabilities $p_{\alpha \in i}$ or the exit probability $q_{i \curvearrowright }$. The boxes corresponding to the same module are collected together and ordered according to their weight; in turn the modules are ordered according to their total weights $p_{\circlearrowright}^i = q_{i \curvearrowright } + \sum_{\alpha \in i} p_\alpha$. The codewords naming the nodes and exit in each module $i$ are the Huffman codes calculated from the probabilities $p_{\alpha \in i}/p_{\circlearrowright}^i$ (nodes) and $q_{i \curvearrowright }/p_{\circlearrowright}^i$ (exit). The length of codewords naming nodes $\alpha \in i$ and exit from module $i$ are approximately $-\log(p_{\alpha \in i} /p_{\circlearrowright}^i)$ (nodes) and $-\log(q_{i \curvearrowright }/p_{\circlearrowright}^i)$ (exit). In the figure, the per step description length of the random walker's movements within modules is the sum of the length of the codewords weighted by their use. This length is bounded below by the limit $-\sum_1^m \left[q_{i \curvearrowright }\log(q_{i \curvearrowright } / p_{\circlearrowright}^i) + \sum_{\alpha \in i}p_\alpha\log(p_\alpha / p_{\circlearrowright}^i)\right]$.\\
The total description length $L(\mathsf{M})$ for the module partition $\mathsf{M}$ of the network is the sum of the contributions from movements between and within modules and it takes a minimum value for the partitioning into four models on the last slide.
%This also highlights the duality between finding community structure in networks and the coding problem of minimizing the expected description length of a random walk on the network.
}
\end{figure*}

\pagestyle{empty}
\clearpage
\begin{center}
\includegraphics[width=1.5\textwidth,angle=90,bb=-784 -614 325 36]{net25.eps}
\end{center}
\clearpage
\begin{center}
\includegraphics[width=1.5\textwidth,angle=90,bb=-784 -614 325 36]{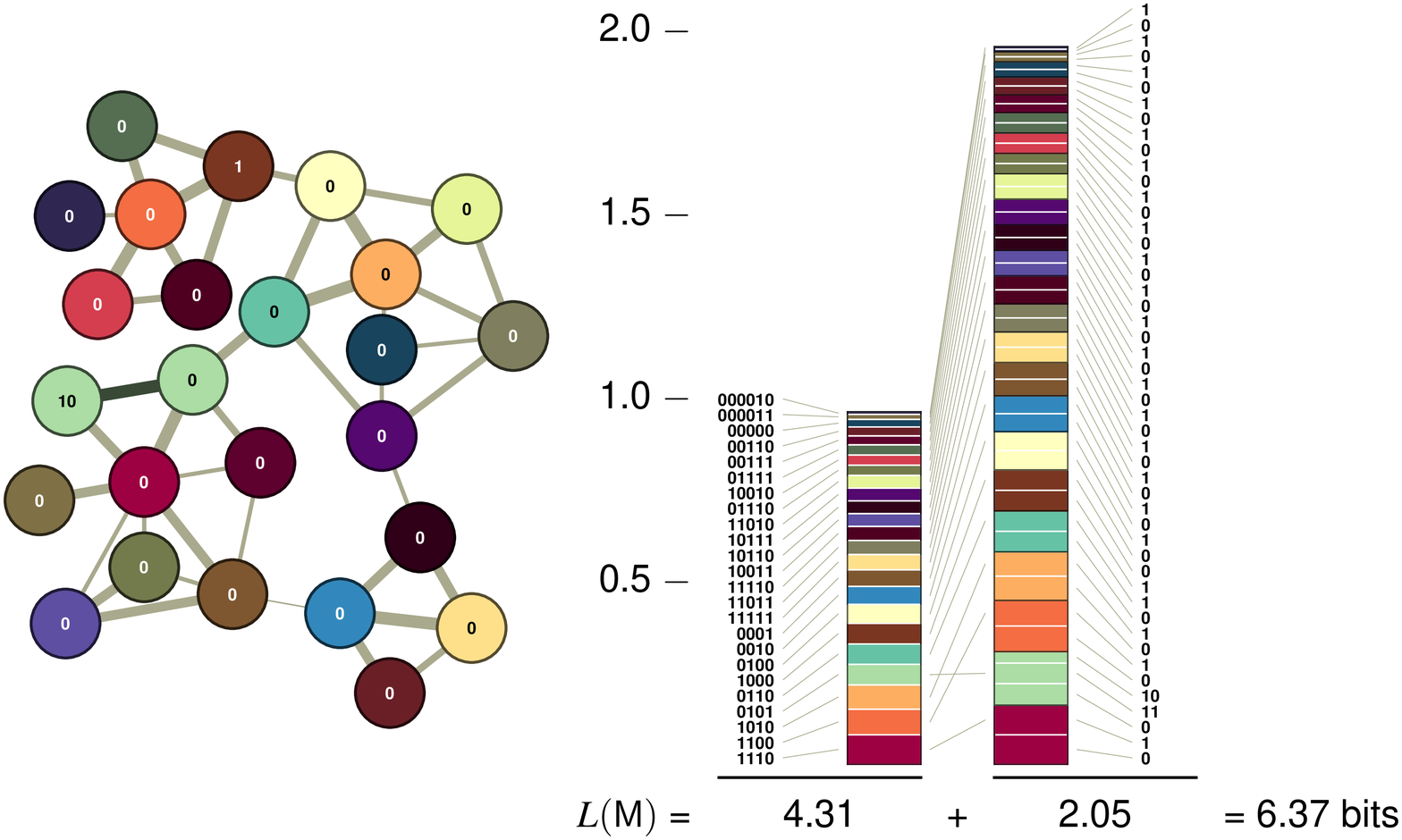}
\end{center}
\clearpage
\begin{center}
\includegraphics[width=1.5\textwidth,angle=90,bb=-784 -614 325 36]{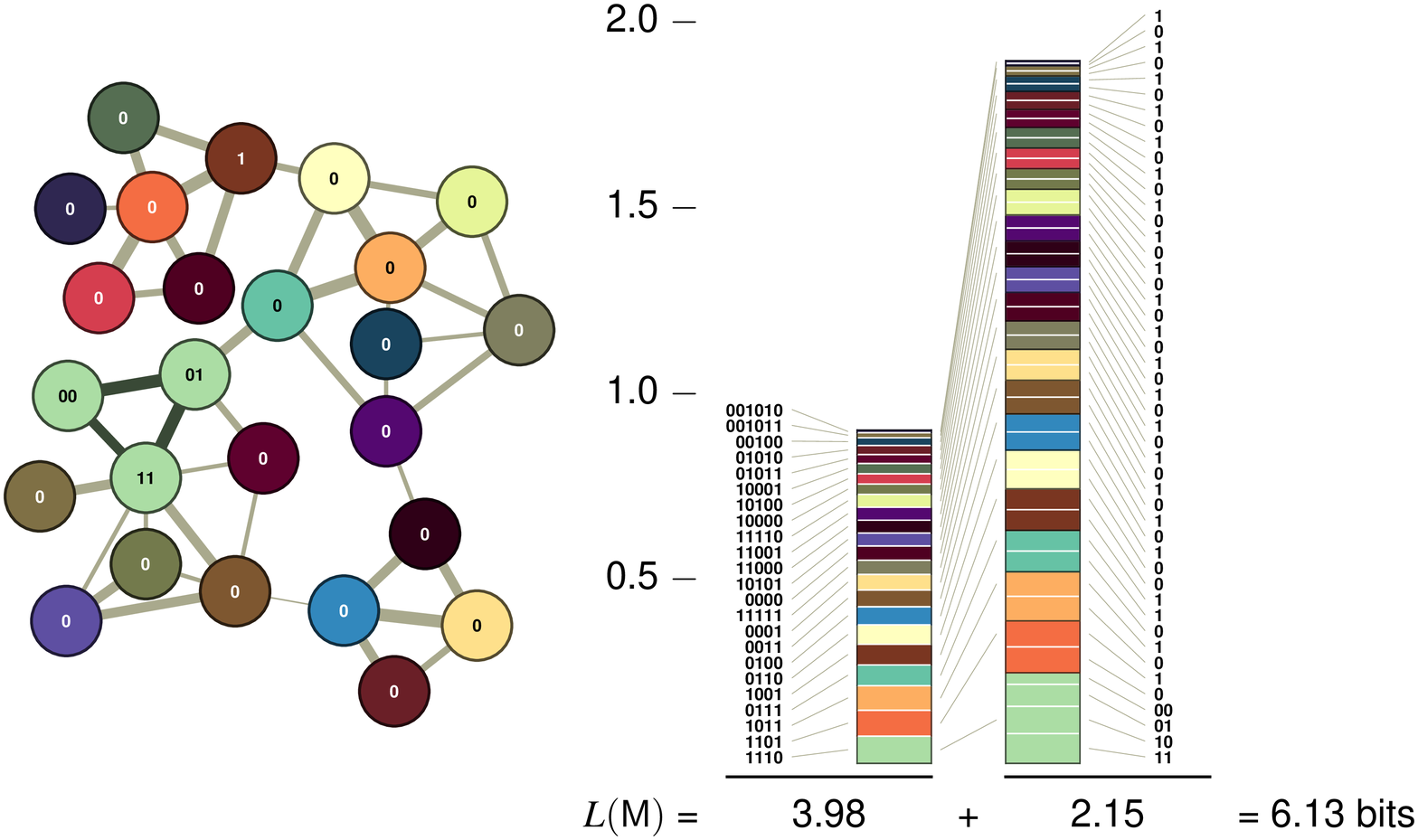}
\end{center}
\clearpage
\begin{center}
\includegraphics[width=1.5\textwidth,angle=90,bb=-784 -614 325 36]{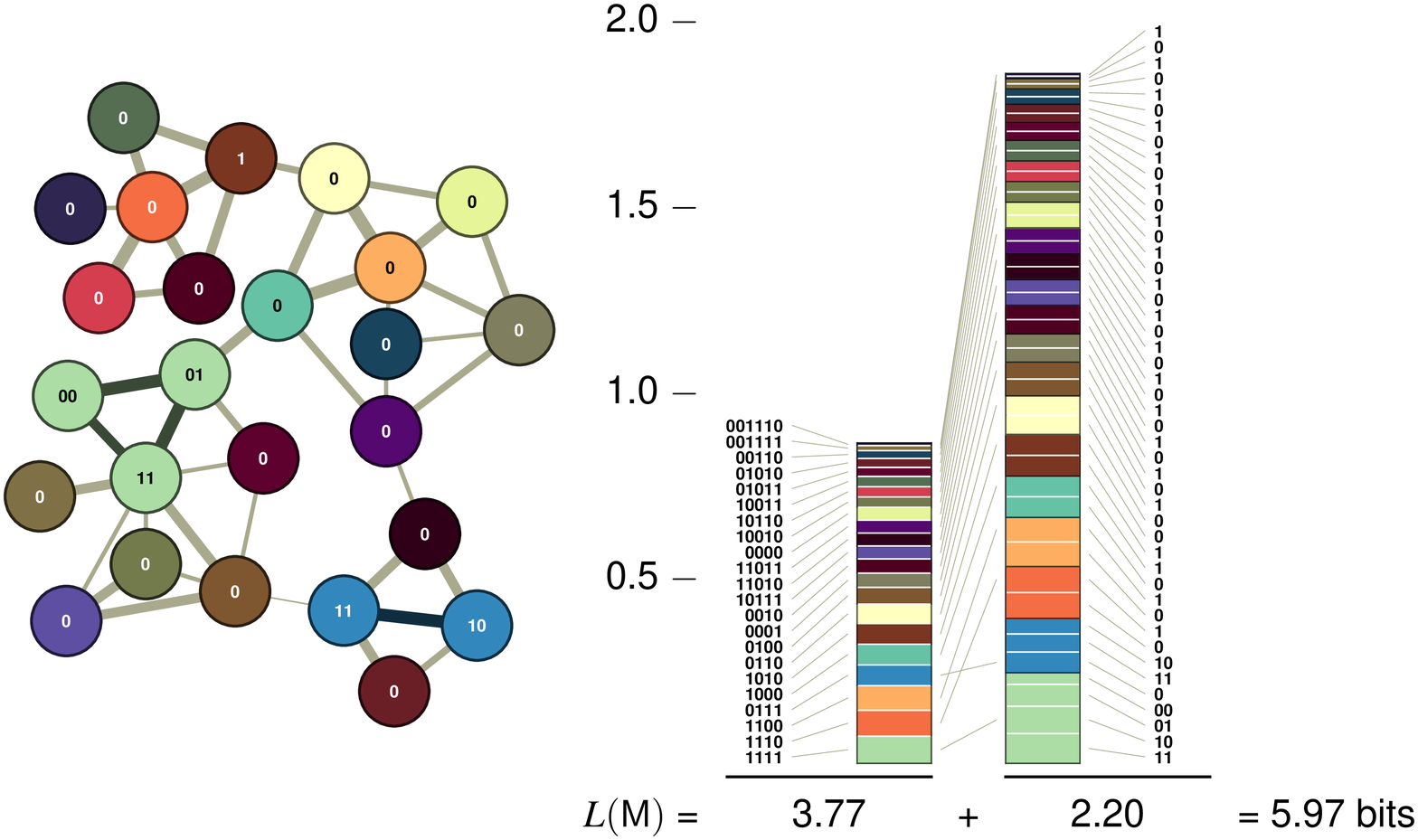}
\end{center}
\clearpage
\begin{center}
\includegraphics[width=1.5\textwidth,angle=90,bb=-784 -614 325 36]{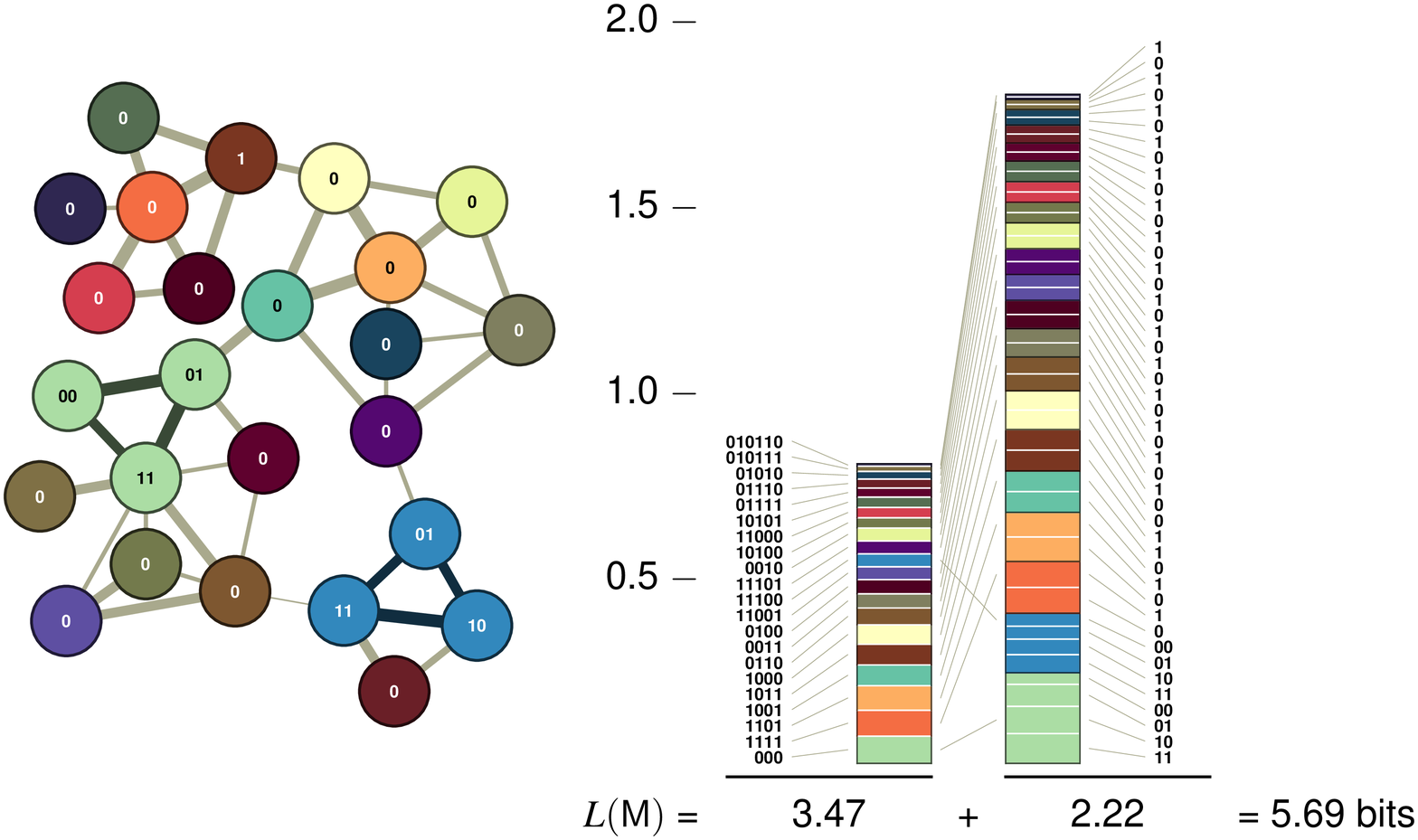}
\end{center}
\clearpage
\begin{center}
\includegraphics[width=1.5\textwidth,angle=90,bb=-784 -614 325 36]{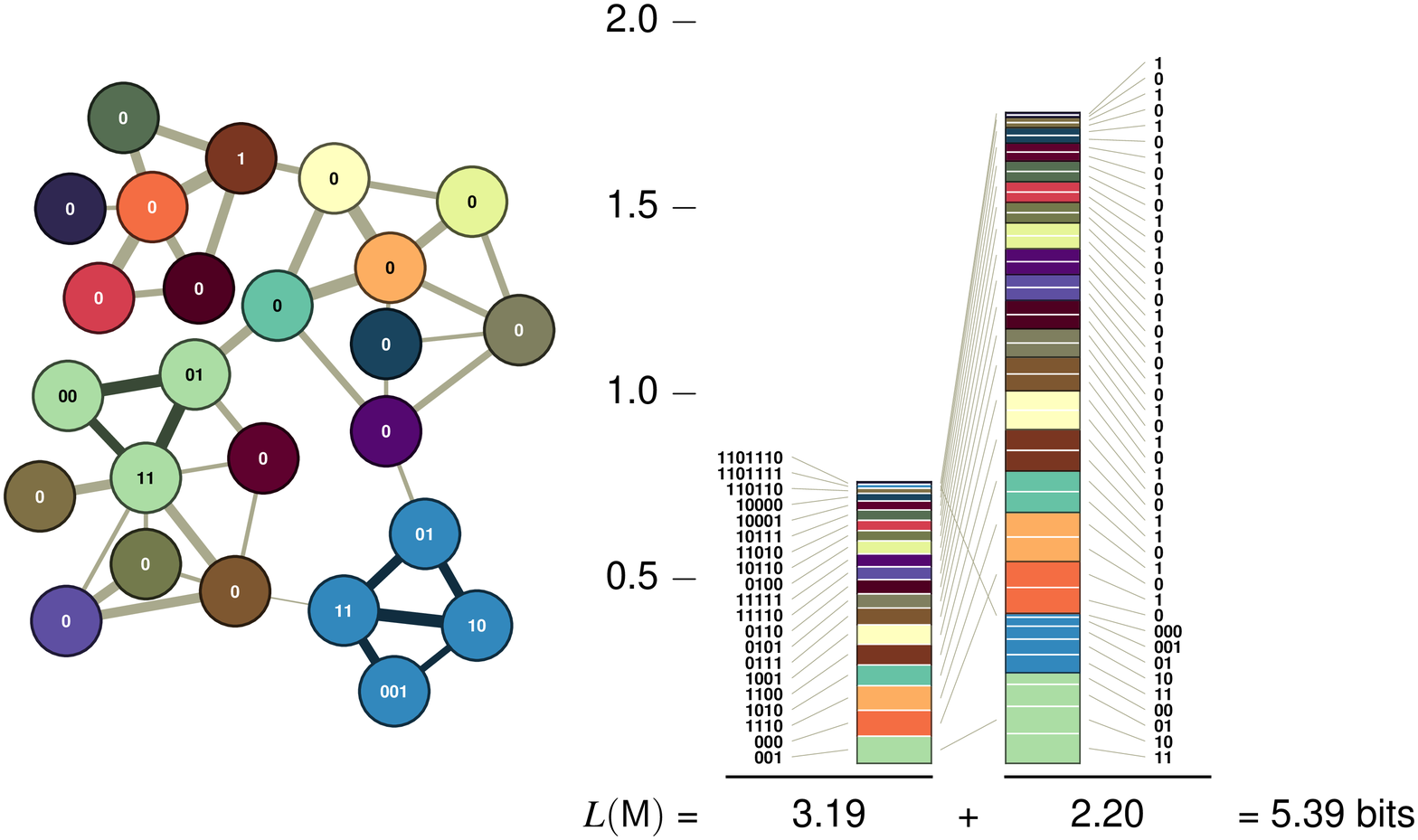}
\end{center}
\clearpage
\begin{center}
\includegraphics[width=1.5\textwidth,angle=90,bb=-784 -614 325 36]{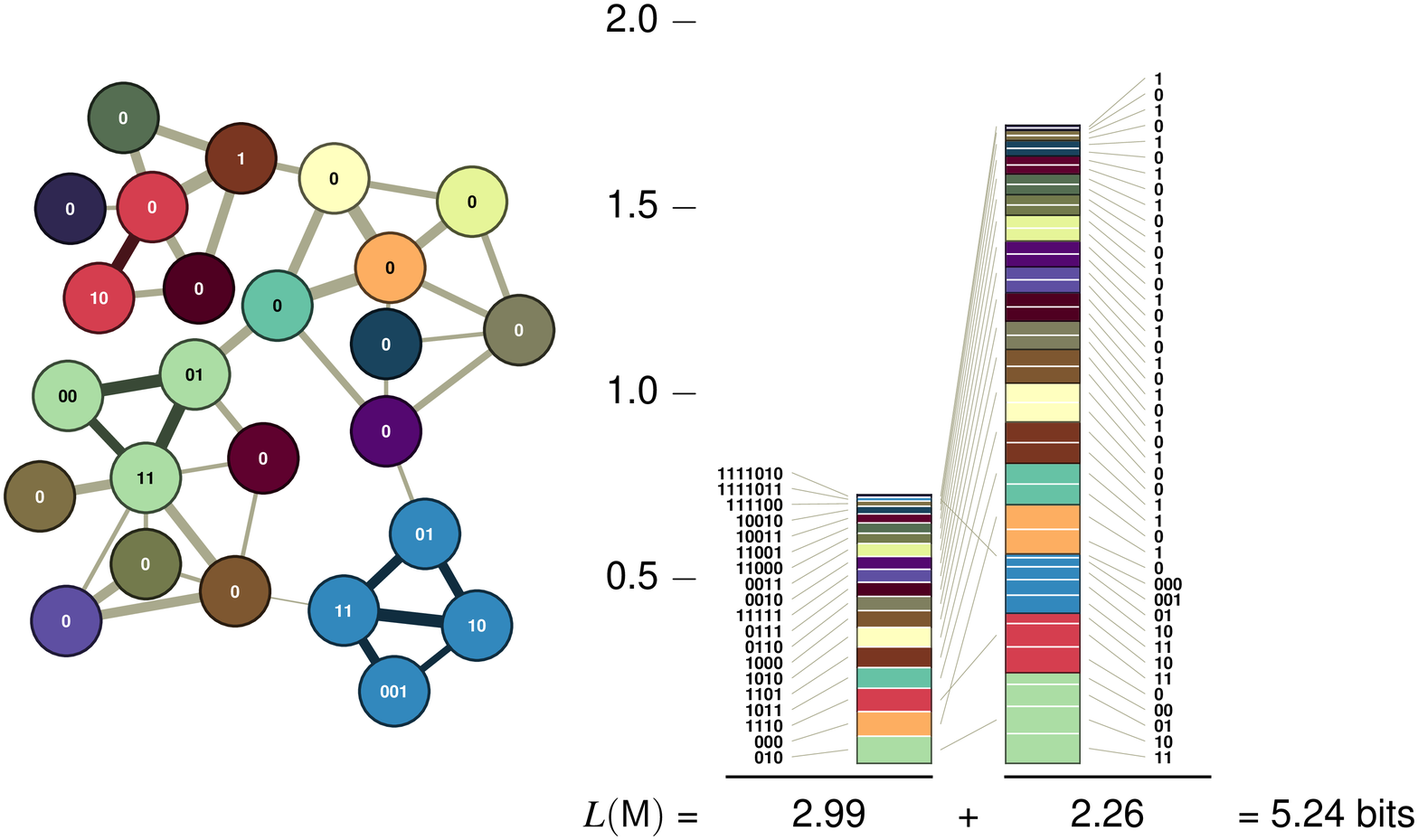}
\end{center}
\clearpage
\begin{center}
\includegraphics[width=1.5\textwidth,angle=90,bb=-784 -614 325 36]{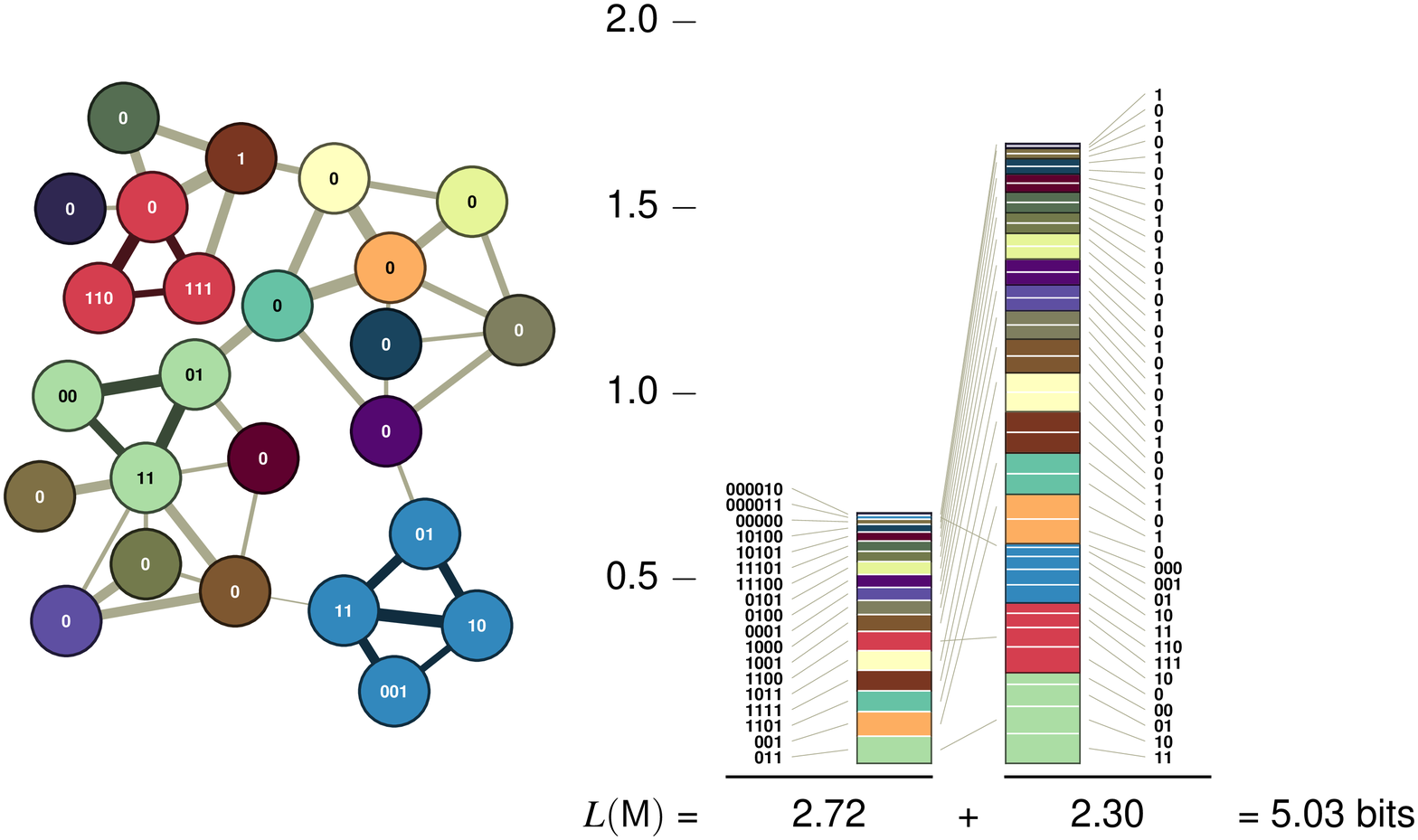}
\end{center}
\clearpage
\begin{center}
\includegraphics[width=1.5\textwidth,angle=90,bb=-784 -614 325 36]{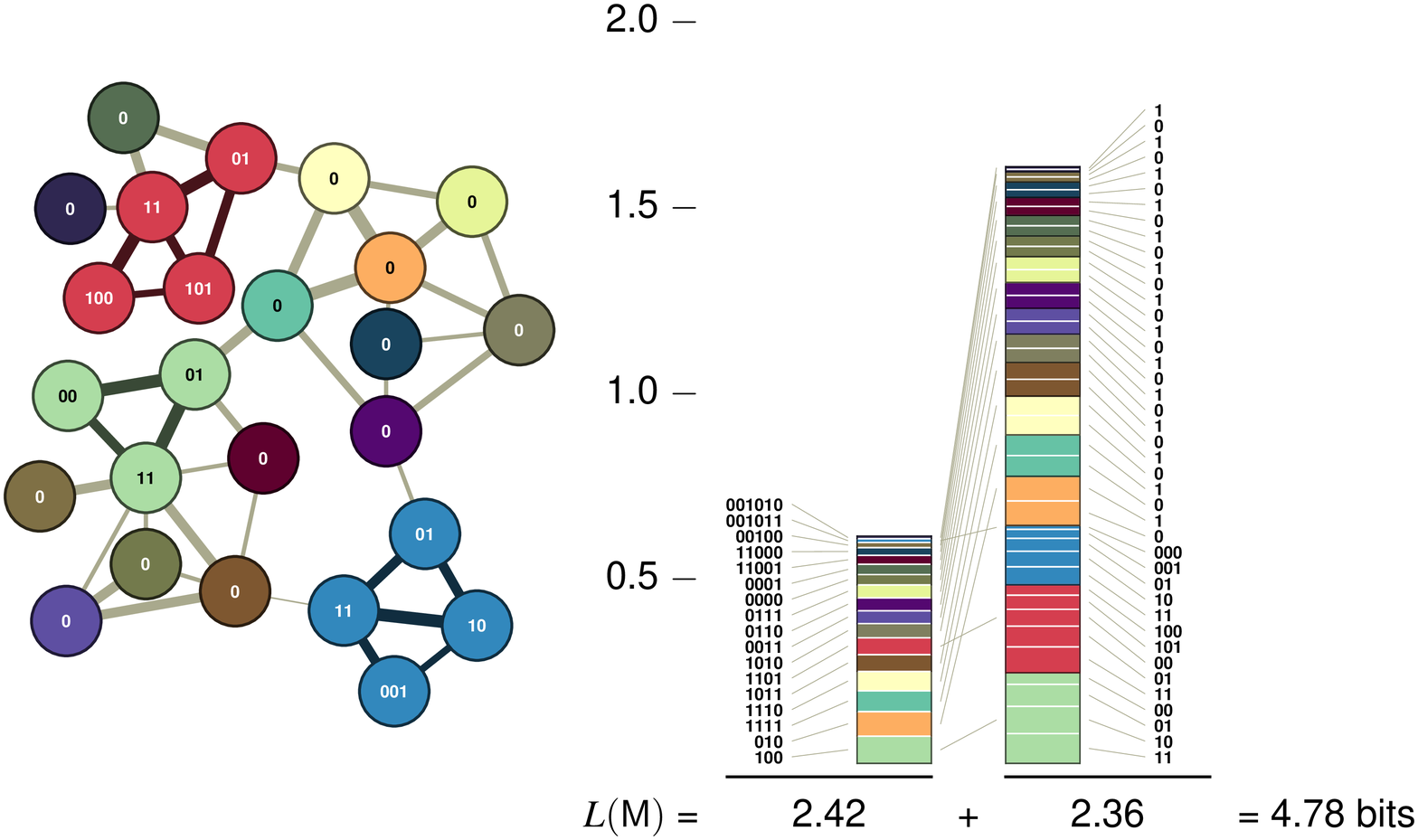}
\end{center}
\clearpage
\begin{center}
\includegraphics[width=1.5\textwidth,angle=90,bb=-784 -614 325 36]{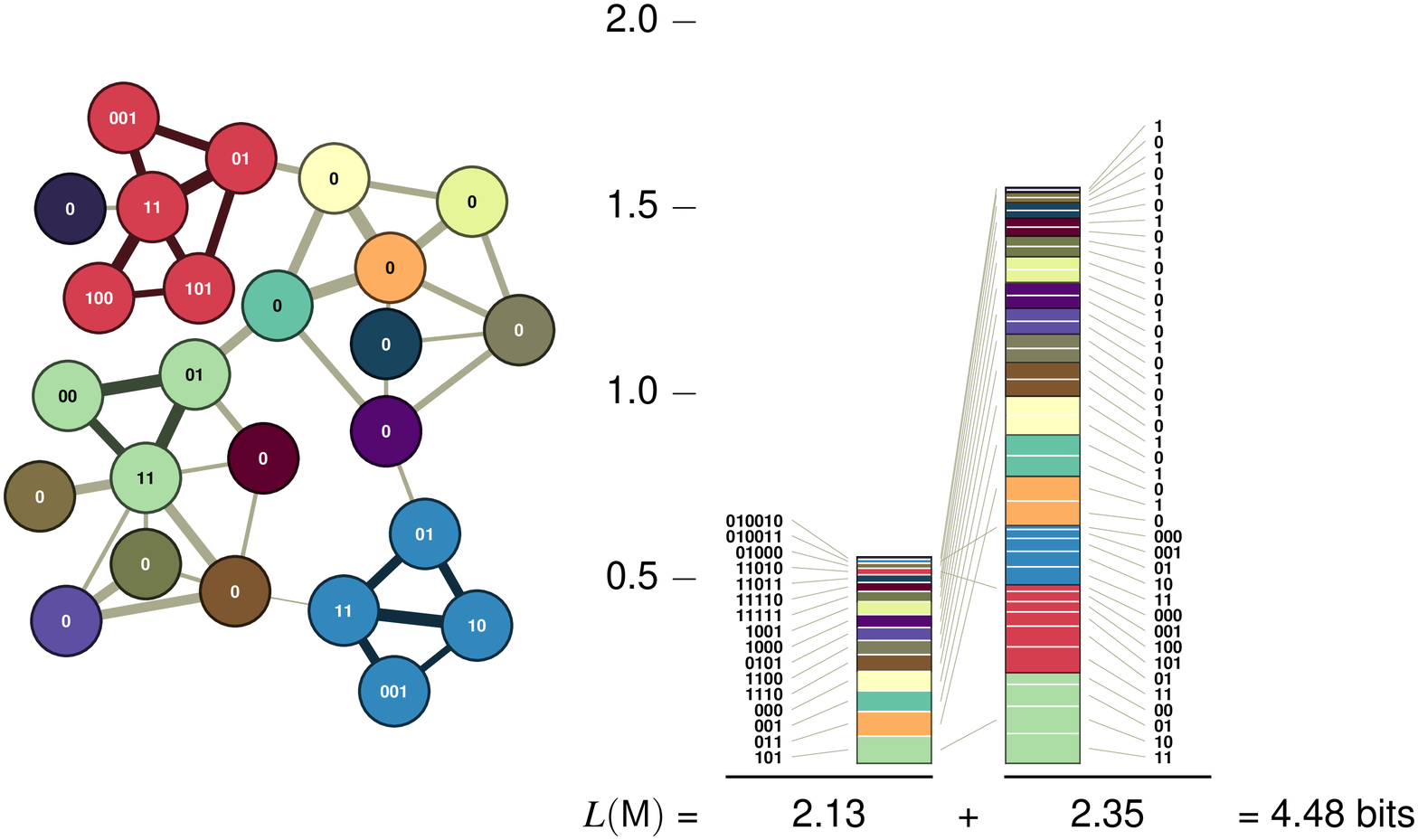}
\end{center}
\clearpage
\begin{center}
\includegraphics[width=1.5\textwidth,angle=90,bb=-784 -614 325 36]{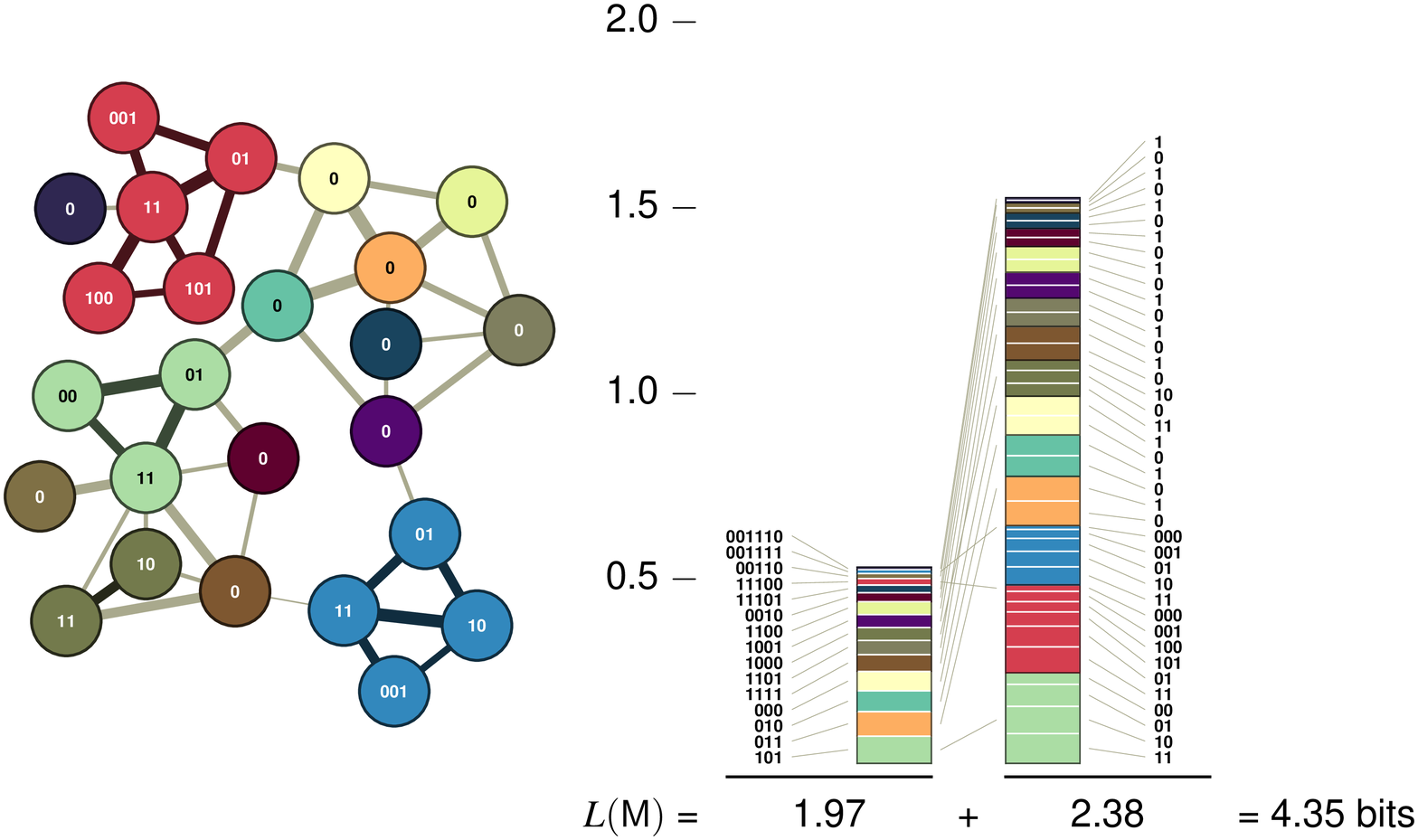}
\end{center}
\clearpage
\begin{center}
\includegraphics[width=1.5\textwidth,angle=90,bb=-784 -614 325 36]{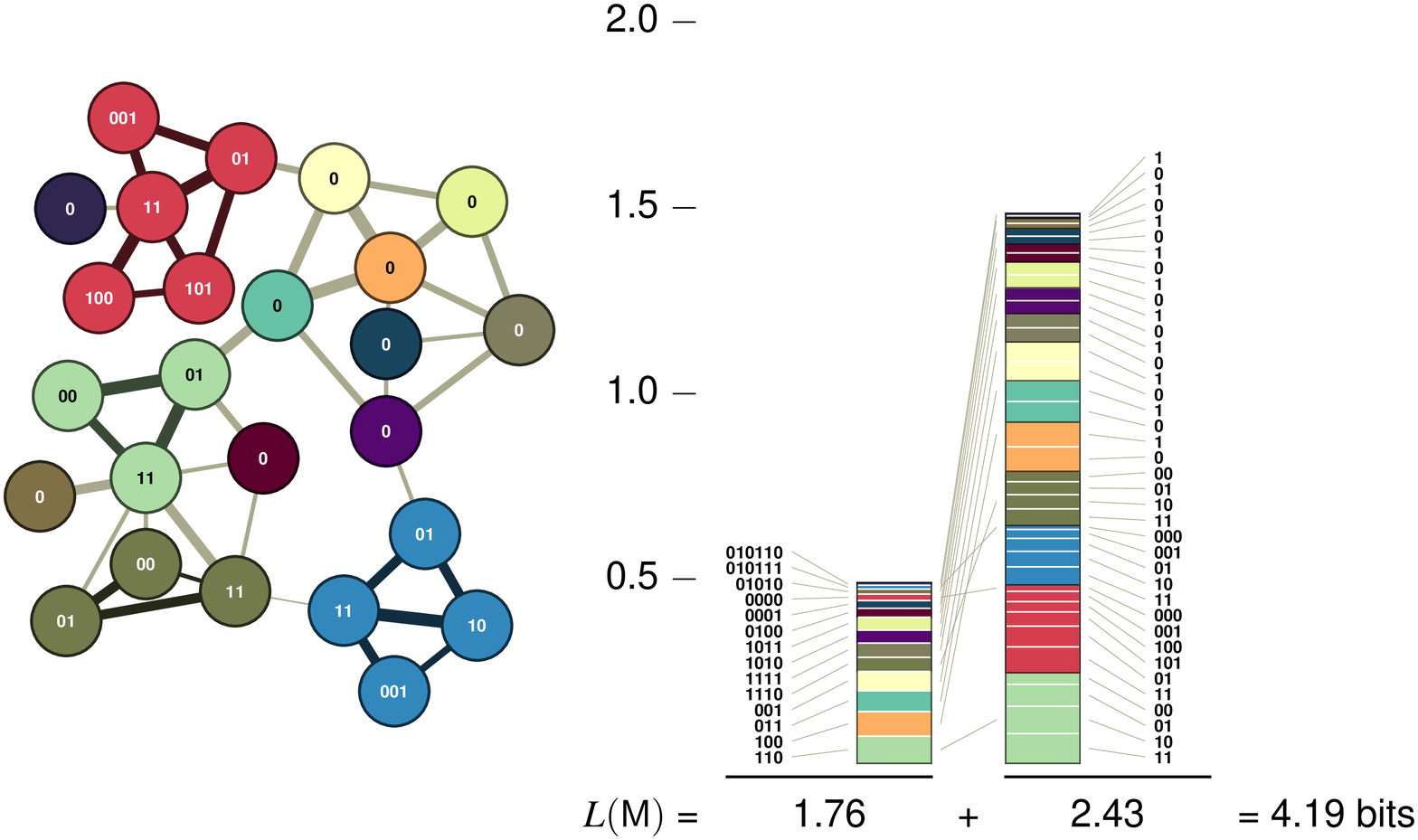}
\end{center}
\clearpage
\begin{center}
\includegraphics[width=1.5\textwidth,angle=90,bb=-784 -614 325 36]{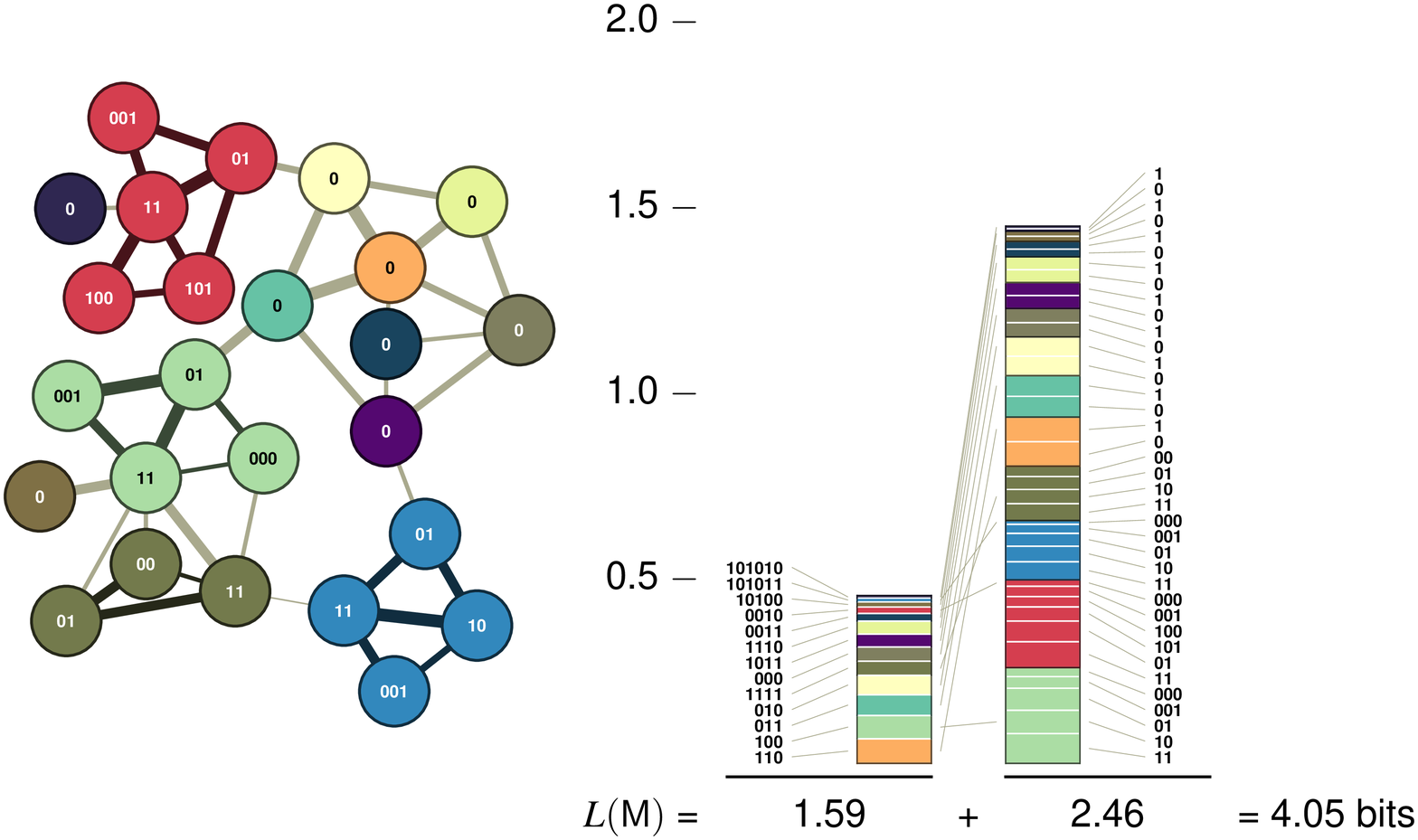}
\end{center}
\clearpage
\begin{center}
\includegraphics[width=1.5\textwidth,angle=90,bb=-784 -614 325 36]{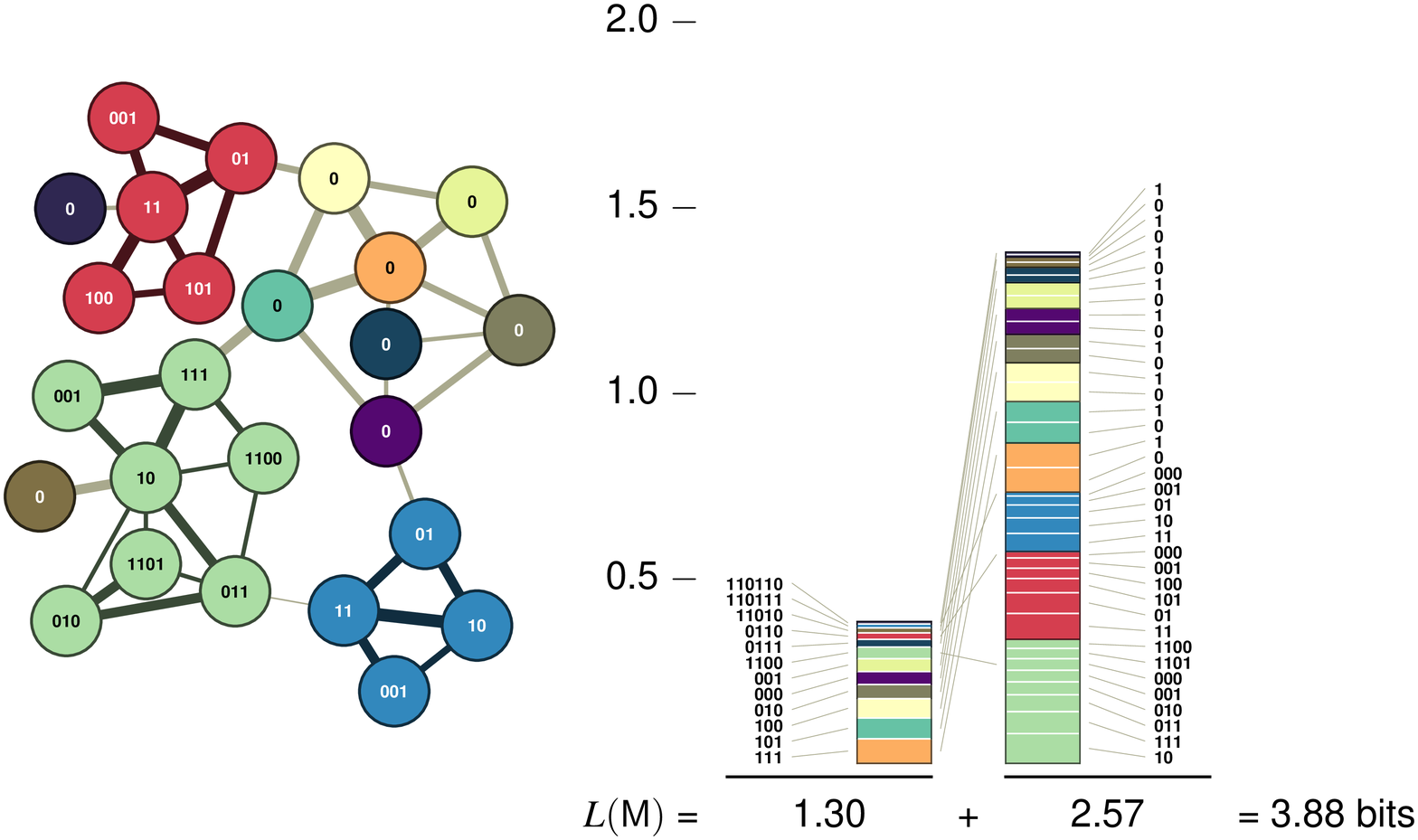}
\end{center}
\clearpage
\begin{center}
\includegraphics[width=1.5\textwidth,angle=90,bb=-784 -614 325 36]{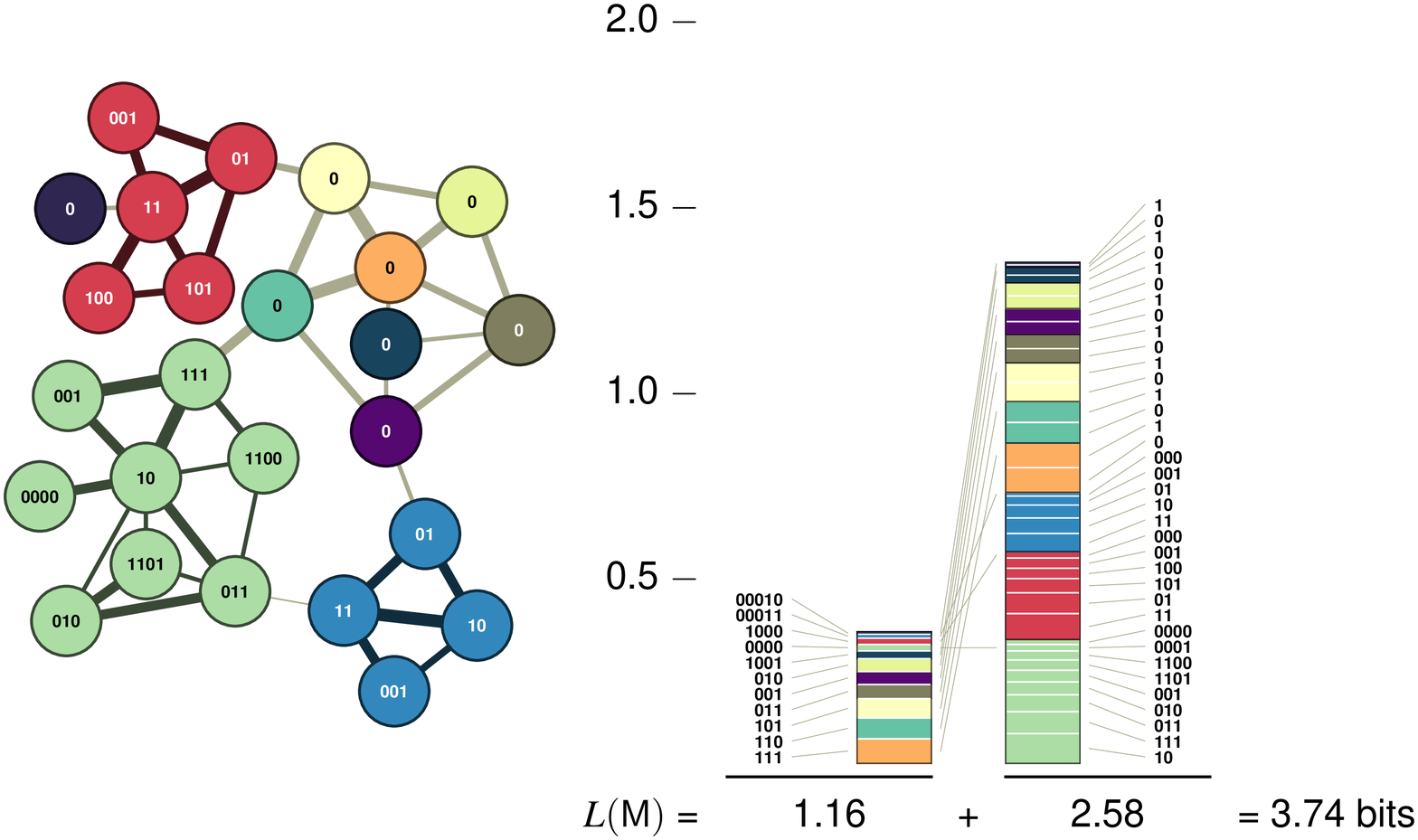}
\end{center}
\clearpage
\begin{center}
\includegraphics[width=1.5\textwidth,angle=90,bb=-784 -614 325 36]{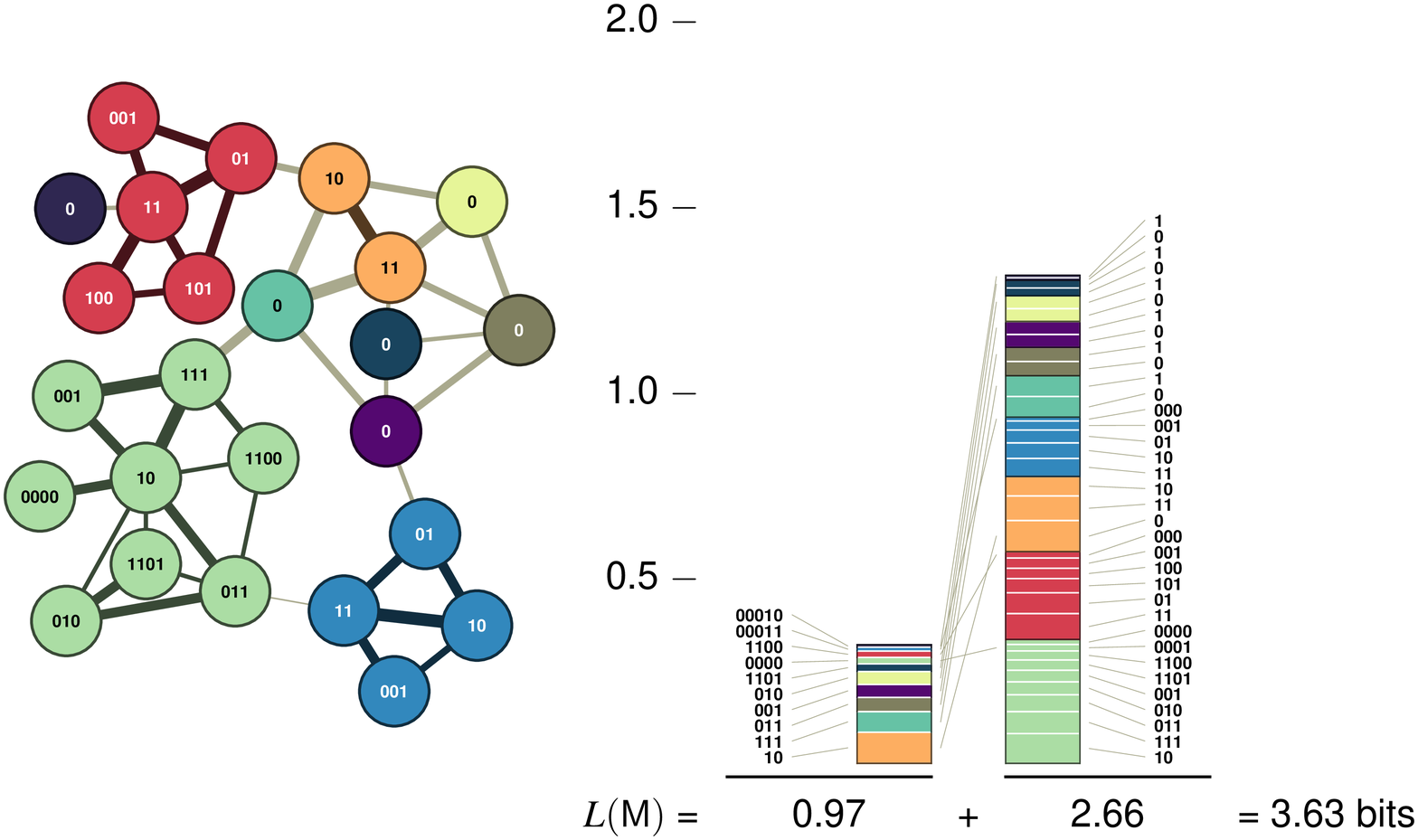}
\end{center}
\clearpage
\begin{center}
\includegraphics[width=1.5\textwidth,angle=90,bb=-784 -614 325 36]{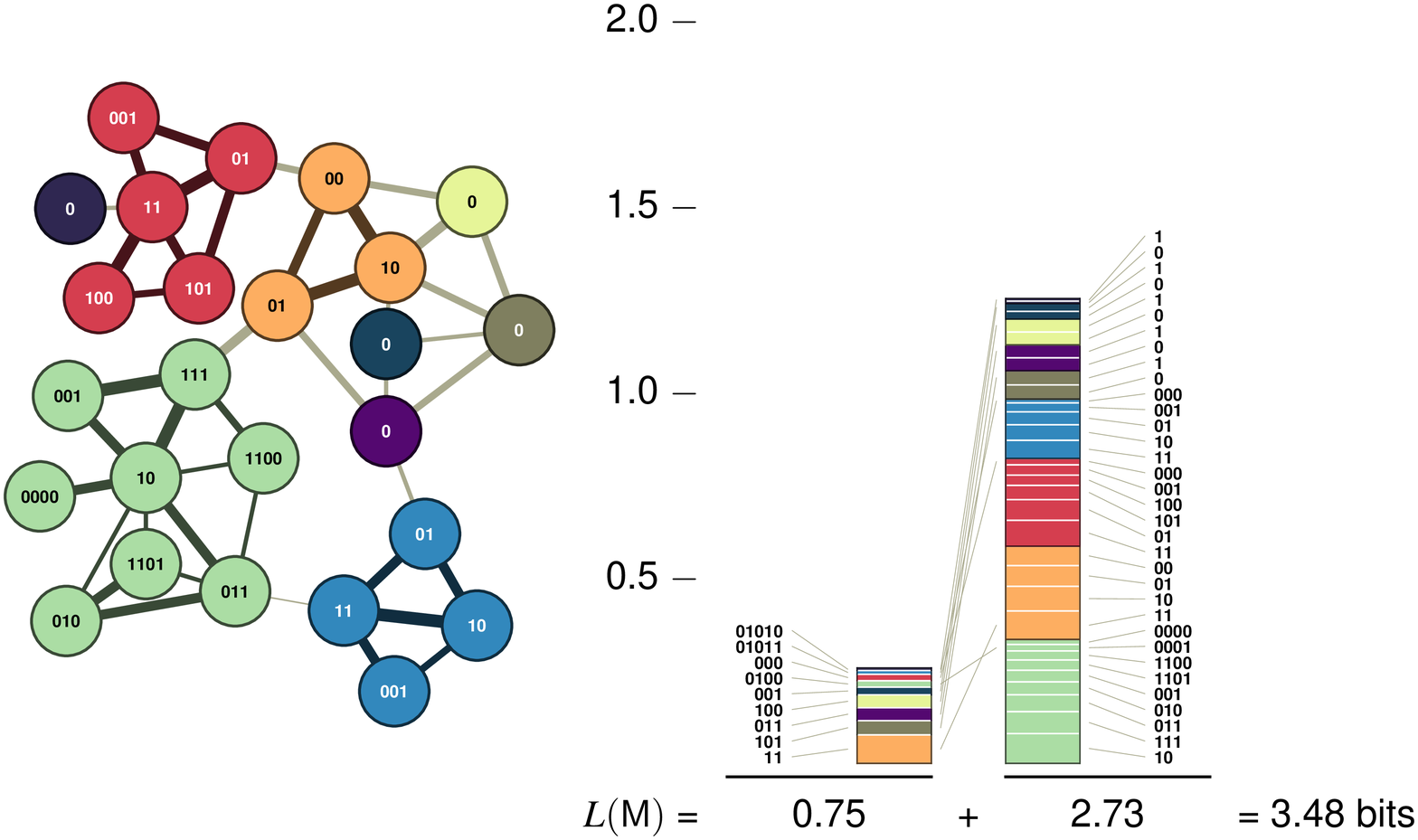}
\end{center}
\clearpage
\begin{center}
\includegraphics[width=1.5\textwidth,angle=90,bb=-784 -614 325 36]{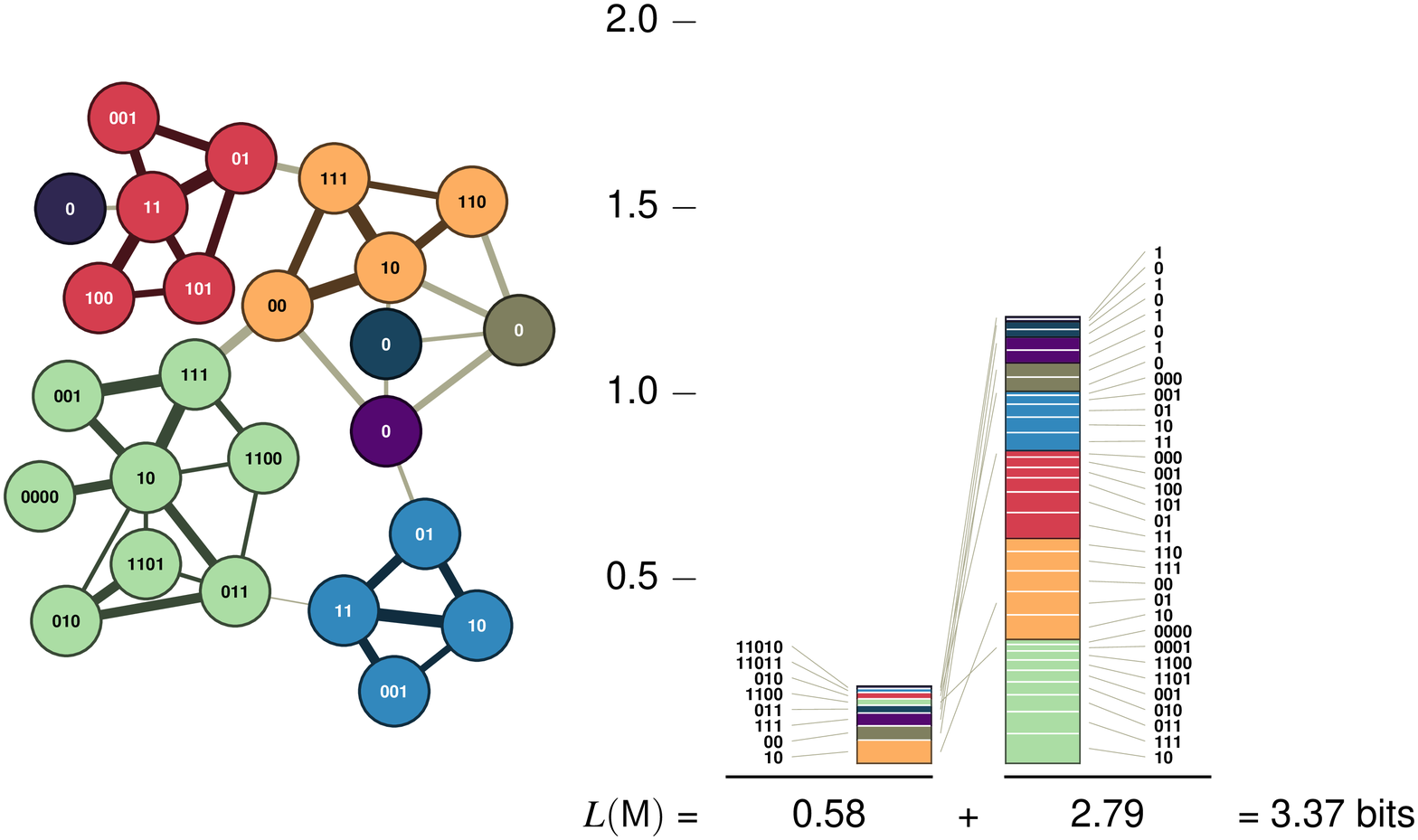}
\end{center}
\clearpage
\begin{center}
\includegraphics[width=1.5\textwidth,angle=90,bb=-784 -614 325 36]{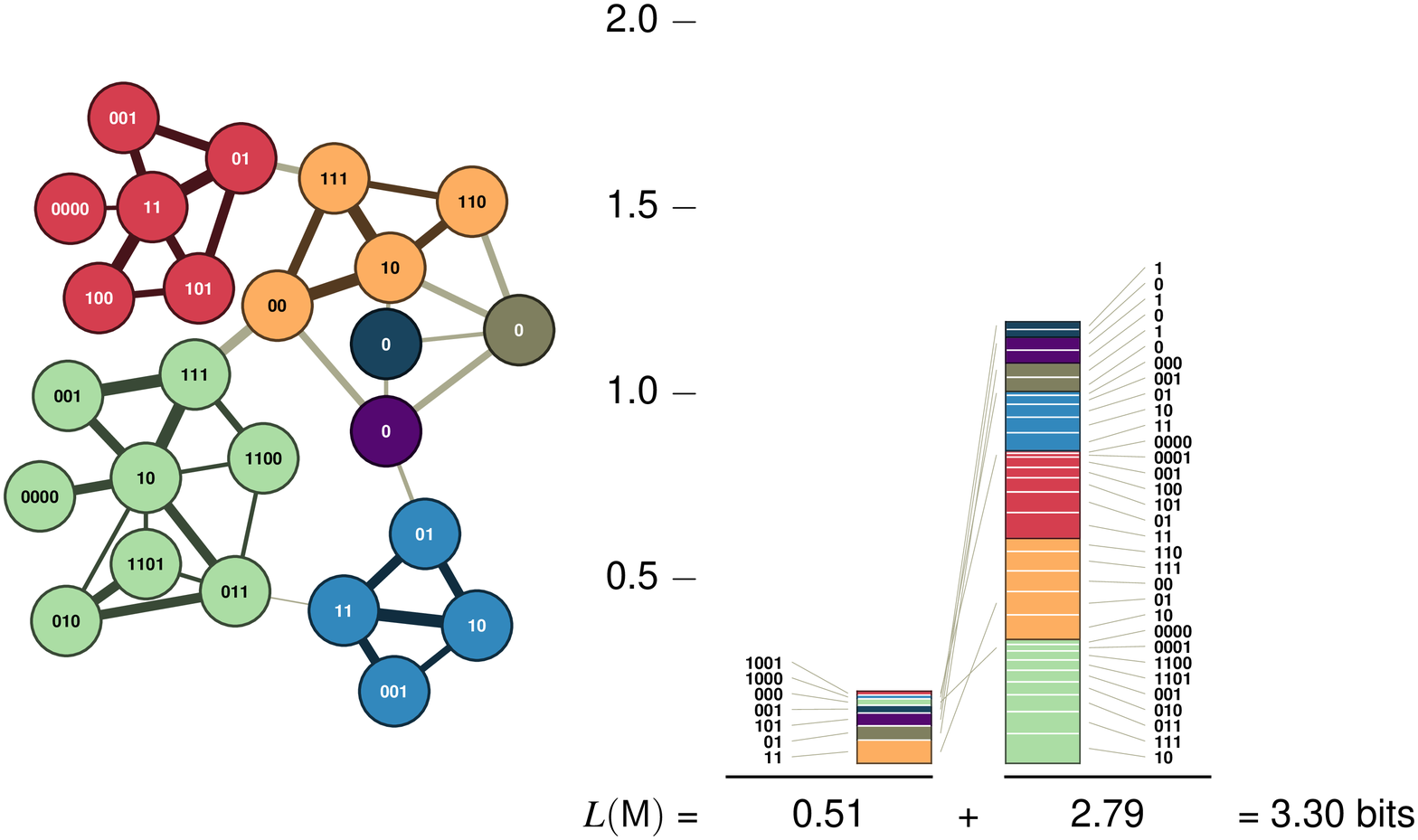}
\end{center}
\clearpage
\begin{center}
\includegraphics[width=1.5\textwidth,angle=90,bb=-784 -614 325 36]{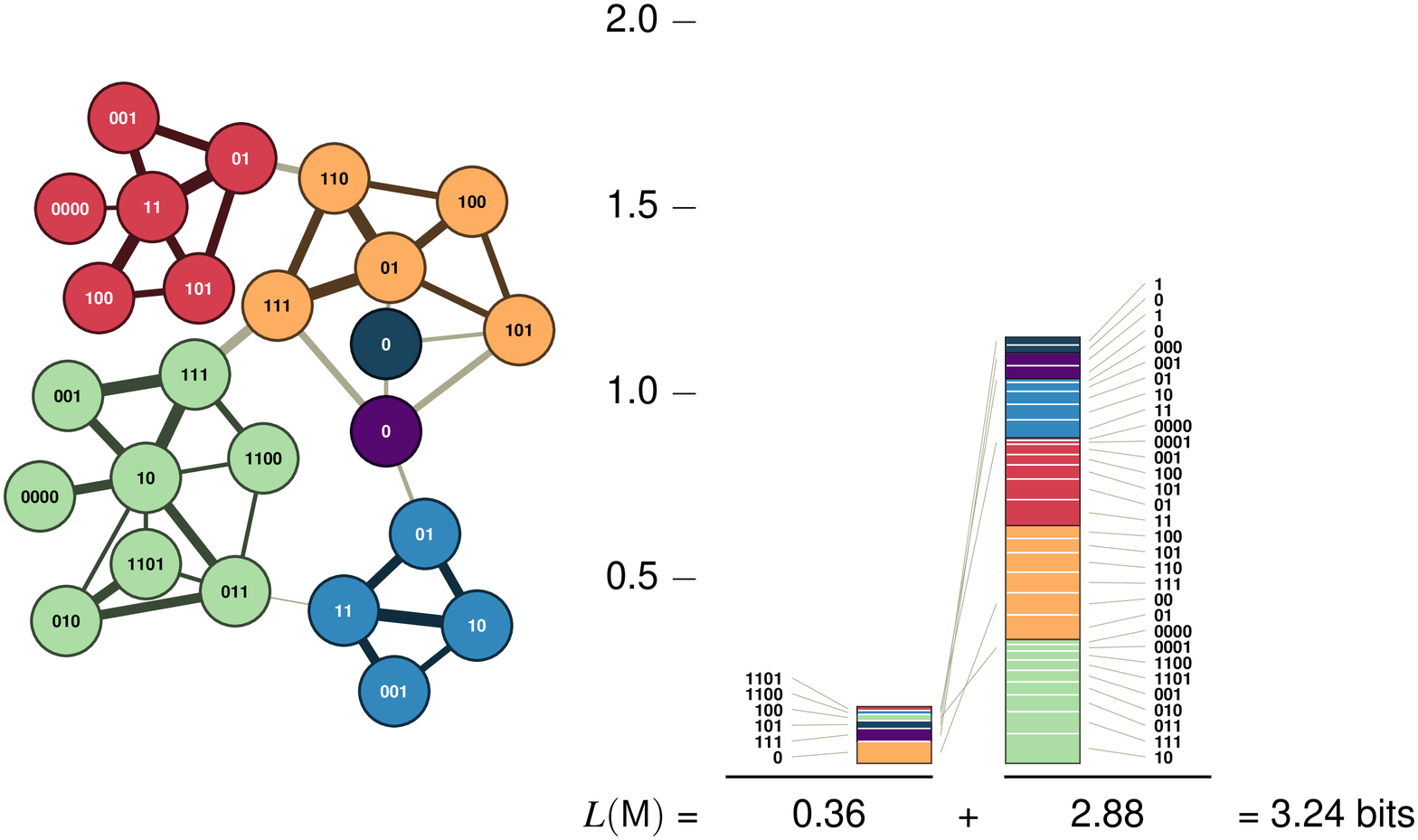}
\end{center}
\clearpage
\begin{center}
\includegraphics[width=1.5\textwidth,angle=90,bb=-784 -614 325 36]{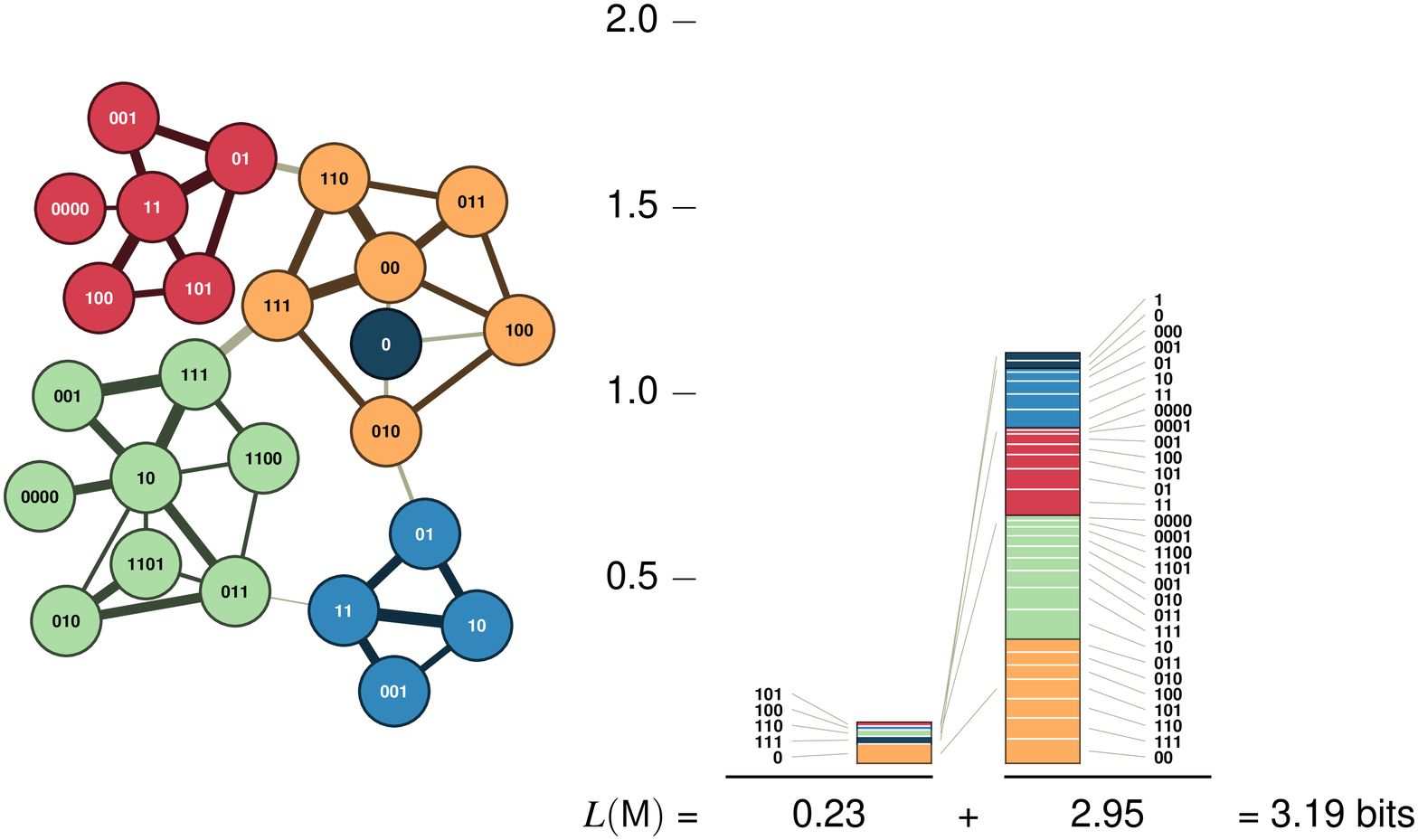}
\end{center}
\clearpage
\begin{center}
\includegraphics[width=1.5\textwidth,angle=90,bb=-784 -614 325 36]{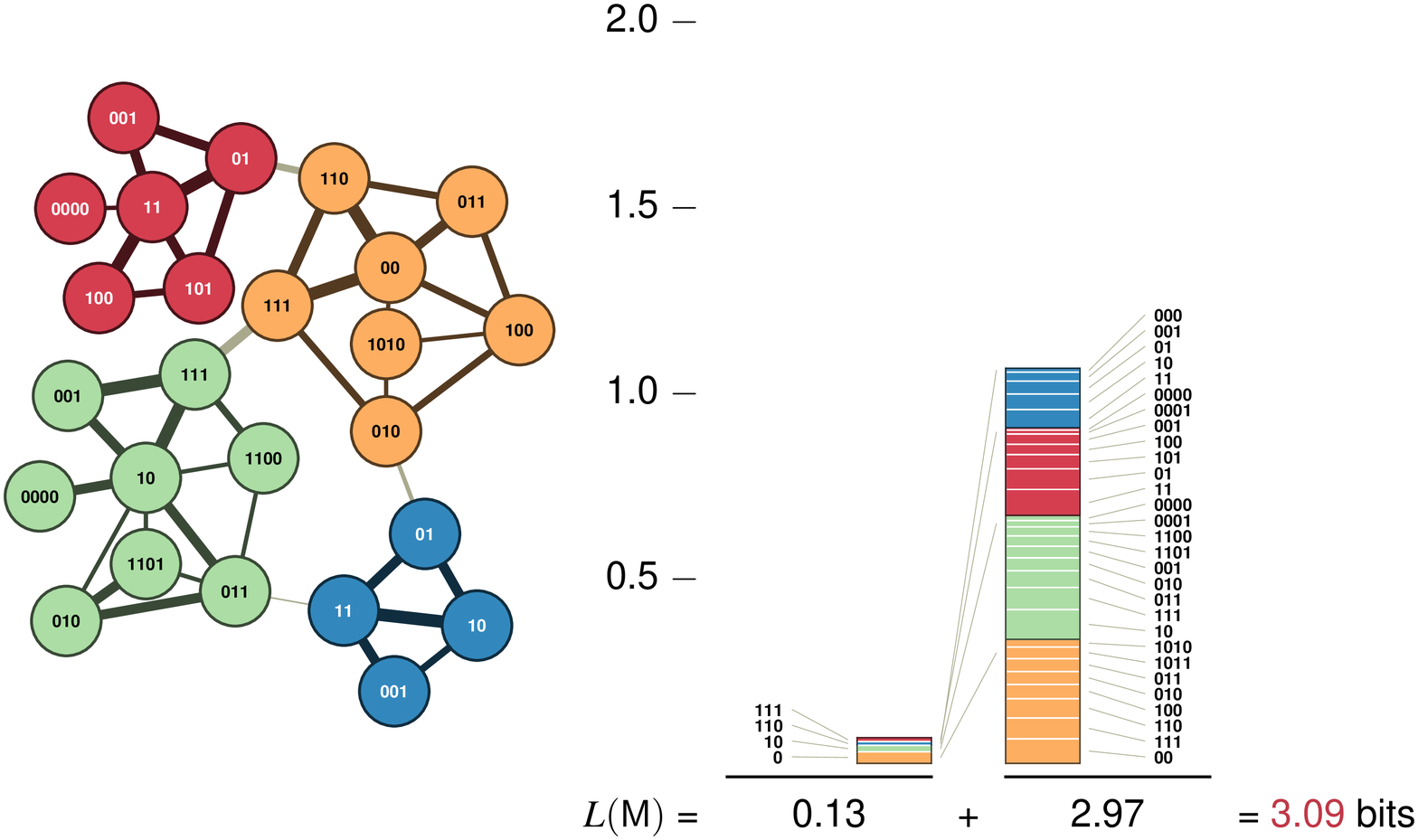}
\end{center}

\begin{figure*}[tbp]
\hspace{-1.5cm}
\input{net_fig_2_ps_in.tex}
\hypersetup{linkcolor=darkred}
\caption{S3 A map of \hyperref[science]{science} with fields according to a partitioning with the presented method. All 88 field names in this caption and the 51 field names in the map are clickable links to lists with the journals in the corresponding fields. The fields and journals are ranked by the fraction of time they are visited by a random surfer. 1 \hyperref[module1]{Molecular \& Cell Biology}, 2 \hyperref[module2]{Medicine}, 3 \hyperref[module3]{Physics}, 4 \hyperref[module4]{Neuroscience}, 5 \hyperref[module5]{Ecology \& Evolution}, 6 \hyperref[module6]{Economics}, 7 \hyperref[module7]{Geosciences}, 8 \hyperref[module8]{Psychology}, 9 \hyperref[module9]{Chemistry}, 10 \hyperref[module10]{Psychiatry}, 11 \hyperref[module11]{Environmental Chemistry \& Microbiology}, 12 \hyperref[module12]{Mathematics}, 13 \hyperref[module13]{Computer Science}, 14 \hyperref[module14]{Analytic Chemistry}, 15 \hyperref[module15]{Business \& Marketing}, 16 \hyperref[module16]{Political Science}, 17 \hyperref[module17]{Fluid Mechanics}, 18 \hyperref[module18]{Medical Imaging}, 19 \hyperref[module19]{Material Engineering}, 20 \hyperref[module20]{Sociology}, 21 \hyperref[module21]{Probability \& Statistics}, 22 \hyperref[module22]{Astronomy \& Astrophysics}, 23 \hyperref[module23]{Gastroenterology}, 24 \hyperref[module24]{Law}, 25 \hyperref[module25]{Chemical Engineering}, 26 \hyperref[module26]{Education}, 27 \hyperref[module27]{Telecommunication}, 28 \hyperref[module28]{Orthopedics}, 29 \hyperref[module29]{Control Theory}, 30 \hyperref[module30]{Environmental Health}, 31 \hyperref[module31]{Operations Research}, 32 \hyperref[module32]{Ophthalmology}, 33 \hyperref[module33]{Crop Science}, 34 \hyperref[module34]{Geography}, 35 \hyperref[module35]{Anthropology}, 36 \hyperref[module36]{Veterinary}, 37 \hyperref[module37]{Computer Imaging}, 38 \hyperref[module38]{Agriculture}, 39 \hyperref[module39]{Parasitology}, 40 \hyperref[module40]{Dentistry}, 41 \hyperref[module41]{Dermatology}, 42 \hyperref[module42]{Urology}, 43 \hyperref[module43]{Rheumatology}, 44 \hyperref[module44]{Applied Acoustics}, 45 \hyperref[module45]{Pharmacology}, 46 \hyperref[module46]{Pathology}, 47 \hyperref[module47]{History \& Philosophy of Science}, 48 \hyperref[module48]{Otolaryngology}, 49 \hyperref[module49]{Electromagnetic Engineering}, 50 \hyperref[module50]{Information Science}, 51 \hyperref[module51]{Circuits}, 52 \hyperref[module52]{Media \& Communication}, 53 \hyperref[module53]{Power Systems}, 54 \hyperref[module54]{Tribology}, 55 \hyperref[module55]{History}, 56 \hyperref[module56]{Geotechnology}, 57 \hyperref[module57]{Wood Products}, 58 \hyperref[module58]{Radiation}, 59 \hyperref[module59]{Linguistics}, 60 \hyperref[module60]{Social Work}, 61 \hyperref[module61]{Psychoanalysis}, 62 \hyperref[module62]{Middle Eastern Studies}, 63 \hyperref[module63]{Forensic Science}, 64 \hyperref[module64]{Transfusion}, 65 \hyperref[module65]{Mycology}, 66 \hyperref[module66]{Nuclear Energy}, 67 \hyperref[module67]{offshore Engineering}, 68 \hyperref[module68]{Environmental Ethics}, 69 \hyperref[module69]{Entomology}, 70 \hyperref[module70]{Higher Education}, 71 \hyperref[module71]{Refineries}, 72 \hyperref[module72]{Reliability Engineering}, 73 \hyperref[module73]{Other}, 74 \hyperref[module74]{Civil Engineering}, 75 \hyperref[module75]{Lab Veterinary}, 76 \hyperref[module76]{Music}, 77 \hyperref[module77]{Tourism}, 78 \hyperref[module78]{Textiles}, 79 \hyperref[module79]{Creativity Research}, 80 \hyperref[module80]{Travel Sociology}, 81 \hyperref[module81]{Medical Informatics}, 82 \hyperref[module82]{Leprosy}, 83 \hyperref[module83]{Sociology (Russian)}, 84 \hyperref[module84]{Cryobiology}, 85 \hyperref[module85]{Death Studies}, 86 \hyperref[module86]{Rehabilitation Counseling}, 87 \hyperref[module87]{Steel}, 88 \hyperref[module88]{Futurist}.}\end{figure*}
\begin{figure*}[tbp]
\hspace{-2.5cm}\input{netSS_fig_2_ps_in.tex}
\hypersetup{linkcolor=darkred}
\caption{S2 A map of \hyperref[socialscience]{social science} with fields according to a partitioning with the presented method. All 54 field names in this caption and the 36 field names in the map are clickable links to lists with the journals in the corresponding fields. The fields and journals are ranked by the fraction of time they are visited by a random surfer. 1 \hyperref[moduleSS1]{Economics}, 2 \hyperref[moduleSS2]{Psychology}, 3 \hyperref[moduleSS3]{Psychiatry}, 4 \hyperref[moduleSS4]{Healthcare}, 5 \hyperref[moduleSS5]{Political Science}, 6 \hyperref[moduleSS6]{Sociology (Behavioral)}, 7 \hyperref[moduleSS7]{Management}, 8 \hyperref[moduleSS8]{Law}, 9 \hyperref[moduleSS9]{Education}, 10 \hyperref[moduleSS10]{Geography}, 11 \hyperref[moduleSS11]{Physical Anthropology}, 12 \hyperref[moduleSS12]{Cultural Anthropology}, 13 \hyperref[moduleSS13]{Marketing}, 14 \hyperref[moduleSS14]{Information Science}, 15 \hyperref[moduleSS15]{Philosophy of Science}, 16 \hyperref[moduleSS16]{Sociology (Institutional)}, 17 \hyperref[moduleSS17]{Communication}, 18 \hyperref[moduleSS18]{Educational Assessment}, 19 \hyperref[moduleSS19]{Educational Psychology}, 20 \hyperref[moduleSS20]{Human-Computer Interface}, 21 \hyperref[moduleSS21]{Applied Linguistics}, 22 \hyperref[moduleSS22]{Experimental Psychology}, 23 \hyperref[moduleSS23]{History}, 24 \hyperref[moduleSS24]{Social Work}, 25 \hyperref[moduleSS25]{Speech And Hearing}, 26 \hyperref[moduleSS26]{Disabilities}, 27 \hyperref[moduleSS27]{Transportation}, 28 \hyperref[moduleSS28]{Psychoanalysis}, 29 \hyperref[moduleSS29]{Guidance Counseling}, 30 \hyperref[moduleSS30]{Middle Eastern Studies}, 31 \hyperref[moduleSS31]{East Asian Studies}, 32 \hyperref[moduleSS32]{Ergonomics}, 33 \hyperref[moduleSS33]{Medical Ethics}, 34 \hyperref[moduleSS34]{Public Administration}, 35 \hyperref[moduleSS35]{Ethics}, 36 \hyperref[moduleSS36]{Economic History}, 37 \hyperref[moduleSS37]{Sport Psychology}, 38 \hyperref[moduleSS38]{Public Affairs}, 39 \hyperref[moduleSS39]{Social Policy}, 40 \hyperref[moduleSS40]{Family Relations}, 41 \hyperref[moduleSS41]{Law And Behavior}, 42 \hyperref[moduleSS42]{Criminology}, 43 \hyperref[moduleSS43]{Sexuality}, 44 \hyperref[moduleSS44]{Higher Education}, 45 \hyperref[moduleSS45]{Leisure Studies}, 46 \hyperref[moduleSS46]{Neurorehabilitation}, 47 \hyperref[moduleSS47]{Pacific Studies}, 48 \hyperref[moduleSS48]{Tourism}, 49 \hyperref[moduleSS49]{Creativity}, 50 \hyperref[moduleSS50]{Death And Dying}, 51 \hyperref[moduleSS51]{Sociology (French)}, 52 \hyperref[moduleSS52]{Sociology (Eastern Europe)}, 53 \hyperref[moduleSS53]{Maritime Law}, 54 \hyperref[moduleSS54]{Hypnosis}}\end{figure*}
\clearpage
\section*{{\Large Fields of science}\label{science}}
\subsection*{{\large 1 Molecular \& Cell Biology (20\%)}\label{module1}}
\noindent 1~J Biol Chem, 2~Cell, 3~J Immunol, 4~Embo J, 5~Blood, 6~Mol Cell Biol, 7~Cancer Res, 8~Nat Genet, 9~J Cell Biol, 10~Gene Dev, 11~Biochemistry-Us, 12~Mol Cell, 13~J Exp Med, 14~Nucleic Acids Res, 15~Oncogene, 16~J Clin Invest, 17~Biochem Bioph Res Co, 18~Curr Biol, 19~J Mol Biol, 20~J Virol, 21~Nat Med, 22~Development, 23~Febs Lett, 24~Nat Cell Biol, 25~Am J Hum Genet, 26~J Cell Sci, 27~Biochem J, 28~Mol Biol Cell, 29~J Bacteriol, 30~Infect Immun, 31~Hum Mol Genet, 32~Nat Immunol, 33~Immunity, 34~Genetics, 35~Plant Physiol, 36~Mol Microbiol, 37~Am J Pathol, 38~Faseb J, 39~Clin Cancer Res, 40~Biophys J, 41~Endocrinology, 42~Nat Biotechnol, 43~Genome Res, 44~Dev Biol, 45~Plant Cell, 46~Nat Rev Mol Cell Bio, 47~Int J Cancer, 48~Bioinformatics, 49~Eur J Biochem, 50~Brit J Cancer, 51~Curr Opin Cell Biol, 52~Eur J Immunol, 53~Plant J, 54~Trends Biochem Sci, 55~Mol Biol Evol, 56~Annu Rev Immunol, 57~Dev Cell, 58~Gene, 59~Free Radical Bio Med, 60~Trends Cell Biol, 61~Trends Genet, 62~Exp Cell Res, 63~Virology, 64~Mol Pharmacol, 65~Brit J Haematol, 66~Nat Rev Genet, 67~Annu Rev Biochem, 68~Physiol Rev, 69~Curr Opin Genet Dev, 70~Am J Physiol-Cell Ph, 71~Biol Reprod, 72~Bioessays, 73~Mech Develop, 74~Vaccine, 75~Biochem Pharmacol, 76~Embo Rep, 77~Arch Biochem Biophys, 78~J Invest Dermatol, 79~Fems Microbiol Lett, 80~Protein Sci, 81~Mol Endocrinol, 82~Carcinogenesis, 83~Curr Opin Immunol, 84~Anal Biochem, 85~Method Enzymol, 86~Genomics, 87~Structure, 88~J Gen Virol, 89~J Leukocyte Biol, 90~Immunol Rev, 91~Proteins, 92~Am J Physiol-Lung C, 93~Leukemia, 94~Cell Mol Life Sci, 95~Rna, 96~Curr Opin Struc Biol, 97~Annu Rev Cell Dev Bi, 98~Plant Mol Biol, 99~J Pathol, 100~Trends Plant Sci, 101~Microbiol-Sgm, 102~Acta Crystallogr D, 103~Cell Death Differ, 104~J Lipid Res, 105~Am J Resp Cell Mol, 106~Cancer Cell, 107~Am J Physiol-Renal, 108~Gene Ther, 109~J Exp Bot, 110~J Cell Physiol, 111~Endocr Rev, 112~J Cell Biochem, 113~Planta, 114~Anticancer Res, 115~Hum Genet, 116~J Med Genet, 117~Biotechniques, 118~Cancer Lett, 119~Curr Opin Plant Biol, 120~Curr Opin Microbiol, 121~Lab Invest, 122~Microbiol Mol Biol R, 123~Trends Immunol, 124~J Mol Evol, 125~Hum Mutat, 126~Bba-Biomembranes, 127~Clin Exp Immunol, 128~Int Immunol, 129~Mol Cell Endocrinol, 130~Trends Microbiol, 131~J Endocrinol, 132~Bone Marrow Transpl, 133~Dev Dynam, 134~Hum Gene Ther, 135~Annu Rev Physiol, 136~Traffic, 137~Microbes Infect, 138~Chem Biol, 139~J Immunol Methods, 140~Methods, 141~Mol Plant Microbe In, 142~Annu Rev Genet, 143~Int J Oncol, 144~Exp Hematol, 145~Mol Ther, 146~J Biochem, 147~Annu Rev Plant Biol, 148~Plant Cell Physiol, 149~Bba-Mol Cell Biol L, 150~Immunology, 151~J Gen Physiol, 152~Front Biosci, 153~Biochem Soc T, 154~Annu Rev Microbiol, 155~Annu Rev Pharmacol, 156~Bba-Bioenergetics, 157~Curr Opin Biotech, 158~Adv Exp Med Biol, 159~Biol Chem, 160~Int J Biochem Cell B, 161~Eur J Hum Genet, 162~Mutat Res-Fund Mol M, 163~Mol Biochem Parasit, 164~Biopolymers, 165~Pflug Arch Eur J Phy, 166~J Struct Biol, 167~Gene Chromosome Canc, 168~Mamm Genome, 169~Genes Cells, 170~Microsc Res Techniq, 171~Trends Endocrin Met, 172~Physiol Plantarum, 173~Biochimie, 174~Prostate, 175~Bba-Mol Cell Res, 176~Cell Signal, 177~Genet Epidemiol, 178~Bba-Gen Subjects, 179~Trends Biotechnol, 180~Pharmacol Therapeut, 181~Plant Sci, 182~J Histochem Cytochem, 183~Bba-Gene Struct Expr, 184~Curr Top Microbiol, 185~Mol Cell Biochem, 186~Proteomics, 187~Clin Immunol, 188~Arch Virol, 189~Cell Microbiol, 190~J Steroid Biochem, 191~Biosci Biotech Bioch, 192~Trends Mol Med, 193~Glycobiology, 194~Immunogenetics, 195~Haematologica, 196~Mol Hum Reprod, 197~Fems Microbiol Rev, 198~Mol Genet Metab, 199~J Mol Med-Jmm, 200~Free Radical Res, 201~Yeast, 202~Annu Rev Bioph Biom, 203~Cytokine, 204~Photochem Photobiol, 205~J Comput Biol, 206~Mol Reprod Dev, 207~Hum Immunol, 208~Genesis, 209~Semin Cancer Biol, 210~Semin Cell Dev Biol, 211~J Membrane Biol, 212~Int Rev Cytol, 213~Exp Gerontol, 214~Cell Calcium, 215~Leukemia Lymphoma, 216~Cytokine Growth F R, 217~Clin Genet, 218~Mol Genet Genomics, 219~Int Arch Allergy Imm, 220~Arch Microbiol, 221~Reproduction, 222~Mol Immunol, 223~Semin Immunol, 224~Curr Pharm Design, 225~Drug Discov Today, 226~J Interf Cytok Res, 227~Immunol Lett, 228~Mech Ageing Dev, 229~Biophys Chem, 230~Ann Hum Genet, 231~Toxicon, 232~Virus Res, 233~Physiol Genomics, 234~Cell Motil Cytoskel, 235~Biol Pharm Bull, 236~Protein Expres Purif, 237~Cell Immunol, 238~Scand J Immunol, 239~Cancer Chemoth Pharm, 240~Stem Cells, 241~Dev Genes Evol, 242~Cancer Genet Cytogen, 243~Matrix Biol, 244~J Plant Physiol, 245~Semin Hematol, 246~Tissue Antigens, 247~Placenta, 248~Genet Res, 249~Int J Mol Med, 250~Genes Immun, 251~Trends Cardiovas Med, 252~J Mol Endocrinol, 253~Mol Med, 254~Bba-Mol Basis Dis, 255~Cancer Gene Ther, 256~Hum Hered, 257~Eur J Cell Biol, 258~Fungal Genet Biol, 259~Am J Hematol, 260~Curr Genet, 261~Oncol Rep, 262~Res Microbiol, 263~J Androl, 264~Antioxid Redox Sign, 265~Int J Dev Biol, 266~Curr Opin Hematol, 267~Chromosoma, 268~Fems Immunol Med Mic, 269~Histochem Cell Biol, 270~Histol Histopathol, 271~Dev Comp Immunol, 272~Leukemia Res, 273~Plant Physiol Bioch, 274~J Photoch Photobio B, 275~J Autoimmun, 276~Photosynth Res, 277~Adv Immunol, 278~Differentiation, 279~Prog Nucleic Acid Re, 280~Blood Cell Mol Dis, 281~Cancer Immunol Immun, 282~Dna Res, 283~Annu Rev Genom Hum G, 284~Mol Carcinogen, 285~Steroids, 286~Int J Hematol, 287~Curr Opin Oncol, 288~J Bioenerg Biomembr, 289~Biol Blood Marrow Tr, 290~Adv Protein Chem, 291~Chromosome Res, 292~Neoplasia, 293~Crit Rev Oncol Hemat, 294~Dna Cell Biol, 295~Iubmb Life, 296~Bba-Rev Cancer, 297~Apmis, 298~Adv Cancer Res, 299~Apoptosis, 300~Immunol Cell Biol, 301~Pigm Cell Res, 302~J Cancer Res Clin, 303~J Mol Microb Biotech, 304~News Physiol Sci, 305~Exp Biol Med, 306~J Inherit Metab Dis, 307~Microbiol Immunol, 308~J Gene Med, 309~Endocr-Relat Cancer, 310~Cold Spring Harb Sym, 311~Clin Exp Metastas, 312~Prog Biophys Mol Bio, 313~Inflamm Res, 314~Chem Phys Lipids, 315~Prog Lipid Res, 316~Protoplasma, 317~Ann Hematol, 318~Eur J Haematol, 319~Evol Dev, 320~Int Immunopharmacol, 321~Anat Embryol, 322~Biochem Cell Biol, 323~Cell Mol Biol, 324~Anti-Cancer Drug, 325~Curr Top Dev Biol, 326~Int J Med Microbiol, 327~Microvasc Res, 328~Mutat Res-Rev Mutat, 329~Mol Membr Biol, 330~Microb Pathogenesis, 331~Immunol Res, 332~Exp Dermatol, 333~Curr Opin Mol Ther, 334~Transgenic Res, 335~J Clin Immunol, 336~Biochemistry-Moscow+, 337~Method Cell Biol, 338~Prostag Oth Lipid M, 339~Recent Prog Horm Res, 340~Glycoconjugate J, 341~Biol Cell, 342~Nitric Oxide-Biol Ch, 343~J Muscle Res Cell M, 344~Cell Struct Funct, 345~Eur Biophys J Biophy, 346~Melanoma Res, 347~Cancer Metast Rev, 348~Cell Stress Chaperon, 349~Dev Growth Differ, 350~Crit Rev Plant Sci, 351~Cell Biol Int, 352~J Pept Res, 353~J Mammary Gland Biol, 354~Immunobiology, 355~Adv Virus Res, 356~Arch Dermatol Res, 357~Eur Cytokine Netw, 358~Rev Med Virol, 359~J Biomol Struct Dyn, 360~J Reprod Immunol, 361~Biofactors, 362~Biosystems, 363~Crit Rev Biochem Mol, 364~Cell Physiol Biochem, 365~J Dermatol Sci, 366~J Plant Growth Regul, 367~Mol Cells, 368~J Endotoxin Res, 369~Biometals, 370~Int J Androl, 371~Drug Resist Update, 372~J Biomed Sci, 373~Micron, 374~Sex Plant Reprod, 375~Crit Rev Immunol, 376~J Mol Recognit, 377~Mol Biotechnol, 378~Amino Acids, 379~Plasmid, 380~Acta Biochim Pol, 381~Vitam Horm, 382~J Mol Model, 383~Virus Genes, 384~Acta Haematol-Basel, 385~Expert Opin Biol Th, 386~Connect Tissue Res, 387~Autoimmunity, 388~J Biomol Screen, 389~Cells Tissues Organs, 390~J Mol Diagn, 391~Neuro-Oncology, 392~Viral Immunol, 393~Exp Mol Pathol, 394~Cell Biochem Biophys, 395~Biochem Soc Symp, 396~Biotechnol Appl Bioc, 397~Gene Expression, 398~Redox Rep, 399~In Vitro Cell Dev-An, 400~Blood Rev, 401~Exp Mol Med, 402~Genes Genet Syst, 403~Tissue Cell, 404~Drug Develop Res, 405~Biomol Eng, 406~Int J Exp Pathol, 407~Eur J Immunogenet, 408~J Pept Sci, 409~Genet Test, 410~Med Microbiol Immun, 411~Plant Mol Biol Rep, 412~Crit Rev Eukar Gene, 413~Exp Lung Res, 414~Zygote, 415~Scientist, 416~Amyloid, 417~Receptor Channel, 418~Cell Res, 419~Bioscience Rep, 420~Springer Semin Immun, 421~Neoplasma, 422~Endocr Res, 423~In Vivo, 424~Tumor Biol, 425~J Invest Derm Symp P, 426~Inflammation, 427~Biogerontology, 428~Oncol Res, 429~Biodrugs, 430~Cytotherapy, 431~Clin Lab Haematol, 432~Asm News, 433~Growth Factors, 434~J Med Primatol, 435~Pathobiology, 436~J Biosciences, 437~J Biol Phys, 438~Arch Histol Cytol, 439~Cell Proliferat, 440~Adv Bot Res, 441~J Biochem Mol Biol, 442~Mol Biol Rep, 443~Endothelium-J Endoth, 444~Cell Biochem Funct, 445~Mol Biol+, 446~Int J Biol Marker, 447~Dis Markers, 448~Mediat Inflamm, 449~Drug News Perspect, 450~J Biol Reg Homeos Ag, 451~Crit Rev Cl Lab Sci, 452~Anat Histol Embryol, 453~Acta Physiol Plant, 454~An Acad Bras Cienc, 455~Acta Virol, 456~Exp Anim Tokyo, 457~Ann Genet-Paris, 458~Russ J Plant Physl+, 459~Arch Immunol Ther Ex, 460~Dna Sequence, 461~Protein Peptide Lett, 462~Trends Glycosci Glyc, 463~Biofizika+, 464~Folia Histochem Cyto, 465~Cell Mol Biol Lett, 466~Clin Dysmorphol, 467~P Jpn Acad B-Phys, 468~Zh Obshch Biol, 469~Acta Histochem, 470~Eur J Histochem, 471~J Liposome Res, 472~Biol Res, 473~Anim Biotechnol, 474~Immunopharm Immunot, 475~Genet Counsel, 476~Gen Physiol Biophys, 477~Sci China Ser C, 478~Immunol Invest, 479~Hemoglobin, 480~Indian J Biochem Bio, 481~Hematol Oncol, 482~J Toxicol-Toxin Rev, 483~Lett Pept Sci, 484~Theor Biosci, 485~Biofutur, 486~Int J Immunopath Ph, 487~J Genet, 488~Biotech Histochem, 489~Folia Biol-Prague, 490~Minerva Biotecnol, 491~Korean J Genetic, 492~Riv Biol-Biol Forum, 493~Acta Biotheor, 494~Prep Biochem Biotech, 495~Biotechnol Genet Eng, 496~Issues Law Med, 497~Biomed Res-Tokyo, 498~Acta Histochem Cytoc, 499~Spectrosc-Int J, 500~Biocell, 501~Natl Acad Sci Lett, 502~M S-Med Sci, 503~Biol Membrany, 504~Pteridines, 505~Period Biol, 506~Prog Biochem Biophys, 507~Lymphology, 508~Gematol Transfuziol, 509~Nippon Nogeik Kaishi, 510~J Exp Anim Sci, 511~Seikagaku, 512~J Clin Biochem Nutr\subsection*{{\large 2 Medicine (14\%)}\label{module2}}
\noindent 1~New Engl J Med, 2~Lancet, 3~Circulation, 4~Jama-J Am Med Assoc, 5~Brit Med J, 6~J Clin Oncol, 7~J Am Coll Cardiol, 8~J Clin Endocr Metab, 9~Am J Resp Crit Care, 10~Circ Res, 11~Diabetes, 12~Ann Intern Med, 13~J Infect Dis, 14~Pediatrics, 15~J Clin Microbiol, 16~Cancer, 17~Arch Intern Med, 18~Clin Infect Dis, 19~Arterioscl Throm Vas, 20~Stroke, 21~J Natl Cancer I, 22~Aids, 23~Am J Cardiol, 24~Diabetes Care, 25~Kidney Int, 26~Am J Public Health, 27~Am J Epidemiol, 28~Chest, 29~Am J Physiol-Heart C, 30~Antimicrob Agents Ch, 31~Hypertension, 32~Am J Clin Nutr, 33~J Nutr, 34~Ann Thorac Surg, 35~Transplantation, 36~J Am Soc Nephrol, 37~J Allergy Clin Immun, 38~Soc Sci Med, 39~J Appl Physiol, 40~Crit Care Med, 41~J Bone Miner Res, 42~Am J Med, 43~Emerg Infect Dis, 44~Am J Obstet Gynecol, 45~Cardiovasc Res, 46~Ann Surg, 47~Hum Reprod, 48~Am J Physiol-Endoc M, 49~Am Heart J, 50~Thromb Haemostasis, 51~Diabetologia, 52~J Am Geriatr Soc, 53~Eur Respir J, 54~Am J Physiol-Reg I, 55~J Pediatr, 56~Health Affair, 57~Obstet Gynecol, 58~Am J Kidney Dis, 59~Anesthesiology, 60~J Thorac Cardiov Sur, 61~Clin Chem, 62~Atherosclerosis, 63~Eur Heart J, 64~Med Sci Sport Exer, 65~Jaids-J Acq Imm Def, 66~Med Care, 67~Fertil Steril, 68~Cancer Epidem Biomar, 69~Int J Obesity, 70~Thorax, 71~Eur J Cancer, 72~Anesth Analg, 73~J Hypertens, 74~J Vasc Surg, 75~Ann Oncol, 76~J Antimicrob Chemoth, 77~Nephrol Dial Transpl, 78~Brit J Surg, 79~Heart, 80~Int J Epidemiol, 81~Arch Pediat Adol Med, 82~Bone, 83~Arch Dis Child, 84~J Epidemiol Commun H, 85~Pediatr Infect Dis J, 86~Pediatr Res, 87~J Trauma, 88~Epidemiology, 89~Am J Prev Med, 90~Clin Exp Allergy, 91~Semin Oncol, 92~Can Med Assoc J, 93~J Mol Cell Cardiol, 94~Surgery, 95~Prev Med, 96~B World Health Organ, 97~Acad Med, 98~Arch Surg-Chicago, 99~J Gen Intern Med, 100~J Clin Epidemiol, 101~Eur J Cardio-Thorac, 102~Am J Surg, 103~Osteoporosis Int, 104~Clin Pharmacol Ther, 105~Intens Care Med, 106~Clin Microbiol Rev, 107~J Gerontol A-Biol, 108~Brit J Nutr, 109~Drugs, 110~Mayo Clin Proc, 111~Obes Res, 112~Surg Endosc, 113~Am J Hypertens, 114~Metabolism, 115~J Med Virol, 116~Dis Colon Rectum, 117~Ca-Cancer J Clin, 118~Clin Endocrinol, 119~Gynecol Oncol, 120~J Adolescent Health, 121~Allergy, 122~J Cardiovasc Electr, 123~Transplant P, 124~Med J Australia, 125~J Am Coll Surgeons, 126~Gerontologist, 127~Breast Cancer Res Tr, 128~Ann Emerg Med, 129~Brit J Anaesth, 130~Eur J Endocrinol, 131~Ann Surg Oncol, 132~Health Serv Res, 133~Acta Paediatr, 134~J Pediatr Surg, 135~J Adv Nurs, 136~J Intern Med, 137~World J Surg, 138~Sex Transm Dis, 139~Pharmacogenetics, 140~Diabetic Med, 141~Aids Res Hum Retrov, 142~Eur J Clin Nutr, 143~J Surg Res, 144~Brit J Clin Pharmaco, 145~Clin Sci, 146~Lipids, 147~J Cardiovasc Pharm, 148~Clin Ther, 149~Cancer Cause Control, 150~Ann Pharmacother, 151~Calcified Tissue Int, 152~Shock, 153~Lung Cancer-J Iaslc, 154~J Heart Lung Transpl, 155~Eur J Appl Physiol, 156~Med Educ, 157~Eur J Clin Invest, 158~Clin Chim Acta, 159~Clin Diagn Lab Immun, 160~Epidemiol Infect, 161~Pharmacotherapy, 162~Eur J Vasc Endovasc, 163~Tob Control, 164~J Am Diet Assoc, 165~Catheter Cardio Inte, 166~Curr Opin Lipidol, 167~Infect Cont Hosp Ep, 168~J Fam Practice, 169~Ultrasound Obst Gyn, 170~Acad Emerg Med, 171~J Pain Symptom Manag, 172~Ann Allerg Asthma Im, 173~Brit J Gen Pract, 174~Int J Tuberc Lung D, 175~Thromb Res, 176~Clin Pharmacokinet, 177~Sex Transm Infect, 178~Acta Physiol Scand, 179~J Med Microbiol, 180~Eur J Clin Microbiol, 181~Nutrition, 182~Thyroid, 183~Am Surgeon, 184~Ann Behav Med, 185~Eur J Pediatr, 186~J Virol Methods, 187~Annu Rev Med, 188~Pace, 189~Prenatal Diag, 190~Int J Cardiol, 191~Ann Epidemiol, 192~Anaesthesia, 193~Ann Med, 194~Expert Opin Inv Drug, 195~Int J Antimicrob Ag, 196~J Am Soc Echocardiog, 197~Pediatr Pulm, 198~Acta Obstet Gyn Scan, 199~Int J Std Aids, 200~Cerebrovasc Dis, 201~J Clin Pharmacol, 202~Clin Exp Pharmacol P, 203~Sports Med, 204~Annu Rev Nutr, 205~Control Clin Trials, 206~Psycho-Oncol, 207~Resp Med, 208~Eur J Clin Pharmacol, 209~Oncologist, 210~Acta Anaesth Scand, 211~Pediatr Nephrol, 212~Eur J Obstet Gyn R B, 213~Drug Safety, 214~Public Health Rep, 215~Diagn Micr Infec Dis, 216~Nutr Cancer, 217~J Hosp Infect, 218~Clin Chem Lab Med, 219~J Hum Hypertens, 220~P Nutr Soc, 221~Clin Microbiol Infec, 222~Qjm-Int J Med, 223~J Endovasc Ther, 224~J Am Med Inform Assn, 225~Qual Life Res, 226~J Gastrointest Surg, 227~Hum Reprod Update, 228~Annu Rev Publ Health, 229~J Card Fail, 230~Horm Res, 231~Oncology-Basel, 232~Fam Pract, 233~Horm Metab Res, 234~Pharmacoeconomics, 235~Am J Manag Care, 236~Nutr Rev, 237~Scand J Infect Dis, 238~Am Fam Physician, 239~Age Ageing, 240~Patient Educ Couns, 241~J Am Coll Nutr, 242~Am J Med Sci, 243~J Lab Clin Med, 244~Int J Sports Med, 245~Ther Drug Monit, 246~Res Q Exercise Sport, 247~Obes Surg, 248~South Med J, 249~J Clin Virol, 250~J Endocrinol Invest, 251~Can J Anaesth, 252~J Surg Oncol, 253~Milbank Q, 254~J Pediatr Endocr Met, 255~Curr Opin Nephrol Hy, 256~Eur J Heart Fail, 257~Med Care Res Rev, 258~Am J Infect Control, 259~Medicine, 260~J Pediatr Psychol, 261~Diabetes Res Clin Pr, 262~Public Health Nutr, 263~Med Decis Making, 264~Antivir Res, 265~Future Child, 266~Surg Clin N Am, 267~Am J Clin Oncol-Canc, 268~Am J Emerg Med, 269~J Dev Behav Pediatr, 270~Diabetes-Metab Res, 271~Aids Care, 272~Health Educ Res, 273~Semin Thromb Hemost, 274~Endocrin Metab Clin, 275~Int J Gynecol Obstet, 276~J Pediat Hematol Onc, 277~J Reprod Med, 278~Genet Med, 279~Contraception, 280~Health Policy, 281~Med Teach, 282~Aust Nz J Publ Heal, 283~Clin Nephrol, 284~Pediatr Clin N Am, 285~Exp Physiol, 286~Clin Cardiol, 287~Blood Coagul Fibrin, 288~Health Policy Plann, 289~Ann Vasc Surg, 290~Eur J Epidemiol, 291~Clin Biochem, 292~Curr Opin Infect Dis, 293~Epidemiol Rev, 294~Maturitas, 295~J Infection, 296~Brit Med Bull, 297~Clin Transplant, 298~Paediatr Perinat Ep, 299~J Health Polit Polic, 300~Int J Clin Pract, 301~J Heart Valve Dis, 302~J Urban Health, 303~Int J Med Inform, 304~Med Hypotheses, 305~J Sport Sci, 306~Am J Health Promot, 307~Med Mycol, 308~Health Educ Behav, 309~Int J Health Serv, 310~Menopause, 311~Endocrine, 312~J School Health, 313~J Med Ethics, 314~Braz J Med Biol Res, 315~Resuscitation, 316~J Nutr Biochem, 317~Basic Res Cardiol, 318~J Paediatr Child H, 319~J Behav Med, 320~Prostag Leukotr Ess, 321~Cancer Invest, 322~Support Care Cancer, 323~Am J Health-Syst Ph, 324~Cell Transplant, 325~Pharmacoepidem Dr S, 326~Eur J Cancer Prev, 327~Artif Organs, 328~Pharmacol Res, 329~Aids Educ Prev, 330~Postgrad Med J, 331~Aviat Space Envir Md, 332~Curr Opin Clin Nutr, 333~Med Clin N Am, 334~Nurs Res, 335~Appetite, 336~J Law Med Ethics, 337~J Vasc Res, 338~J Emerg Med, 339~Drug Aging, 340~Hastings Cent Rep, 341~Hematol Oncol Clin N, 342~Curr Opin Cardiol, 343~J Cardiovasc Surg, 344~Exp Clin Endocr Diab, 345~Can J Cardiol, 346~Transplant Int, 347~Palliative Med, 348~Res Aging, 349~J Toxicol-Clin Toxic, 350~Haemophilia, 351~Res Nurs Health, 352~Semin Perinatol, 353~Qual Health Res, 354~Surg Today, 355~Early Hum Dev, 356~Infect Dis Clin N Am, 357~J Roy Soc Med, 358~Hypertens Res, 359~J Natl Med Assoc, 360~Clin Chest Med, 361~J Soc Gynecol Invest, 362~Diabetes Metab, 363~Region Anesth Pain M, 364~Method Inform Med, 365~Inquiry-J Health Car, 366~Curr Opin Pediatr, 367~J Nephrol, 368~J Cardiothor Vasc An, 369~Mol Cell Probe, 370~Cardiology, 371~Pediatr Cardiol, 372~Microb Drug Resist, 373~J Asthma, 374~Coronary Artery Dis, 375~Microcirculation, 376~Women Health, 377~J Strength Cond Res, 378~Nutr Res, 379~Scand J Clin Lab Inv, 380~Health Care Manage R, 381~Clin Pediatr, 382~Crit Care, 383~Periton Dialysis Int, 384~Biol Neonate, 385~Cancer Treat Rev, 386~Clin Perinatol, 387~Jpen-Parenter Enter, 388~Curr Med Res Opin, 389~Asaio J, 390~Pediatr Allergy Immu, 391~Stud Family Plann, 392~Internal Med, 393~J Public Health Med, 394~J Assist Reprod Gen, 395~Angiology, 396~Biomed Pharmacother, 397~J Diabetes Complicat, 398~J Med Screen, 399~Pediatr Surg Int, 400~Chinese Med J-Peking, 401~Int J Qual Health C, 402~J Nurs Admin, 403~Fam Med, 404~Quest, 405~Heart Surg Forum, 406~Infection, 407~J Clin Nurs, 408~Eur J Anaesth, 409~Curr Opin Obstet Gyn, 410~Clin Nutr, 411~Gerontology, 412~Surg Laparo Endo Per, 413~Mycopathologia, 414~Am J Nephrol, 415~Ann Roy Coll Surg, 416~J Clin Anesth, 417~Growth Horm Igf Res, 418~Int J Gynecol Cancer, 419~Digest Surg, 420~Semin Nephrol, 421~Semin Dialysis, 422~Int J Technol Assess, 423~Chirurg, 424~Mycoses, 425~Prog Cardiovasc Dis, 426~Gynecol Obstet Inves, 427~Ann Clin Biochem, 428~Aids Patient Care St, 429~Cancer Nurs, 430~Eur J Public Health, 431~J Clin Densitom, 432~Diabetes Obes Metab, 433~Langenbeck Arch Surg, 434~Tumori, 435~Arch Med Res, 436~Arzneimittel-Forsch, 437~J Laparoendosc Adv A, 438~Invest New Drug, 439~Can J Public Health, 440~Pediatr Transplant, 441~Health Place, 442~Eur J Pediatr Surg, 443~J Invest Med, 444~Teach Learn Med, 445~Anaesth Intens Care, 446~Thorac Cardiov Surg, 447~J Telemed Telecare, 448~Postgrad Med, 449~Eur Heart J Suppl, 450~Pediatr Int, 451~J Aging Health, 452~Jpn J Clin Oncol, 453~Endocr J, 454~J Sport Med Phys Fit, 455~Respiration, 456~Bioethics, 457~Z Kardiol, 458~Generations, 459~Presse Med, 460~J Chemotherapy, 461~Birth-Iss Perinat C, 462~Int J Clin Pharm Th, 463~Who Tech Rep Ser, 464~Public Health, 465~Clin Geriatr Med, 466~J Altern Complem Med, 467~Pediatr Anesth, 468~Int J Artif Organs, 469~Platelets, 470~Med Clin-Barcelona, 471~J Am Assoc Gyn Lap, 472~Scand J Public Healt, 473~Health Soc Care Comm, 474~Cancer Detect Prev, 475~J Biosoc Sci, 476~J Aerosol Med, 477~Echocardiogr-J Card, 478~Deut Med Wochenschr, 479~Pulm Pharmacol Ther, 480~Eur J Nutr, 481~Blood Press Monit, 482~Int Fam Plan Perspec, 483~Heart Lung, 484~Med Anthropol Q, 485~Croat Med J, 486~J Bone Miner Metab, 487~Public Health Nurs, 488~Am J Health Behav, 489~Cardiol Young, 490~Perspect Biol Med, 491~J Travel Med, 492~J Nurs Scholarship, 493~Crit Care Clin, 494~Europace, 495~Herz, 496~Fetal Diagn Ther, 497~Pediatr Emerg Care, 498~Obstet Gynecol Surv, 499~Blood Pressure, 500~Cardiovasc Drug Ther, 501~J Nurs Educ, 502~Adv Nurs Sci, 503~Nutr Metab Cardiovas, 504~Vasc Med, 505~Cardiovasc Pathol, 506~Transpl Immunol, 507~Can J Appl Physiol, 508~Kennedy Inst Ethic J, 509~Physiol Meas, 510~Am J Perinat, 511~Can J Surg, 512~Xenotransplantation, 513~Western J Nurs Res, 514~Int J Nurs Stud, 515~Fund Clin Pharmacol, 516~Physiol Res, 517~J Eval Clin Pract, 518~Best Pract Res Cl En, 519~J Rural Health, 520~Gynecol Endocrinol, 521~Israel Med Assoc J, 522~Int J Sport Nutr Exe, 523~J Aging Stud, 524~Obstet Gyn Clin N Am, 525~Diabetes Educator, 526~J Exp Clin Canc Res, 527~J Interv Card Electr, 528~Pediatr Exerc Sci, 529~J Korean Med Sci, 530~Child Care Hlth Dev, 531~Drug Inf J, 532~J Perinat Med, 533~Dan Med Bull, 534~J Thromb Thrombolys, 535~Clev Clin J Med, 536~Ann Nutr Metab, 537~Semin Reprod Med, 538~J Healthc Manag, 539~Int J Behav Med, 540~Disasters, 541~Blood Purificat, 542~Wien Klin Wochenschr, 543~Altern Ther Health M, 544~Mt Sinai J Med, 545~Am J Nurs, 546~J Med Philos, 547~Complement Ther Med, 548~Clin Exp Hypertens, 549~Indian J Med Res, 550~Int Angiol, 551~Clin Drug Invest, 552~J Crit Care, 553~Jpn J Physiol, 554~J Palliative Care, 555~Am J Med Qual, 556~J Health Care Poor U, 557~J Nutr Sci Vitaminol, 558~Arch Mal Coeur Vaiss, 559~Chemotherapy, 560~J Trop Pediatrics, 561~Breast, 562~J Formos Med Assoc, 563~J Clin Ethic, 564~Arch Pediatrie, 565~J Health Commun, 566~Int J Vitam Nutr Res, 567~Med Oncol, 568~J Clin Pharm Ther, 569~Nurs Ethics, 570~Anaesthesist, 571~Eur J Gynaecol Oncol, 572~Nurs Educ Today, 573~J Neurosurg Anesth, 574~Renal Failure, 575~Cult Med Psychiat, 576~Chem Immunol, 577~Pathol Biol, 578~J Psychosoc Oncol, 579~Clin Rev Allerg Immu, 580~Can J Aging, 581~Emerg Med Clin N Am, 582~Surg Oncol, 583~Health Promot Int, 584~Nurs Outlook, 585~Zbl Chir, 586~Scand J Prim Health, 587~Clin Auton Res, 588~Onkologie, 589~Neth J Med, 590~Health Commun, 591~Ann Chir, 592~J Psychosom Obst Gyn, 593~Method Find Exp Clin, 594~Rev Saude Publ, 595~Sarcoidosis Vasc Dif, 596~Yonsei Med J, 597~J Commun Health, 598~Scand Cardiovasc J, 599~Allergy Asthma Proc, 600~Can Fam Physician, 601~Pediatr Hemat Oncol, 602~J Appl Gerontol, 603~J Electrocardiol, 604~Scand J Caring Sci, 605~Health Care Anal, 606~J Prof Nurs, 607~Eur Surg Res, 608~J Teach Phys Educ, 609~Rev Esp Cardiol, 610~Perfusion-Uk, 611~Adv Renal Replace Th, 612~Clin Anat, 613~Hypertens Pregnancy, 614~Best Pract Res Cl Ob, 615~Ann Trop Paediatr, 616~Women Health Iss, 617~Ann Clin Lab Sci, 618~Sci Eng Ethics, 619~Jpn J Infect Dis, 620~Tohoku J Exp Med, 621~Lung, 622~Clin Appl Thromb-Hem, 623~Acta Cardiol, 624~Geriatrics, 625~Behav Med, 626~J Aging Phys Activ, 627~J Clin Lab Anal, 628~Int Surg, 629~Diabetes Nutr Metab, 630~Eval Health Prof, 631~Saudi Med J, 632~Rev Med Interne, 633~Gesundheitswesen, 634~Drugs Today, 635~J Invest Allerg Clin, 636~J Cardiac Surg, 637~Primary Care, 638~Arch Gerontol Geriat, 639~Acta Chir Belg, 640~J Elder Abuse Negl, 641~Negotiation J, 642~Geriatr Nurs, 643~Vasa-J Vascular Dis, 644~J Hum Nutr Diet, 645~J Hum Movement Stud, 646~Adapt Phys Act Q, 647~Nutr Res Rev, 648~J Women Aging, 649~Ann Endocrinol-Paris, 650~Nurs Clin N Am, 651~Educ Gerontol, 652~Microbiologica, 653~Camb Q Healthc Ethic, 654~Jpn Heart J, 655~Int J Health Plan M, 656~Contrib Nephrol, 657~Hosp Med, 658~Midwifery, 659~Pharm World Sci, 660~Tex Heart I J, 661~Appl Nurs Res, 662~Acta Diabetol, 663~Trop Doct, 664~Med Lett Drugs Ther, 665~Clin Invest Med, 666~Kidney Blood Press R, 667~Phlebology, 668~Salud Publica Mexico, 669~Klin Padiatr, 670~Curr Ther Res Clin E, 671~Medicina-Buenos Aire, 672~J Public Health Pol, 673~Ann Fr Anesth, 674~B Cancer, 675~Monatsschr Kinderh, 676~Drug Exp Clin Res, 677~J Invest Surg, 678~Fam Community Health, 679~Natl Med J India, 680~J Int Med Res, 681~Brit J Biomed Sci, 682~Nurs Sci Quart, 683~Clin Lab Med, 684~Rev Med Chile, 685~Clin Hemorheol Micro, 686~Therapie, 687~Adv Ther, 688~Int J Obstet Anesth, 689~Pediatr Ann, 690~Ecol Food Nutr, 691~J Nurs Care Qual, 692~Endocrinologist, 693~Heart Vessels, 694~Med Klin, 695~Cardiovasc Drug Rev, 696~Rev Epidemiol Sante, 697~J Midwifery Wom Heal, 698~Anasth Intensiv Notf, 699~Ernahrungs-Umschau, 700~Sem Resp Crit Care M, 701~Med Inform Internet, 702~Int J Fertil Women M, 703~Rev Med Microbiol, 704~Acad Psychiatr, 705~Rev Mal Respir, 706~Soz Praventiv Med, 707~Theor Med Bioeth, 708~J Environ Health, 709~Wild Environ Med, 710~Undersea Hyperbar M, 711~Nefrologia, 712~Arch Latinoam Nutr, 713~Ann Biol Clin-Paris, 714~J Physiol Biochem, 715~J Trace Elem Exp Med, 716~J Music Ther, 717~Turkish J Pediatr, 718~J Mal Vascul, 719~Child Health Care, 720~Asian Pac J Allergy, 721~Anasth Intensivmed, 722~Geburtsh Frauenheilk, 723~Minim Invasiv Ther, 724~Rev Invest Clin, 725~Allergologie, 726~Magnesium Res, 727~Infect Med, 728~Internist, 729~Med Maladies Infect, 730~Strength Cond J, 731~Int J Clin Pharm Res, 732~Z Gerontol Geriatr, 733~Panminerva Med, 734~Acta Clin Belg, 735~Kardiologiya, 736~Upsala J Med Sci, 737~Sci Sport, 738~Terapevt Arkh, 739~J Mycol Med, 740~Nephrology, 741~Scot Med J, 742~Formulary, 743~Irish J Med Sci, 744~Lab Med, 745~Rev Clin Esp, 746~J Perinat Neonat Nur, 747~B Acad Nat Med Paris, 748~Ethiopian Med J, 749~Nephrologie, 750~Med Prin Pract, 751~Dialysis Transplant, 752~Chem Unserer Zeit, 753~Biol Sport, 754~Top Geriatr Rehabil, 755~Dm-Dis Mon, 756~Ann Saudi Med, 757~Chir Gastroenterol, 758~J Chir-Paris, 759~Acta Med Aust, 760~Rev Fr Allergol, 761~W Indian Med J, 762~Jpn J Phys Fit Sport, 763~S Afr J Surg, 764~Ejso, 765~Med Sport, 766~Man India\subsection*{{\large 3 Physics (10\%)}\label{module3}}
\noindent 1~Phys Rev Lett, 2~Phys Rev B, 3~Appl Phys Lett, 4~J Appl Phys, 5~J Chem Phys, 6~Phys Rev E, 7~J Phys Chem B, 8~Phys Rev D, 9~Langmuir, 10~Phys Rev A, 11~Macromolecules, 12~Chem Phys Lett, 13~Phys Lett B, 14~J Phys Chem A, 15~Adv Mater, 16~Chem Mater, 17~J Phys-Condens Mat, 18~Opt Lett, 19~Electron Lett, 20~Jpn J Appl Phys, 21~Polymer, 22~Nucl Phys B, 23~Acta Mater, 24~J Electrochem Soc, 25~Mat Sci Eng A-Struct, 26~Thin Solid Films, 27~Surf Sci, 28~Ieee Photonic Tech L, 29~J Cryst Growth, 30~J Appl Polym Sci, 31~Rev Mod Phys, 32~Appl Optics, 33~J Phys A-Math Gen, 34~J Colloid Interf Sci, 35~Phys Chem Chem Phys, 36~Europhys Lett, 37~Phys Lett A, 38~J Magn Magn Mater, 39~J Mater Chem, 40~Phys Rev C, 41~Nucl Instrum Meth A, 42~J Lightwave Technol, 43~J Am Ceram Soc, 44~Phys Plasmas, 45~Appl Surf Sci, 46~Phys Rep, 47~Rev Sci Instrum, 48~Nano Lett, 49~Surf Coat Tech, 50~Electrochim Acta, 51~Opt Commun, 52~Ieee T Electron Dev, 53~Physica B, 54~Synthetic Met, 55~J Alloy Compd, 56~J Power Sources, 57~J Non-Cryst Solids, 58~Nucl Phys A, 59~Physica A, 60~J Mater Res, 61~Nucl Instrum Meth B, 62~Scripta Mater, 63~J Phys D Appl Phys, 64~Physica C, 65~Ieee T Magn, 66~J Mater Sci, 67~J Vac Sci Technol B, 68~J Phys Soc Jpn, 69~Solid State Ionics, 70~Colloid Surface A, 71~Eur Phys J B, 72~J Electroanal Chem, 73~J Math Phys, 74~Physica D, 75~J Stat Phys, 76~J Phys B-At Mol Opt, 77~J Polym Sci Pol Chem, 78~Sensor Actuat B-Chem, 79~J Opt Soc Am B, 80~Ieee Electr Device L, 81~Mater Sci Forum, 82~Carbon, 83~Solid State Commun, 84~J Nucl Mater, 85~Metall Mater Trans A, 86~Appl Phys A-Mater, 87~Ieee J Sel Top Quant, 88~Chem Phys, 89~J Vac Sci Technol A, 90~Ieee J Quantum Elect, 91~J Eur Ceram Soc, 92~Opt Express, 93~Appl Phys B-Lasers O, 94~Classical Quant Grav, 95~Ieee T Appl Supercon, 96~Eur Phys J C, 97~J Solid State Chem, 98~Sensor Actuat A-Phys, 99~J Appl Crystallogr, 100~Sci Am, 101~Phys Status Solidi B, 102~Diam Relat Mater, 103~J Polym Sci Pol Phys, 104~Mat Sci Eng B-Solid, 105~Mater Lett, 106~Phys Status Solidi A, 107~J Comput Chem, 108~Electrochem Solid St, 109~Philos T Roy Soc A, 110~Comput Phys Commun, 111~Macromol Chem Phys, 112~Corros Sci, 113~Macromol Rapid Comm, 114~Nucl Fusion, 115~Mater Trans, 116~Nucl Phys B-Proc Sup, 117~Solid State Electron, 118~Nanotechnology, 119~Rep Prog Phys, 120~Eur Phys J D, 121~Meas Sci Technol, 122~J Mol Struc-Theochem, 123~Ieee T Nucl Sci, 124~Mol Phys, 125~Mrs Bull, 126~J Phys Chem Solids, 127~Opt Eng, 128~Supercond Sci Tech, 129~Polym Eng Sci, 130~Phys Today, 131~Jetp Lett+, 132~Microelectron Eng, 133~Physica E, 134~Fusion Eng Des, 135~Mater Chem Phys, 136~Prog Polym Sci, 137~J Microelectromech S, 138~J Photoch Photobio A, 139~Int J Quantum Chem, 140~J Lumin, 141~Eur Polym J, 142~Plasma Phys Contr F, 143~Adv Funct Mater, 144~Semicond Sci Tech, 145~Biomacromolecules, 146~Electrochem Commun, 147~J Quant Spectrosc Ra, 148~J Electron Mater, 149~Sol Energ Mat Sol C, 150~Int J Bifurcat Chaos, 151~Mater Res Bull, 152~Int J Mod Phys A, 153~Spectrochim Acta A, 154~Polym Degrad Stabil, 155~Chaos, 156~J Rheol, 157~Intermetallics, 158~Macromol Symp, 159~Ieee T Instrum Meas, 160~Ieee T Plasma Sci, 161~Ultramicroscopy, 162~Eur Phys J E, 163~Chemphyschem, 164~Isij Int, 165~Mater Sci Tech-Lond, 166~J Appl Electrochem, 167~Annu Rev Phys Chem, 168~J Non-Newton Fluid, 169~Corrosion, 170~J Exp Theor Phys+, 171~Ibm J Res Dev, 172~J Microsc-Oxford, 173~J Mod Optic, 174~Phys Scripta, 175~Surf Interface Anal, 176~J Micromech Microeng, 177~J Phys Iv, 178~Adv Colloid Interfac, 179~Curr Opin Colloid In, 180~Radiat Phys Chem, 181~Polym Int, 182~Eur Phys J A, 183~Key Eng Mater, 184~J Electron Spectrosc, 185~Mod Phys Lett A, 186~Comp Mater Sci, 187~Vacuum, 188~J Low Temp Phys, 189~Int J Hydrogen Energ, 190~Ferroelectrics, 191~Chaos Soliton Fract, 192~J Phys G Nucl Partic, 193~Opt Mater, 194~Mat Sci Eng R, 195~Int J Mod Phys B, 196~Radiat Meas, 197~Ann Phys-New York, 198~J Opt B-Quantum S O, 199~Astropart Phys, 200~J Raman Spectrosc, 201~Metall Mater Trans B, 202~Colloid Polym Sci, 203~Colloid Surface B, 204~Faraday Discuss, 205~Microelectron Reliab, 206~Phys Solid State+, 207~Prog Theor Phys, 208~Z Metallkd, 209~Acta Phys Pol B, 210~J Sol-Gel Sci Techn, 211~J Mol Spectrosc, 212~Plasma Sources Sci T, 213~Ceram Int, 214~J Synchrotron Radiat, 215~Phys World, 216~Mat Sci Eng C-Bio S, 217~Theor Chem Acc, 218~Appl Clay Sci, 219~Liq Cryst, 220~J Biomed Opt, 221~Adv Eng Mater, 222~Hyperfine Interact, 223~Am J Phys, 224~Cryst Res Technol, 225~Phil Mag Lett, 226~Curr Opin Solid St M, 227~Solid State Sci, 228~Chinese Phys Lett, 229~Adv Polym Sci, 230~Can J Phys, 231~React Funct Polym, 232~Oxid Met, 233~J Opt A-Pure Appl Op, 234~Rheol Acta, 235~Acta Crystallogr A, 236~Cryogenics, 237~Polym J, 238~J Ceram Soc Jpn, 239~Int J Theor Phys, 240~Polym Test, 241~Macromol Mater Eng, 242~Theor Math Phys+, 243~Prog Surf Sci, 244~Polym Composite, 245~Gen Relat Gravit, 246~Z Kristallogr, 247~J Mol Liq, 248~Surf Rev Lett, 249~Jom-Us, 250~Phys Atom Nucl+, 251~Polym Bull, 252~J Korean Phys Soc, 253~Chem J Chinese U, 254~Prog Part Nucl Phys, 255~Quantum Electron+, 256~Prog Org Coat, 257~Superlattice Microst, 258~Model Simul Mater Sc, 259~J Adhes Sci Technol, 260~Polym Advan Technol, 261~Rubber Chem Technol, 262~J Supercond, 263~Int J Mod Phys D, 264~Russ J Phys Chem+, 265~Opt Spectrosc+, 266~J Mater Sci-Mater El, 267~Tetsu To Hagane, 268~Laser Phys, 269~Metrologia, 270~Surf Eng, 271~Ieee T Semiconduct M, 272~Plasma Phys Rep+, 273~Integr Ferroelectr, 274~Opt Quant Electron, 275~Int J Chem Kinet, 276~Int J Mod Phys C, 277~X-Ray Spectrom, 278~J Therm Spray Techn, 279~Opt Laser Eng, 280~Macromol Theor Simul, 281~Tech Phys Lett+, 282~Int Rev Phys Chem, 283~Mater Charact, 284~Weld J, 285~Semiconductors+, 286~Prog Photovoltaics, 287~J Jpn I Met, 288~Solid State Phenom, 289~Microsc Microanal, 290~Prog Theor Phys Supp, 291~Tech Phys+, 292~Fortschr Phys, 293~Braz J Phys, 294~Eur Phys J-Appl Phys, 295~Acta Phys Sin-Ch Ed, 296~J Macromol Sci Pure, 297~J Electron Microsc, 298~Iee P-Optoelectron, 299~J Solid State Electr, 300~Electrochemistry, 301~J Electroceram, 302~Microelectron J, 303~Mater Corros, 304~Czech J Phys, 305~Mrs Internet J N S R, 306~J Nanosci Nanotechno, 307~Int J Adhes Adhes, 308~Found Phys, 309~Z Phys Chem, 310~Opt Laser Technol, 311~Z Naturforsch A, 312~Low Temp Phys+, 313~J Macromol Sci Phys, 314~Plasma Chem Plasma P, 315~Mater High Temp, 316~Calphad, 317~Infrared Phys Techn, 318~J New Mat Electr Sys, 319~Ann Phys-Berlin, 320~Mat Sci Semicon Proc, 321~Chem Vapor Depos, 322~J Mater Eng Perform, 323~Atom Data Nucl Data, 324~Phys Met Metallogr+, 325~Ironmak Steelmak, 326~Solid State Technol, 327~Top Appl Phys, 328~Int Mater Rev, 329~Laser Part Beams, 330~Mater Res Innov, 331~Am Ceram Soc Bull, 332~Solid State Nucl Mag, 333~Mod Phys Lett B, 334~Phys Chem Glasses, 335~Acta Phys Pol A, 336~Fractals, 337~Pramana-J Phys, 338~Scand J Metall, 339~Inst Phys Conf Ser, 340~Mol Simulat, 341~Wave Random Media, 342~Contrib Plasm Phys, 343~Defect Diffus Forum, 344~B Mater Sci, 345~Opt Rev, 346~J Res Natl Inst Stan, 347~Interface Sci, 348~Plast Rubber Compos, 349~Int J Refract Met H, 350~Radiat Eff Defect S, 351~Appl Magn Reson, 352~Opt Fiber Technol, 353~Russ J Appl Chem+, 354~High Perform Polym, 355~Crystallogr Rep+, 356~J Surfactants Deterg, 357~Colloid J+, 358~Sci Technol Weld Joi, 359~Semiconduct Semimet, 360~Chinese Phys, 361~High Temp+, 362~Stud Appl Math, 363~Commun Theor Phys, 364~J Plasma Phys, 365~J Adhesion, 366~Adv Quantum Chem, 367~Phase Transit, 368~Polym Sci Ser A+, 369~Contemp Phys, 370~Kaut Gummi Kunstst, 371~Iee P-Circ Dev Syst, 372~T Nonferr Metal Soc, 373~Inorg Mater+, 374~Iee P-Sci Meas Tech, 375~J Struct Chem+, 376~Ieee Circuits Device, 377~Microsyst Technol, 378~Optik, 379~Polym Polym Compos, 380~Int J Powder Metall, 381~High Pressure Res, 382~Int Polym Proc, 383~Found Phys Lett, 384~J Disper Sci Technol, 385~High Temp-High Press, 386~Few-Body Syst, 387~Instrum Exp Tech+, 388~Powder Metall, 389~J Cell Plast, 390~Ieee T Compon Pack T, 391~J Mater Sci Technol, 392~Brit Ceram T, 393~Scanning, 394~Israel J Chem, 395~Materialwiss Werkst, 396~Russ J Electrochem+, 397~Ann Chim-Sci Mat, 398~J Math Chem, 399~Sci China Ser B, 400~Res Chem Intermediat, 401~Part Part Syst Char, 402~Adv Polym Tech, 403~Int J Mod Phys E, 404~Dokl Phys, 405~Powder Diffr, 406~Indian J Pure Ap Phy, 407~J Nonlinear Opt Phys, 408~Acta Polym Sin, 409~Sci China Ser E, 410~Polym-Plast Technol, 411~Polimery-W, 412~J Optoelectron Adv M, 413~Acta Phys-Chim Sin, 414~Displays, 415~Int J Cast Metal Res, 416~Mater Manuf Process, 417~B Electrochem, 418~Laser Focus World, 419~J Ind Eng Chem, 420~New Diam Front C Tec, 421~J Laser Appl, 422~Springer Tr Mod Phys, 423~Compos Interface, 424~Int J Appl Electrom, 425~J Non-Equil Thermody, 426~Prog Cryst Growth Ch, 427~J Inorg Mater, 428~Adv Imag Elect Phys, 429~Solder Surf Mt Tech, 430~Tenside Surfact Det, 431~Micro, 432~Phys Low-Dimens Str, 433~J Adv Mater-Covina, 434~Glass Sci Technol, 435~Acta Phys Slovaca, 436~Zkg Int, 437~High Temp Mat Pr-Isr, 438~Glass Phys Chem+, 439~Rev Mex Fis, 440~J Vinyl Addit Techn, 441~Cell Polym, 442~Kobunshi Ronbunshu, 443~Mater Performance, 444~Rev Metall-Paris, 445~Chinese J Polym Sci, 446~J Opt Technol+, 447~Fiber Integrated Opt, 448~Sensor Mater, 449~Prot Met+, 450~Rev Metal Madrid, 451~T I Met Finish, 452~Plat Surf Finish, 453~Bol Soc Esp Ceram V, 454~Polym-Korea, 455~Photon Spectra, 456~Des Monomers Polym, 457~Phys Part Nuclei+, 458~Glass Technol, 459~T Indian I Metals, 460~J Polym Eng, 461~Stud Conserv, 462~High Energ Phys Nuc, 463~Anti-Corros Method M, 464~Int J Polym Anal Ch, 465~Kovove Mater, 466~Iran Polym J, 467~Ind Ceram, 468~Heat Treat Met-Uk, 469~High Energ Chem+, 470~Chinese J Phys, 471~J Rare Earth, 472~J Infrared Millim W, 473~High Temp Mater P-Us, 474~Laser Eng, 475~J Polym Res, 476~Tech Mess, 477~Corros Prevent Contr, 478~Electron Comm Jpn 2, 479~Opt Appl, 480~Nukleonika, 481~Acta Phys Hung Ns-H, 482~Meas Tech+, 483~Ceram-Silikaty, 484~Rare Metal Mat Eng, 485~Mater World, 486~Indian J Eng Mater S, 487~Metallofiz Nov Tekh+, 488~J Wuhan Univ Technol, 489~J Russ Laser Res, 490~Theor Found Chem Eng, 491~Chem Res Chinese U, 492~Met Sci Heat Treat+, 493~Ferroelectrics Lett, 494~Cfi-Ceram Forum Int, 495~J Plast Film Sheet, 496~J Elastom Plast, 497~Metall, 498~Mater Plast, 499~J Fusion Energ, 500~Ann Phys-Paris, 501~B Chem Soc Ethiopia, 502~Silic Ind\subsection*{{\large 4 Neuroscience (5.8\%)}\label{module4}}
\noindent 1~J Neurosci, 2~Neuron, 3~Neurology, 4~Nat Neurosci, 5~J Neurophysiol, 6~Brain Res, 7~J Physiol-London, 8~Ann Ny Acad Sci, 9~J Neurochem, 10~Neuroscience, 11~Eur J Neurosci, 12~Ann Neurol, 13~Neuroimage, 14~Brain, 15~J Pharmacol Exp Ther, 16~Neurosci Lett, 17~Neuroreport, 18~J Comp Neurol, 19~Nat Rev Neurosci, 20~Brit J Pharmacol, 21~Eur J Pharmacol, 22~Trends Neurosci, 23~Curr Opin Neurobiol, 24~Exp Brain Res, 25~Pain, 26~Cereb Cortex, 27~Arch Neurol-Chicago, 28~J Neurol Neurosur Ps, 29~Neuropsychologia, 30~Annu Rev Neurosci, 31~Psychopharmacology, 32~J Cognitive Neurosci, 33~J Neurosci Res, 34~Life Sci, 35~Neuropsychopharmacol, 36~Exp Neurol, 37~Epilepsia, 38~J Cerebr Blood F Met, 39~Trends Pharmacol Sci, 40~Neuropharmacology, 41~Behav Brain Res, 42~Neural Comput, 43~Clin Neurophysiol, 44~Movement Disord, 45~Prog Neurobiol, 46~Mol Cell Neurosci, 47~Pharmacol Rev, 48~J Neuroimmunol, 49~Physiol Behav, 50~Pharmacol Biochem Be, 51~Brain Res Rev, 52~Mol Brain Res, 53~J Neurobiol, 54~Hum Brain Mapp, 55~Muscle Nerve, 56~Psychophysiology, 57~Brain Res Bull, 58~J Neurol, 59~Glia, 60~Sleep, 61~J Neuropath Exp Neur, 62~Hippocampus, 63~Neurobiol Aging, 64~Behav Neurosci, 65~J Neurol Sci, 66~Neurosci Biobehav R, 67~Cell Tissue Res, 68~Cognitive Brain Res, 69~Prog Brain Res, 70~Neural Networks, 71~Peptides, 72~J Neurosci Meth, 73~Acta Neuropathol, 74~Neurobiol Dis, 75~Synapse, 76~Dev Med Child Neurol, 77~Horm Behav, 78~Ieee T Bio-Med Eng, 79~Curr Opin Neurol, 80~J Neurotraum, 81~J Neuroendocrinol, 82~Neurochem Res, 83~Cephalalgia, 84~Learn Memory, 85~Dev Brain Res, 86~Brain Cognition, 87~Neuropsychology, 88~Epilepsy Res, 89~J Child Neurol, 90~Headache, 91~Neurochem Int, 92~J Comp Physiol A, 93~Neurosci Res, 94~Psychoneuroendocrino, 95~J Int Neuropsych Soc, 96~N-S Arch Pharmacol, 97~Biol Cybern, 98~Acta Neurol Scand, 99~Regul Peptides, 100~Chem Senses, 101~Brain Pathol, 102~J Neural Transm, 103~J Anat, 104~Int J Psychophysiol, 105~Neuromuscular Disord, 106~Biol Psychol, 107~Neurobiol Learn Mem, 108~Eur J Neurol, 109~J Clin Exp Neuropsyc, 110~Mult Scler, 111~Neurocomputing, 112~Neuroendocrinology, 113~Visual Neurosci, 114~Pediatr Neurol, 115~J Neurovirol, 116~Cortex, 117~Dev Psychobiol, 118~Clin J Pain, 119~Neurotoxicology, 120~J Biol Rhythm, 121~Neurotoxicol Teratol, 122~Dement Geriatr Cogn, 123~Neuropath Appl Neuro, 124~J Neurocytol, 125~J Sleep Res, 126~Can J Physiol Pharm, 127~Dev Neuropsychol, 128~Brain Dev-Jpn, 129~J Comput Neurosci, 130~J Clin Neurophysiol, 131~Eur Neurol, 132~Neurol Res, 133~Behav Pharmacol, 134~J Mol Neurosci, 135~Ment Retard Dev D R, 136~Int J Dev Neurosci, 137~Seizure-Eur J Epilep, 138~Neuropediatrics, 139~Mol Neurobiol, 140~Front Neuroendocrin, 141~Brain Behav Evolut, 142~Neuroscientist, 143~J Pineal Res, 144~J Chem Neuroanat, 145~Neurocase, 146~Cell Mol Neurobiol, 147~Dev Neurosci-Basel, 148~Clin Neuropharmacol, 149~Hum Movement Sci, 150~Brain Behav Immun, 151~Neuroepidemiology, 152~Acta Pharmacol Sin, 153~Arch Clin Neuropsych, 154~Chronobiol Int, 155~Sleep Med Rev, 156~Can J Neurol Sci, 157~Auton Neurosci-Basic, 158~Alz Dis Assoc Dis, 159~Eur J Pain, 160~Neuropeptides, 161~Int Rev Neurobiol, 162~Neurol Sci, 163~J Physiol-Paris, 164~Network-Comp Neural, 165~Rev Neurosci, 166~Soc Res, 167~Pharmacology, 168~Parkinsonism Relat D, 169~J Neuroimaging, 170~Int J Neurosci, 171~Neurol Clin, 172~J Pain, 173~Brain Res Protoc, 174~Clin Neurol Neurosur, 175~Rev Neurol-France, 176~Brain Topogr, 177~Semin Neurol, 178~Amyotroph Lateral Sc, 179~Somatosens Mot Res, 180~J Physiol Pharmacol, 181~J Peripher Nerv Syst, 182~Neurophysiol Clin, 183~Rev Neurologia, 184~Neurogenetics, 185~Neuropathology, 186~Child Neuropsychol, 187~Arq Neuro-Psiquiat, 188~Epileptic Disord, 189~Pol J Pharmacol, 190~Motor Control, 191~Neuroimmunomodulat, 192~Neuroendocrinol Lett, 193~Restor Neurol Neuros, 194~J Vestibul Res-Equil, 195~Psychol Belg, 196~Arch Ital Biol, 197~Behav Neurol, 198~J Psychophysiol, 199~Acta Neurobiol Exp, 200~Schmerz, 201~Neuromodulation, 202~J Recept Sig Transd, 203~Int J Hum-Comput Int, 204~B Exp Biol Med+, 205~Funct Neurol, 206~Neurologist, 207~Biol Rhythm Res, 208~Nervenheilkunde, 209~Integr Phys Beh Sci, 210~Acta Neurol Belg, 211~Acta Med Okayama, 212~Salud Ment, 213~Adv Anat Embryol Cel, 214~Folia Neuropathol, 215~Acta Biol Hung, 216~Zh Vyssh Nerv Deyat+, 217~Neurosci Res Commun, 218~Aktuel Neurol, 219~J Evol Biochem Phys+, 220~Rev Neuropsychol, 221~Chinese J Physiol, 222~Acupuncture Electro, 223~Zh Nevropatol Psikh, 224~Cesk Slov Neurol N\subsection*{{\large 5 Ecology \& Evolution (4.3\%)}\label{module5}}
\noindent 1~Ecology, 2~P Roy Soc B-Biol Sci, 3~Mar Ecol-Prog Ser, 4~Oecologia, 5~Trends Ecol Evol, 6~Evolution, 7~Mol Ecol, 8~Oikos, 9~Conserv Biol, 10~Am Nat, 11~Ecol Appl, 12~Limnol Oceanogr, 13~J Exp Biol, 14~Forest Ecol Manag, 15~Philos T Roy Soc B, 16~Anim Behav, 17~Mol Phylogenet Evol, 18~Am J Bot, 19~Can J Fish Aquat Sci, 20~Biol Conserv, 21~New Phytol, 22~Hydrobiologia, 23~Aquaculture, 24~Mar Biol, 25~Syst Biol, 26~J Anim Ecol, 27~Ecol Lett, 28~Can J Zool, 29~Global Change Biol, 30~Plant Cell Environ, 31~Can J Forest Res, 32~Behav Ecol Sociobiol, 33~J Ecol, 34~J Theor Biol, 35~J Appl Ecol, 36~J Fish Biol, 37~Freshwater Biol, 38~J Exp Mar Biol Ecol, 39~Bioscience, 40~J Evolution Biol, 41~Ecol Model, 42~Ann Bot-London, 43~J Econ Entomol, 44~Heredity, 45~J Phycol, 46~J Wildlife Manage, 47~Behav Ecol, 48~Biol J Linn Soc, 49~J Biogeogr, 50~Funct Ecol, 51~Tree Physiol, 52~Auk, 53~J Chem Ecol, 54~Ices J Mar Sci, 55~Estuar Coast Shelf S, 56~Ecography, 57~Annu Rev Entomol, 58~Ecol Monogr, 59~J Veg Sci, 60~Environ Entomol, 61~Plant Ecol, 62~Insect Biochem Molec, 63~Comp Biochem Phys A, 64~Condor, 65~J Insect Physiol, 66~Ecosystems, 67~Biodivers Conserv, 68~Int J Plant Sci, 69~J Zool, 70~Gen Comp Endocr, 71~J Plankton Res, 72~Can J Bot, 73~Aquat Microb Ecol, 74~J Mammal, 75~T Am Fish Soc, 76~J Hered, 77~Fish Res, 78~Wildlife Soc B, 79~Entomol Exp Appl, 80~Comp Biochem Phys B, 81~Ann Entomol Soc Am, 82~Estuaries, 83~Genetica, 84~Arch Hydrobiol, 85~Ecol Entomol, 86~J Mar Biol Assoc Uk, 87~Ambio, 88~Bot J Linn Soc, 89~Syst Bot, 90~B Mar Sci, 91~Theor Popul Biol, 92~Insect Mol Biol, 93~Ethology, 94~Naturwissenschaften, 95~Environ Biol Fish, 96~Landscape Ecol, 97~Plant Syst Evol, 98~Biol Control, 99~J N Am Benthol Soc, 100~Copeia, 101~Evol Ecol Res, 102~Biol Rev, 103~J Avian Biol, 104~Mar Freshwater Res, 105~Polar Biol, 106~Ibis, 107~Cladistics, 108~Environ Manage, 109~J Arid Environ, 110~Aquat Bot, 111~Biotropica, 112~Plant Biology, 113~J Morphol, 114~Behaviour, 115~Biol Bull-Us, 116~Physiol Biochem Zool, 117~Coral Reefs, 118~Wetlands, 119~Global Ecol Biogeogr, 120~J Forest, 121~Aquac Res, 122~N Am J Fish Manage, 123~Phycologia, 124~J Trop Ecol, 125~Fish B-Noaa, 126~Insect Soc, 127~Bird Study, 128~Forest Sci, 129~Ann Mo Bot Gard, 130~Apidologie, 131~Mar Mammal Sci, 132~Evol Ecol, 133~J Invertebr Pathol, 134~J Herpetol, 135~Ecoscience, 136~Fish Oceanogr, 137~J Crustacean Biol, 138~Landscape Urban Plan, 139~Restor Ecol, 140~J Shellfish Res, 141~J Comp Physiol B, 142~Zool Sci, 143~Fisheries Sci, 144~Zool J Linn Soc-Lond, 145~J Math Biol, 146~Eur J Phycol, 147~J Range Manage, 148~Scand J Forest Res, 149~Anim Conserv, 150~Math Biosci, 151~Taxon, 152~Austral Ecol, 153~Am Midl Nat, 154~J Sea Res, 155~B Entomol Res, 156~Mycorrhiza, 157~Bot Mar, 158~Aust J Bot, 159~J Nat Hist, 160~J Insect Behav, 161~Trees-Struct Funct, 162~Fla Entomol, 163~Can Entomol, 164~Ecol Res, 165~Acta Oecol, 166~Ann Forest Sci, 167~Arch Insect Biochem, 168~Biocontrol Sci Techn, 169~Sci Mar, 170~Physiol Entomol, 171~New Zeal J Mar Fresh, 172~Environ Exp Bot, 173~Environ Conserv, 174~Mar Biotechnol, 175~J Arachnol, 176~Fisheries, 177~Funct Plant Biol, 178~J Appl Phycol, 179~J Great Lakes Res, 180~Wildlife Res, 181~Ocean Coast Manage, 182~Eur J Entomol, 183~Rev Fish Biol Fisher, 184~Bot Rev, 185~P Biol Soc Wash, 186~B Math Biol, 187~S Afr J Sci, 188~Zool Scr, 189~Bryologist, 190~Mar Policy, 191~Fish Physiol Biochem, 192~Wilson Bull, 193~Ann Zool Fenn, 194~Herpetologica, 195~J Field Ornithol, 196~Nova Hedwigia, 197~J Plant Res, 198~Crustaceana, 199~J World Aquacult Soc, 200~Southwest Nat, 201~Forest Chron, 202~Appl Entomol Zool, 203~Aquat Living Resour, 204~Sociobiology, 205~Aquat Conserv, 206~Arctic, 207~J Stored Prod Res, 208~Oryx, 209~Aust J Zool, 210~Int J Wildland Fire, 211~Acta Theriol, 212~Nat Area J, 213~Ethol Ecol Evol, 214~Biocontrol, 215~Flora, 216~J Mollus Stud, 217~Photosynthetica, 218~Econ Bot, 219~Aust Syst Bot, 220~Silva Fenn, 221~Int Rev Hydrobiol, 222~J Entomol Sci, 223~Syst Entomol, 224~Can Field Nat, 225~Wildlife Biol, 226~Helgoland Mar Res, 227~Aquat Sci, 228~Waterbirds, 229~J Freshwater Ecol, 230~J Appl Ichthyol, 231~Aquacult Nutr, 232~Ecol Freshw Fish, 233~J Therm Biol, 234~Afr J Ecol, 235~J Zool Syst Evol Res, 236~Acta Zool-Stockholm, 237~Forestry, 238~Veliger, 239~Sarsia, 240~Zool Anz, 241~Mammal Rev, 242~Ardea, 243~New Forest, 244~Novon, 245~Folia Geobot, 246~Zoo Biol, 247~Rev Chil Hist Nat, 248~Folia Zool, 249~J Kansas Entomol Soc, 250~Emu, 251~Nippon Suisan Gakk, 252~Coast Manage, 253~Northwest Sci, 254~J Roy Soc New Zeal, 255~New Zeal J Bot, 256~Cah Biol Mar, 257~Aquacult Eng, 258~Nord J Bot, 259~Ophelia, 260~Am Bee J, 261~Amphibia-Reptilia, 262~Invertebr Biol, 263~Ann Soc Entomol Fr, 264~Int J Pest Manage, 265~Rev Suisse Zool, 266~Rev Biol Trop, 267~J Raptor Res, 268~Ital J Zool, 269~Phys Geogr, 270~Silvae Genet, 271~Raffles B Zool, 272~J Apicult Res, 273~Grana, 274~J Torrey Bot Soc, 275~Coleopts Bull, 276~Brittonia, 277~Israel J Zool, 278~Ann Limnol-Int J Lim, 279~Fisheries Manag Ecol, 280~Mammalia, 281~Malacologia, 282~Aquat Insect, 283~N Am J Aquacult, 284~New Zeal J Zool, 285~Zool Stud, 286~Ichthyol Res, 287~Zool Zh, 288~Aquacult Int, 289~Cryptogamie Algol, 290~Southwest Entomol, 291~Biologia, 292~Zoomorphology, 293~S Afr J Bot, 294~Symbiosis, 295~Vie Milieu, 296~Belg J Zool, 297~Ann Bot Fenn, 298~Ornis Fennica, 299~New Zeal J Ecol, 300~Cienc Mar, 301~J Bryol, 302~Forstwiss Centralbl, 303~Am Fern J, 304~Cybium, 305~Herpetol J, 306~Ostrich, 307~Aust J Entomol, 308~Deut Entomol Z, 309~Ocean Dev Int Law, 310~Allg Forst Jagdztg, 311~Israel J Plant Sci, 312~Stud Neotrop Fauna E, 313~Biochem Genet, 314~Rev Ecol-Terre Vie, 315~Arch Fish Mar Res, 316~J Ethol, 317~Nautilus, 318~Great Lakes Entomol, 319~S Afr J Wildl Res, 320~Phytocoenologia, 321~Zoology, 322~Amazoniana, 323~Trop Zool, 324~Folia Biol-Krakow, 325~Bee World, 326~Acta Soc Bot Pol, 327~Phyton-Ann Rei Bot A, 328~Tex J Sci, 329~Afr Entomol, 330~Ekol Bratislava, 331~J Aquat Plant Manage, 332~Jpn J Appl Entomol Z, 333~Interciencia, 334~Bot Helv, 335~Entomol Gen, 336~J Conchol, 337~Rhodora, 338~Calif Fish Game, 339~Blumea, 340~J Adv Zool, 341~Bothalia, 342~Acta Bot Gallica, 343~Ohio J Sci, 344~Isr J Aquacult-Bamid, 345~Belg J Bot, 346~Russ J Ecol+, 347~Acta Zool Acad Sci H, 348~Odonatologica, 349~Entomol Fennica\subsection*{{\large 6 Economics (3.5\%)}\label{module6}}
\noindent 1~Am Econ Rev, 2~J Financ, 3~Q J Econ, 4~Econometrica, 5~J Polit Econ, 6~J Financ Econ, 7~Econ J, 8~J Econometrics, 9~J Econ Perspect, 10~J Monetary Econ, 11~Rev Econ Stud, 12~Rev Econ Stat, 13~J Econ Lit, 14~Eur Econ Rev, 15~Rev Financ Stud, 16~J Int Econ, 17~J Econ Theory, 18~J Public Econ, 19~World Dev, 20~Rand J Econ, 21~J Labor Econ, 22~J Dev Econ, 23~Int Econ Rev, 24~Econ Lett, 25~J Hum Resour, 26~J Bus Econ Stat, 27~J Health Econ, 28~J Econ Behav Organ, 29~J Financ Quant Anal, 30~J Urban Econ, 31~Ind Labor Relat Rev, 32~Game Econ Behav, 33~J Law Econ Organ, 34~J Bus, 35~Natl Tax J, 36~Am J Agr Econ, 37~J Law Econ, 38~J Money Credit Bank, 39~Economet Theor, 40~J Econ Growth, 41~Ecol Econ, 42~J Bank Financ, 43~World Bank Econ Rev, 44~J Dev Stud, 45~J Account Econ, 46~J Appl Econom, 47~Public Choice, 48~J Environ Econ Manag, 49~Econ Theor, 50~Int J Ind Organ, 51~J Accounting Res, 52~Can J Econ, 53~J Econ Dyn Control, 54~Brookings Pap Eco Ac, 55~Math Financ, 56~J Ind Econ, 57~Appl Econ, 58~Ind Relat, 59~J Risk Uncertainty, 60~Econ Inq, 61~Energ Policy, 62~South Econ J, 63~Oxford B Econ Stat, 64~Econ Educ Rev, 65~Health Econ, 66~J Policy Anal Manag, 67~Reg Sci Urban Econ, 68~Land Econ, 69~J Econ Manage Strat, 70~Soc Choice Welfare, 71~J Int Money Financ, 72~Oxford Econ Pap, 73~Account Rev, 74~Economica, 75~Econ Dev Cult Change, 76~J Comp Econ, 77~Financ Manage, 78~Oxford Rev Econ Pol, 79~Scand J Econ, 80~Cambridge J Econ, 81~Environ Resour Econ, 82~World Econ, 83~J Math Econ, 84~Ids Bull-I Dev Stud, 85~Mon Labor Rev, 86~Dev Change, 87~Int J Game Theory, 88~Math Soc Sci, 89~Energy J, 90~J Popul Econ, 91~Brit J Ind Relat, 92~J Financ Intermed, 93~Econ Transit, 94~Agr Econ, 95~Small Bus Econ, 96~Int Tax Public Finan, 97~Rev Ind Organ, 98~J Regional Sci, 99~Energ Econ, 100~Resour Energy Econ, 101~Real Estate Econ, 102~Int J Forecasting, 103~J Econ Psychol, 104~Theor Decis, 105~J Labor Res, 106~J Forecasting, 107~Macroecon Dyn, 108~Kyklos, 109~J Real Estate Financ, 110~J Econ Educ, 111~Int Rev Law Econ, 112~Scot J Polit Econ, 113~J Regul Econ, 114~J Futures Markets, 115~J Afr Econ, 116~Eur Rev Agric Econ, 117~J Portfolio Manage, 118~J Hous Econ, 119~J Econ, 120~Appl Econ Lett, 121~Food Policy, 122~J Jpn Int Econ, 123~J Inst Theor Econ, 124~J Risk Insur, 125~J Agr Econ, 126~China Econ Rev, 127~Can Public Pol, 128~J Econ Issues, 129~Resour Policy, 130~Int Labour Rev, 131~Jpn World Econ, 132~Int J Manpower, 133~J Policy Model, 134~Eur J Ind Relat, 135~J Macroecon, 136~Relat Ind-Ind Relat, 137~Contemp Econ Policy, 138~Econ Philos, 139~J Evol Econ, 140~S Afr J Econ, 141~Am J Econ Sociol, 142~Econ Model, 143~Auditing-J Pract Th, 144~Econ Rec, 145~Aust J Agr Resour Ec, 146~B Indones Econ Stud, 147~Can J Dev Stud, 148~Can J Agr Econ, 149~Jahrb Natl Stat, 150~Eastern Eur Econ, 151~J Post Keynesian Ec, 152~Economist-Netherland, 153~Int J Sust Dev World, 154~Betrieb Forsch Prax, 155~Polit Ekon, 156~Dev Econ, 157~Open Econ Rev, 158~Ekon Cas, 159~Trimest Econ\subsection*{{\large 7 Geosciences (3.4\%)}\label{module7}}
\noindent 1~J Geophys Res, 2~Geophys Res Lett, 3~Earth Planet Sc Lett, 4~Geology, 5~J Climate, 6~Geochim Cosmochim Ac, 7~Atmos Environ, 8~J Atmos Sci, 9~Chem Geol, 10~Tectonophysics, 11~Palaeogeogr Palaeocl, 12~Quaternary Sci Rev, 13~Mon Weather Rev, 14~Geophys J Int, 15~B Am Meteorol Soc, 16~J Phys Oceanogr, 17~Remote Sens Environ, 18~Mar Geol, 19~Geol Soc Am Bull, 20~Ieee T Geosci Remote, 21~Int J Remote Sens, 22~Q J Roy Meteor Soc, 23~Am Mineral, 24~Global Biogeochem Cy, 25~Icarus, 26~Clim Dynam, 27~Contrib Mineral Petr, 28~B Seismol Soc Am, 29~J Petrol, 30~Geophysics, 31~J Appl Meteorol, 32~Climatic Change, 33~Quaternary Res, 34~Precambrian Res, 35~Agr Forest Meteorol, 36~J Atmos Ocean Tech, 37~J Volcanol Geoth Res, 38~Paleoceanography, 39~Sediment Geol, 40~Int J Climatol, 41~Deep-Sea Res Pt Ii, 42~Deep-Sea Res Pt I, 43~Lithos, 44~J Struct Geol, 45~Tectonics, 46~Phys Earth Planet In, 47~Geomorphology, 48~J Air Waste Manage, 49~Cont Shelf Res, 50~Org Geochem, 51~Prog Oceanogr, 52~J Geol Soc London, 53~Sedimentology, 54~Adv Space Res, 55~Holocene, 56~Global Planet Change, 57~Mar Chem, 58~J Sediment Res, 59~Meteorit Planet Sci, 60~Can J Earth Sci, 61~Photogramm Eng Rem S, 62~Bound-Lay Meteorol, 63~Rev Geophys, 64~Aapg Bull, 65~Space Sci Rev, 66~Earth-Sci Rev, 67~Ann Glaciol, 68~J Aerosol Sci, 69~Tellus B, 70~Aerosol Sci Tech, 71~Econ Geol, 72~J Quaternary Sci, 73~Curr Sci India, 74~Ann Geophys-Germany, 75~J Metamorph Geol, 76~Earth Surf Proc Land, 77~Chinese Sci Bull, 78~J Glaciol, 79~Quatern Int, 80~J Marine Syst, 81~J Geol, 82~Eur J Mineral, 83~Pure Appl Geophys, 84~J Atmos Sol-Terr Phy, 85~Terra Nova, 86~Paleobiology, 87~Am J Sci, 88~J Vertebr Paleontol, 89~Int J Earth Sci, 90~Planet Space Sci, 91~J Paleontol, 92~Annu Rev Earth Pl Sc, 93~Mar Micropaleontol, 94~B Volcanol, 95~Comput Geosci-Uk, 96~Radio Sci, 97~Climate Res, 98~Can Mineral, 99~J Coastal Res, 100~Palaios, 101~J Hydrometeorol, 102~Aust J Earth Sci, 103~J Paleolimnol, 104~Mar Petrol Geol, 105~J Meteorol Soc Jpn, 106~Mineral Mag, 107~Phys Chem Miner, 108~Geol Mag, 109~Tellus A, 110~Miner Deposita, 111~Clay Clay Miner, 112~J Geodyn, 113~Weather Forecast, 114~Radiocarbon, 115~J Atmos Chem, 116~Rev Palaeobot Palyno, 117~Geophys Prospect, 118~Arct Antarct Alp Res, 119~Palaeontology, 120~Earth Planets Space, 121~Global Environ Chang, 122~J Asian Earth Sci, 123~Boreas, 124~Theor Appl Climatol, 125~J Appl Geophys, 126~J Afr Earth Sci, 127~Atmos Res, 128~Int Geol Rev, 129~J Mar Res, 130~J S Am Earth Sci, 131~Meteorol Atmos Phys, 132~Basin Res, 133~Antarct Sci, 134~Geobios-Lyon, 135~Cretaceous Res, 136~Miner Petrol, 137~J Geochem Explor, 138~Isprs J Photogramm, 139~New Zeal J Geol Geop, 140~B Soc Geol Fr, 141~Sci China Ser D, 142~Permafrost Periglac, 143~Geo-Mar Lett, 144~Clay Miner, 145~Isl Arc, 146~Prog Phys Geog, 147~Petrol Geosci, 148~Atmos Ocean, 149~Dynam Atmos Oceans, 150~Schweiz Miner Petrog, 151~J Foramin Res, 152~Int J Biometeorol, 153~Episodes, 154~Polar Res, 155~Cold Reg Sci Technol, 156~J Seismol, 157~Lethaia, 158~Geodin Acta, 159~Gondwana Res, 160~Micropaleontology, 161~J Geol Soc India, 162~T Roy Soc Edin-Earth, 163~Int J Coal Geol, 164~Geol Geofiz+, 165~Z Geomorphol, 166~B Can Petrol Geol, 167~Nonlinear Proc Geoph, 168~Nat Hazards, 169~J Geodesy, 170~Acta Palaeontol Pol, 171~Acta Geol Sin-Engl, 172~Norw J Geol, 173~J Petrol Geol, 174~Cim Bull, 175~Facies, 176~Geol J, 177~S Afr J Geol, 178~Ore Geol Rev, 179~Meteorol Appl, 180~Ameghiniana, 181~Eclogae Geol Helv, 182~Acta Petrol Sin, 183~Terr Atmos Ocean Sci, 184~Phys Chem Earth, 185~Surv Geophys, 186~Meteorol Z, 187~Geochem J, 188~Veg Hist Archaeobot, 189~Neues Jahrb Geol P-A, 190~Prog Nat Sci, 191~Gff, 192~Geol Carpath, 193~Adv Atmos Sci, 194~P Geologist Assoc, 195~J Seism Explor, 196~Riv Ital Paleontol S, 197~Alcheringa, 198~Geothermics, 199~Chinese J Geophys-Ch, 200~Neues Jb Miner Monat, 201~Stud Geophys Geod, 202~Photogramm Rec, 203~P Indian As-Earth, 204~Aquat Geochem, 205~Petrology+, 206~Nuovo Cimento C, 207~Aust Meteorol Mag, 208~Rev Geol Chile, 209~Int J Environ Pollut, 210~Carbonate Evaporite, 211~Stratigr Geo Correl+, 212~J Cold Reg Eng, 213~Neth J Geosci, 214~J Micropalaeontol, 215~Space Policy, 216~Neues Jb Miner Abh, 217~Atmosfera, 218~Scot J Geol, 219~Chem Erde-Geochem, 220~Resour Geol, 221~Geosci Can, 222~Mar Georesour Geotec, 223~Indian J Mar Sci\subsection*{{\large 8 Psychology (2.8\%)}\label{module8}}
\noindent 1~J Pers Soc Psychol, 2~Am Psychol, 3~Psychol Bull, 4~Child Dev, 5~Pers Soc Psychol B, 6~Psychol Sci, 7~Psychol Rev, 8~Dev Psychol, 9~Trends Cogn Sci, 10~Annu Rev Psychol, 11~J Exp Psychol Learn, 12~Vision Res, 13~Cognition, 14~Pers Indiv Differ, 15~J Exp Psychol Human, 16~Mem Cognition, 17~Psychon B Rev, 18~Behav Brain Sci, 19~J Mem Lang, 20~J Exp Soc Psychol, 21~Dev Psychopathol, 22~Percept Psychophys, 23~Health Psychol, 24~J Exp Psychol Gen, 25~Organ Behav Hum Dec, 26~J Soc Issues, 27~Psychol Methods, 28~J Pers, 29~Psychol Aging, 30~Brain Lang, 31~Cognition Emotion, 32~J Gerontol B-Psychol, 33~Eur J Soc Psychol, 34~J Appl Soc Psychol, 35~Cognitive Psychol, 36~Curr Dir Psychol Sci, 37~J Exp Child Psychol, 38~Pers Soc Psychol Rev, 39~Q J Exp Psychol-A, 40~J Fam Psychol, 41~Psychol Rep, 42~Sex Roles, 43~J Couns Psychol, 44~Educ Psychol Meas, 45~Psychol Inq, 46~J Cross Cult Psychol, 47~Perception, 48~Psychometrika, 49~J Res Pers, 50~Cognitive Sci, 51~Behav Res Meth Ins C, 52~Appl Cognitive Psych, 53~Acta Psychol, 54~Evol Hum Behav, 55~Eur J Personality, 56~Int J Behav Dev, 57~Soc Dev, 58~Brit J Soc Psychol, 59~Am J Commun Psychol, 60~Percept Motor Skill, 61~Intelligence, 62~J Youth Adolescence, 63~Lang Cognitive Proc, 64~Memory, 65~Appl Psych Meas, 66~Cogn Neuropsychol, 67~J Adolescence, 68~J Exp Anal Behav, 69~Vis Cogn, 70~J Exp Psychol Anim B, 71~Soc Cognition, 72~Behav Process, 73~J Soc Psychol, 74~Pers Relationship, 75~Cognitive Dev, 76~Psychol Women Quart, 77~J Math Psychol, 78~Psychol Res-Psych Fo, 79~Hum Factors, 80~J Adolescent Res, 81~Psychol Health, 82~J Couns Dev, 83~J Educ Behav Stat, 84~Couns Psychol, 85~Multivar Behav Res, 86~J Educ Meas, 87~Brit J Dev Psychol, 88~J Soc Pers Relat, 89~J Exp Psychol-Appl, 90~Adolescence, 91~Brit J Psychol, 92~J Psycholinguist Res, 93~J Res Adolescence, 94~Arch Sex Behav, 95~J Behav Decis Making, 96~J Community Psychol, 97~J Comp Psychol, 98~Merrill Palmer Quart, 99~Infant Behav Dev, 100~Aggressive Behav, 101~J Motor Behav, 102~J Early Adolescence, 103~J Soc Clin Psychol, 104~Comput Hum Behav, 105~Inf Mental Hlth J, 106~Environ Behav, 107~J Marital Fam Ther, 108~J Sex Res, 109~Dev Rev, 110~Basic Appl Soc Psych, 111~Conscious Cogn, 112~J Child Lang, 113~Fam Relat, 114~Fam Process, 115~J Sport Exercise Psy, 116~Cyberpsychol Behav, 117~J Psychol, 118~Language, 119~Motiv Emotion, 120~Int J Intercult Rel, 121~J Environ Psychol, 122~Int J Psychol, 123~J Appl Dev Psychol, 124~Eur J Cogn Psychol, 125~Theor Psychol, 126~Can J Exp Psychol, 127~Q J Exp Psychol-B, 128~Soc Indic Res, 129~Discourse Process, 130~J Consciousness Stud, 131~Appl Meas Educ, 132~Scand J Psychol, 133~Am J Psychol, 134~Youth Soc, 135~Hum Dev, 136~Hispanic J Behav Sci, 137~Psychol Rec, 138~Int J Aging Hum Dev, 139~J Homosexual, 140~Hum Nature-Int Bios, 141~J Genet Psychol, 142~Soc Sci Comput Rev, 143~Teach Psychol, 144~Ethos, 145~Meas Eval Couns Dev, 146~J Nonverbal Behav, 147~J Appl Sport Psychol, 148~Spatial Vision, 149~New Ideas Psychol, 150~Eur J Psychol Assess, 151~Brit J Health Psych, 152~Anxiety Stress Copin, 153~Exp Aging Res, 154~Soc Behav Personal, 155~J Theor Soc Behav, 156~Learn Motiv, 157~Behav Analyst, 158~Can J Behav Sci, 159~Ann Psychol, 160~J Multicult Couns D, 161~Fem Psychol, 162~Aging Neuropsychol C, 163~Mind Lang, 164~Sport Psychol, 165~Linguistics, 166~Cult Psychol, 167~J Neurolinguist, 168~Lang Speech, 169~Int J Sport Psychol, 170~J Gen Psychol, 171~J Community Appl Soc, 172~J Classif, 173~Ecol Psychol, 174~Genet Soc Gen Psych, 175~Am J Fam Ther, 176~J Psychol Theol, 177~Swiss J Psychol, 178~Psychologist, 179~Soc Sci Inform, 180~J Adult Dev, 181~Psicothema, 182~Color Res Appl, 183~J Fam Ther, 184~Connect Sci, 185~Z Entwickl Padagogis, 186~Ethics Behav, 187~Curr Psychol, 188~Psychol Rundsch, 189~Lingua, 190~Contemp Fam Ther, 191~Z Sozialpsychol, 192~Cah Psychol Cogn, 193~Aust J Psychol, 194~Philos Psychol, 195~Psychol Erz Unterr, 196~Humor, 197~Zygon, 198~J Mind Behav, 199~Z Psychol, 200~J Constr Psychol, 201~New Zeal J Psychol, 202~Women Ther, 203~J Relig Health, 204~J Humanist Psychol, 205~Psychologia, 206~J Parapsychol, 207~Jpn Psychol Res, 208~Stud Psychol, 209~Cesk Psychol, 210~Rev Lat Am Psicol\subsection*{{\large 9 Chemistry (2.7\%)}\label{module9}}
\noindent 1~J Am Chem Soc, 2~Angew Chem Int Edit, 3~Chem Commun, 4~Chem Rev, 5~J Org Chem, 6~Tetrahedron Lett, 7~Org Lett, 8~Inorg Chem, 9~Tetrahedron, 10~J Med Chem, 11~Chem-Eur J, 12~Accounts Chem Res, 13~Organometallics, 14~Bioorg Med Chem Lett, 15~J Organomet Chem, 16~Synlett, 17~Coordin Chem Rev, 18~Curr Opin Chem Biol, 19~Chem Lett, 20~Eur J Org Chem, 21~J Mol Catal A-Chem, 22~Tetrahedron-Asymmetr, 23~J Nat Prod, 24~Bioorgan Med Chem, 25~Eur J Inorg Chem, 26~J Magn Reson, 27~Synthesis-Stuttgart, 28~J Biomol Nmr, 29~Inorg Chim Acta, 30~J Mol Struct, 31~Curr Med Chem, 32~Bioconjugate Chem, 33~New J Chem, 34~Carbohyd Res, 35~Polyhedron, 36~Pure Appl Chem, 37~J Chem Inf Comp Sci, 38~Helv Chim Acta, 39~Chem Soc Rev, 40~B Chem Soc Jpn, 41~J Inorg Biochem, 42~Chembiochem, 43~Chem Pharm Bull, 44~Synthetic Commun, 45~Green Chem, 46~J Biol Inorg Chem, 47~Adv Synth Catal, 48~Acta Crystallogr B, 49~Acta Crystallogr C, 50~Heterocycles, 51~Z Anorg Allg Chem, 52~Nat Prod Rep, 53~Inorg Chem Commun, 54~Can J Chem, 55~J Comb Chem, 56~J Fluorine Chem, 57~J Comput Aid Mol Des, 58~Chirality, 59~J Antibiot, 60~Magn Reson Chem, 61~Med Res Rev, 62~Appl Organomet Chem, 63~Prog Nucl Mag Res Sp, 64~Eur J Med Chem, 65~Monatsh Chem, 66~B Kor Chem Soc, 67~J Mol Graph Model, 68~Acta Crystallogr E, 69~Dyes Pigments, 70~J Heterocyclic Chem, 71~Aust J Chem, 72~Indian J Chem B, 73~Curr Org Chem, 74~Usp Khim+, 75~Russ Chem B+, 76~Nucleos Nucleot Nucl, 77~J Phys Org Chem, 78~J Chem Educ, 79~Org Process Res Dev, 80~Collect Czech Chem C, 81~Cryst Growth Des, 82~J Porphyr Phthalocya, 83~Transit Metal Chem, 84~J Incl Phenom Macro, 85~Acta Chim Sinica, 86~J Chem Res-S, 87~J Labelled Compd Rad, 88~Expert Opin Ther Pat, 89~Origins Life Evol B, 90~Phosphorus Sulfur, 91~Pol J Chem, 92~Z Naturforsch B, 93~Comb Chem High T Scr, 94~Chinese J Chem, 95~Russ J Gen Chem+, 96~Supramol Chem, 97~J Indian Chem Soc, 98~Chimia, 99~Russ J Org Chem+, 100~Syn React Inorg Met, 101~Chinese Chem Lett, 102~J Chin Chem Soc-Taip, 103~Mendeleev Commun, 104~J Coord Chem, 105~J Carbohyd Chem, 106~Russ J Coord Chem+, 107~Heteroatom Chem, 108~J Syn Org Chem Jpn, 109~Molecules, 110~Z Krist-New Cryst St, 111~J Fluoresc, 112~Bioorg Chem, 113~Chinese J Inorg Chem, 114~Org Prep Proced Int, 115~Asian J Chem, 116~Arkivoc, 117~J Chem Crystallogr, 118~Main Group Met Chem, 119~Arch Pharm, 120~Chinese J Org Chem, 121~Turk J Chem, 122~Croat Chem Acta, 123~Struct Chem, 124~J Clust Sci, 125~Drug Future, 126~Propell Explos Pyrot, 127~Heterocycl Commun, 128~J Serb Chem Soc, 129~Khim Geterotsikl+, 130~Indian J Heterocy Ch, 131~Yakugaku Zasshi, 132~Comment Inorg Chem, 133~Chinese J Struc Chem, 134~Adv Heterocycl Chem, 135~Rev Roum Chim, 136~J Asian Nat Prod Res, 137~Oxid Commun, 138~J Inorg Organomet P, 139~Med Chem Res, 140~Acta Chim Slov, 141~Actual Chimique, 142~Chim Oggi, 143~Afinidad, 144~J Chem Soc Pakistan, 145~J Chem Sci\subsection*{{\large 10 Psychiatry (2.4\%)}\label{module10}}
\noindent 1~Am J Psychiat, 2~Biol Psychiat, 3~Arch Gen Psychiat, 4~J Consult Clin Psych, 5~J Clin Psychiat, 6~J Am Acad Child Psy, 7~Brit J Psychiat, 8~Psychol Med, 9~Addiction, 10~J Abnorm Psychol, 11~J Child Psychol Psyc, 12~Schizophr Res, 13~Alcohol Clin Exp Res, 14~Psychosom Med, 15~Psychiat Serv, 16~Mol Psychiatr, 17~Behav Res Ther, 18~J Affect Disorders, 19~Acta Psychiat Scand, 20~J Stud Alcohol, 21~Drug Alcohol Depen, 22~J Nerv Ment Dis, 23~J Psychosom Res, 24~Schizophrenia Bull, 25~Int J Eat Disorder, 26~J Abnorm Child Psych, 27~J Clin Psychopharm, 28~Psychiat Res, 29~Clin Psychol Rev, 30~Psychol Assessment, 31~Int J Geriatr Psych, 32~Addict Behav, 33~Clin Psychol-Sci Pr, 34~Child Abuse Neglect, 35~J Clin Psychol, 36~J Interpers Violence, 37~J Autism Dev Disord, 38~J Subst Abuse Treat, 39~Soc Psych Psych Epid, 40~J Trauma Stress, 41~Law Human Behav, 42~Aust Nz J Psychiat, 43~Psychol Addict Behav, 44~Am J Geriat Psychiat, 45~Behav Genet, 46~Prof Psychol-Res Pr, 47~Compr Psychiat, 48~Cognitive Ther Res, 49~Prog Neuro-Psychoph, 50~Am J Orthopsychiat, 51~J Psychiat Res, 52~Bipolar Disord, 53~Alcohol Alcoholism, 54~Psychosomatics, 55~J Pers Assess, 56~Subst Use Misuse, 57~Eur Neuropsychopharm, 58~Exp Clin Psychopharm, 59~Am J Ment Retard, 60~Int Clin Psychopharm, 61~Psychiat Res-Neuroim, 62~Psychiat Clin N Am, 63~Psychother Psychosom, 64~Behav Ther, 65~Cns Drugs, 66~Gen Hosp Psychiat, 67~Crim Justice Behav, 68~Depress Anxiety, 69~Am J Drug Alcohol Ab, 70~Psychol Public Pol L, 71~J Psychopharmacol, 72~J Appl Behav Anal, 73~Neuropsychobiology, 74~J Am Coll Health, 75~J Neuropsych Clin N, 76~J Intell Disabil Res, 77~J Anxiety Disord, 78~J Child Adol Psychop, 79~J Pers Disord, 80~Eur Psychiat, 81~Suicide Life-Threat, 82~Eur Arch Psy Clin N, 83~Brit J Clin Psychol, 84~Alcohol, 85~Child Adol Psych Cl, 86~Int J Neuropsychoph, 87~Behav Modif, 88~J Drug Issues, 89~Behav Sci Law, 90~Eur Child Adoles Psy, 91~Community Ment Hlt J, 92~Aggress Violent Beh, 93~Am J Addiction, 94~Assessment, 95~Psychiat Clin Neuros, 96~Psychother Res, 97~Psychiatry, 98~Pharmacopsychiatry, 99~J Psychoactive Drugs, 100~Ment Retard, 101~Nervenarzt, 102~Res Dev Disabil, 103~Int Psychogeriatr, 104~J Fam Violence, 105~Alcohol Res Health, 106~Hum Psychopharm Clin, 107~Aging Ment Health, 108~J Psychiatr Neurosci, 109~Harvard Rev Psychiat, 110~Int J Offender Ther, 111~J Psychopathol Behav, 112~Psychiat Genet, 113~Psychother Psych Med, 114~J Addict Dis, 115~Eval Program Plann, 116~Eur Eat Disord Rev, 117~Curr Opin Psychiatr, 118~Int J Soc Psychiatr, 119~Int J Law Psychiat, 120~Evaluation Rev, 121~Drug Alcohol Rev, 122~Psychopathology, 123~Clin Psychol Psychot, 124~Int J Psychiat Med, 125~J Geriatr Psych Neur, 126~Psychiatr Rehabil J, 127~Psychiat Ann, 128~J Am Acad Psychiatry, 129~Int Rev Psychiatr, 130~Psychiat Prax, 131~Psychiat Quart, 132~Psychotherapeut, 133~B Menninger Clin, 134~Psychol Crime Law, 135~Eur Addict Res, 136~J Child Adoles Subst, 137~Diagnostica, 138~Nord J Psychiat, 139~J Behav Ther Exp Psy, 140~J Ect, 141~J Appl Res Intellect, 142~Z Psychosom Med Psyc, 143~Child Psychiat Hum D, 144~J Drug Educ, 145~Aust Psychol, 146~J Dev Phys Disabil, 147~Addict Biol, 148~Adm Policy Ment Hlth, 149~Ann Med Interne, 150~Encephale, 151~Fortschr Neurol Psyc, 152~Am J Eval, 153~Child Fam Behav Ther, 154~Behav Change, 155~J Clin Psychol Med S, 156~Int J Group Psychoth, 157~Z Kl Psych Psychoth, 158~Drug-Educ Prev Polic, 159~Int J Clin Exp Hyp, 160~Israel J Psychiat, 161~Arch Psychiat Nurs, 162~Appl Psychophys Biof, 163~Indian J Soc Work, 164~Can Psychol, 165~Verhaltenstherapie, 166~Z Kinder Jug-Psych, 167~Prax Kinderpsychol K, 168~Art Psychother, 169~Evol Psychiatr, 170~Gruppenpsychother Gr, 171~Acta Neuropsychiatr, 172~Eur J Psychiat, 173~Psychopharmakotherap, 174~Neuropsychiatrie, 175~Z Klin Psych Psychia, 176~Nord Psykol, 177~Neurol Psychiat Br, 178~Am J Clin Hypn\subsection*{{\large 11 Environmental Chemistry \& Microbiology (2.3\%)}\label{module11}}
\noindent 1~Appl Environ Microb, 2~Environ Sci Technol, 3~Water Res, 4~Chemosphere, 5~Water Resour Res, 6~Environ Toxicol Chem, 7~Sci Total Environ, 8~Soil Biol Biochem, 9~Soil Sci Soc Am J, 10~J Environ Qual, 11~Water Sci Technol, 12~Plant Soil, 13~Environ Pollut, 14~J Hydrol, 15~Biotechnol Bioeng, 16~Appl Microbiol Biot, 17~Int J Food Microbiol, 18~J Food Protect, 19~J Appl Microbiol, 20~Hydrol Process, 21~Int J Syst Evol Micr, 22~Fems Microbiol Ecol, 23~Agr Ecosyst Environ, 24~Mar Pollut Bull, 25~Biol Fert Soils, 26~Bioresource Technol, 27~Agron J, 28~Water Air Soil Poll, 29~Environ Microbiol, 30~J Biotechnol, 31~Enzyme Microb Tech, 32~Geoderma, 33~Biogeochemistry, 34~Microbial Ecol, 35~Aquat Toxicol, 36~J Microbiol Meth, 37~Appl Geochem, 38~J Contam Hydrol, 39~Arch Environ Con Tox, 40~T Asae, 41~Biotechnol Progr, 42~Lett Appl Microbiol, 43~J Hydraul Eng-Asce, 44~Soil Till Res, 45~J Hazard Mater, 46~Process Biochem, 47~Biotechnol Lett, 48~Adv Water Resour, 49~Eur J Soil Sci, 50~Field Crop Res, 51~Appl Soil Ecol, 52~Catena, 53~J Chem Technol Biot, 54~Soil Sci, 55~J Environ Eng-Asce, 56~Anton Leeuw Int J G, 57~Ground Water, 58~Can J Microbiol, 59~Comp Biochem Phys C, 60~Mar Environ Res, 61~Syst Appl Microbiol, 62~J Biosci Bioeng, 63~Curr Microbiol, 64~J Am Water Resour As, 65~Ecotox Environ Safe, 66~Commun Soil Sci Plan, 67~B Environ Contam Tox, 68~J Soil Water Conserv, 69~Environ Int, 70~Transport Porous Med, 71~J Environ Monitor, 72~Environ Monit Assess, 73~Aust J Soil Res, 74~J Mol Catal B-Enzym, 75~Agr Water Manage, 76~Eur J Agron, 77~Environ Geol, 78~Agr Syst, 79~J Plant Nutr Soil Sc, 80~Environ Technol, 81~Waste Manage, 82~J Ind Microbiol Biot, 83~Nutr Cycl Agroecosys, 84~Agroforest Syst, 85~J Environ Manage, 86~Adv Agron, 87~Can J Soil Sci, 88~J Hydraul Res, 89~J Am Water Works Ass, 90~Appl Biochem Biotech, 91~Food Microbiol, 92~Hum Ecol Risk Assess, 93~J Plant Nutr, 94~Ecol Eng, 95~Extremophiles, 96~Pedobiologia, 97~J Agr Sci, 98~Soil Use Manage, 99~Hydrogeol J, 100~Hydrolog Sci J, 101~Biochem Eng J, 102~Water Environ Res, 103~Geomicrobiol J, 104~Hydrol Earth Syst Sc, 105~Math Geol, 106~Biodegradation, 107~World J Microb Biot, 108~Agronomie, 109~Land Degrad Dev, 110~J Water Res Pl-Asce, 111~Comput Electron Agr, 112~Environ Toxicol, 113~Int Biodeter Biodegr, 114~Soil Sci Plant Nutr, 115~Environ Modell Softw, 116~Resour Conserv Recy, 117~Eur J Soil Biol, 118~Ecotoxicology, 119~Appl Eng Agric, 120~Waste Manage Res, 121~Crit Rev Microbiol, 122~J Irrig Drain E-Asce, 123~Environ Eng Sci, 124~Environ Sci Pollut R, 125~Microbiology+, 126~Biofouling, 127~Biotechnol Adv, 128~Water Qual Res J Can, 129~Irrigation Sci, 130~Fresen Environ Bull, 131~New Zeal J Agr Res, 132~Compost Sci Util, 133~Water Sa, 134~Acta Hydroch Hydrob, 135~J Microbiol Biotechn, 136~Ground Water Monit R, 137~Rev Environ Contam T, 138~Adsorpt Sci Technol, 139~Water Int, 140~J Environ Sci Heal B, 141~Microbiol Res, 142~J Gen Appl Microbiol, 143~Sar Qsar Environ Res, 144~Ozone-Sci Eng, 145~Adsorption, 146~J Agron Crop Sci, 147~Folia Microbiol, 148~Anaerobe, 149~J Water Supply Res T, 150~Biocatal Biotransfor, 151~Zuckerindustrie, 152~Cytotechnology, 153~J Basic Microb, 154~J Food Safety, 155~Nord Hydrol, 156~Environ Geochem Hlth, 157~Environ Prog, 158~Adv Appl Microbiol, 159~Isot Environ Healt S, 160~J Sustain Agr, 161~Biol Agric Hortic, 162~Appl Biochem Micro+, 163~Agrochimica, 164~Landbauforsch Volk, 165~Int Sugar J, 166~Food Biotechnol, 167~Food Technol Biotech, 168~Arch Lebensmittelhyg, 169~J Environ Biol, 170~Acta Agr Scand B-S P, 171~J Rapid Meth Aut Mic, 172~Bodenkultur, 173~J Microbiol, 174~Eurasian Soil Sci+, 175~Indian J Agron, 176~J Agr U Puerto Rico, 177~Seibutsu-Kogaku Kais, 178~J Fac Agr Kyushu U, 179~Discov Innovat\subsection*{{\large 12 Mathematics (2.0\%)}\label{module12}}
\noindent 1~Commun Math Phys, 2~P Am Math Soc, 3~J Algebra, 4~T Am Math Soc, 5~Invent Math, 6~J Math Anal Appl, 7~J Funct Anal, 8~Duke Math J, 9~Ann Math, 10~J Differ Equations, 11~Math Ann, 12~J Reine Angew Math, 13~Nonlinear Anal-Theor, 14~Commun Algebra, 15~J Am Math Soc, 16~Commun Pur Appl Math, 17~Lect Notes Math, 18~Math Z, 19~Arch Ration Mech An, 20~Adv Math, 21~J Lond Math Soc, 22~J Pure Appl Algebra, 23~Commun Part Diff Eq, 24~Israel J Math, 25~Pac J Math, 26~Siam J Math Anal, 27~Compos Math, 28~Int Math Res Notices, 29~Indiana U Math J, 30~Am J Math, 31~Math Res Lett, 32~J Differ Geom, 33~Geom Funct Anal, 34~Topology, 35~P Lond Math Soc, 36~Ergod Theor Dyn Syst, 37~Stud Math, 38~Nonlinearity, 39~Topol Appl, 40~Manuscripta Math, 41~Comput Math Appl, 42~Mem Am Math Soc, 43~Ann I Fourier, 44~J Number Theory, 45~Ann Sci Ecole Norm S, 46~B Lond Math Soc, 47~Acta Arith, 48~Comment Math Helv, 49~Asterisque, 50~Acta Math-Djursholm, 51~Math Comput Model, 52~Appl Comput Harmon A, 53~Appl Math Comput, 54~Can J Math, 55~Russ Math Surv+, 56~J Math Pure Appl, 57~P Roy Soc Edinb A, 58~Math Nachr, 59~Math Proc Cambridge, 60~Mich Math J, 61~Discrete Cont Dyn S, 62~Integr Equat Oper Th, 63~J Anal Math, 64~Calc Var Partial Dif, 65~Lett Math Phys, 66~Rev Math Phys, 67~Geometriae Dedicata, 68~Ann I H Poincare-An, 69~Arch Math, 70~J Operat Theor, 71~Appl Math Lett, 72~Illinois J Math, 73~Sb Math+, 74~J Geom Phys, 75~Fund Math, 76~J Approx Theory, 77~Math Notes+, 78~K-Theory, 79~J Fourier Anal Appl, 80~Sci China Ser A, 81~Infin Dimens Anal Qu, 82~J Algebraic Geom, 83~Commun Anal Geom, 84~Int J Math, 85~B Aust Math Soc, 86~Constr Approx, 87~Acta Math Sin, 88~Forum Math, 89~J Math Soc Jpn, 90~Potential Anal, 91~Math Method Appl Sci, 92~Monatsh Math, 93~J Knot Theor Ramif, 94~Osaka J Math, 95~Rev Mat Iberoam, 96~Q J Math, 97~Can Math Bull, 98~Nagoya Math J, 99~Ann Henri Poincare, 100~Z Angew Math Phys, 101~Semigroup Forum, 102~Diff Equat+, 103~B Soc Math Fr, 104~Asymptotic Anal, 105~Rocky Mt J Math, 106~Houston J Math, 107~Q Appl Math, 108~Exp Math, 109~Funct Anal Appl+, 110~Ann Math Stud, 111~P Edinburgh Math Soc, 112~J Group Theory, 113~Tohoku Math J, 114~Acta Math Hung, 115~Int J Algebr Comput, 116~Siberian Math J+, 117~Commun Contemp Math, 118~Publ Res I Math Sci, 119~Ann Glob Anal Geom, 120~Acta Appl Math, 121~Math Scand, 122~Rep Math Phys, 123~J Nonlinear Sci, 124~Ann Acad Sci Fenn-M, 125~J Differ Equ Appl, 126~Glasgow Math J, 127~Eur J Appl Math, 128~Algebra Univ, 129~B Sci Math, 130~Taiwan J Math, 131~Indagat Math New Ser, 132~Differ Geom Appl, 133~Math Inequal Appl, 134~Chinese Ann Math B, 135~P Jpn Acad A-Math, 136~J Math Kyoto U, 137~Publ Math-Debrecen, 138~Algebr Colloq, 139~Indian J Pure Ap Mat, 140~P Indian As-Math Sci, 141~J Nonlinear Math Phy, 142~Positivity, 143~B Belg Math Soc-Sim, 144~Czech Math J, 145~Appl Categor Struct, 146~Russ J Math Phys, 147~Acta Math Sci, 148~Integr Transf Spec F, 149~Abh Math Sem Hamburg\subsection*{{\large 13 Computer Science (1.4\%)}\label{module13}}
\noindent 1~Lect Notes Comput Sc, 2~Commun Acm, 3~Lect Notes Artif Int, 4~Ieee T Comput, 5~Computer, 6~Theor Comput Sci, 7~Discrete Math, 8~Artif Intell, 9~Siam J Comput, 10~Mach Learn, 11~Discrete Appl Math, 12~J Acm, 13~Ieee T Parall Distr, 14~Ieee T Knowl Data En, 15~Discrete Comput Geom, 16~Inform Syst, 17~J Comb Theory B, 18~Ieee T Software Eng, 19~Ieee Intell Syst, 20~J Comput Syst Sci, 21~Acm Comput Surv, 22~J Algorithm, 23~Ieee Software, 24~J Comb Theory A, 25~Ieee Micro, 26~Ieee Internet Comput, 27~Random Struct Algor, 28~Algorithmica, 29~Inform Process Lett, 30~Eur J Combin, 31~J Graph Theor, 32~Design Code Cryptogr, 33~J Symbolic Logic, 34~Ieee T Evolut Comput, 35~J Symb Comput, 36~Inform Comput, 37~Adv Appl Math, 38~Ibm Syst J, 39~Acm T Inform Syst, 40~J Syst Software, 41~Combinatorica, 42~Int J Hum-Comput St, 43~Parallel Comput, 44~J Artif Intell Res, 45~Vldb J, 46~Comp Geom-Theor Appl, 47~Acm T Database Syst, 48~Ann Pure Appl Logic, 49~J Parallel Distr Com, 50~Siam J Discrete Math, 51~Acm T Comput Syst, 52~Data Knowl Eng, 53~Comb Probab Comput, 54~Ai Mag, 55~Inform Software Tech, 56~J Comb Des, 57~Data Min Knowl Disc, 58~J Cryptol, 59~Acm T Progr Lang Sys, 60~Expert Syst Appl, 61~Ann Math Artif Intel, 62~J Algebr Comb, 63~Ars Combinatoria, 64~J Complexity, 65~Future Gener Comp Sy, 66~Graph Combinator, 67~Comput J, 68~J Logic Comput, 69~Software Pract Exper, 70~Int J High Perform C, 71~Artif Intell Med, 72~Artif Life, 73~Comput Intell, 74~Comput Linguist, 75~B Symb Log, 76~Real-Time Syst, 77~Comput Secur, 78~Acm Sigplan Notices, 79~Acta Inform, 80~Appl Algebr Eng Comm, 81~Appl Artif Intell, 82~Math Logic Quart, 83~Theor Comput Syst, 84~Int J Comput Geom Ap, 85~J Autom Reasoning, 86~Distrib Parallel Dat, 87~Knowl-Based Syst, 88~Utilitas Mathematica, 89~J Comb Optim, 90~Artif Intell Rev, 91~Int J Coop Inf Syst, 92~J Intell Inf Syst, 93~Sci Comput Program, 94~Knowl Eng Rev, 95~Eng Appl Artif Intel, 96~Interact Comput, 97~Order, 98~New Generat Comput, 99~Comput Humanities, 100~Bt Technol J, 101~Distrib Comput, 102~Int J Softw Eng Know, 103~Appl Intell, 104~Adv Comput, 105~Rairo-Theor Inf Appl, 106~J Supercomput, 107~Iee P-Comput Dig T, 108~Form Method Syst Des, 109~J Syst Architect, 110~J Exp Theor Artif In, 111~J Visual Lang Comput, 112~Ai Commun, 113~Microprocess Microsy, 114~Simul-T Soc Mod Sim, 115~Int J Parallel Prog, 116~Software Qual J, 117~Wirtschaftsinf, 118~Fibonacci Quart, 119~Ieee Ann Hist Comput, 120~Integr Comput-Aid E, 121~Fujitsu Sci Tech J, 122~Comput Stand Inter, 123~Expert Syst, 124~Smpte Motion Imag J\subsection*{{\large 14 Analytic Chemistry (1.4\%)}\label{module14}}
\noindent 1~Anal Chem, 2~J Agr Food Chem, 3~J Chromatogr A, 4~Anal Chim Acta, 5~Electrophoresis, 6~J Chromatogr B, 7~Phytochemistry, 8~Analyst, 9~Rapid Commun Mass Sp, 10~Talanta, 11~J Food Sci, 12~Food Chem, 13~J Anal Atom Spectrom, 14~J Sci Food Agr, 15~Appl Spectrosc, 16~Int J Mass Spectrom, 17~Biosens Bioelectron, 18~J Pharmaceut Biomed, 19~J Am Soc Mass Spectr, 20~Planta Med, 21~Electroanal, 22~Spectrochim Acta B, 23~J Mass Spectrom, 24~J Food Eng, 25~Carbohyd Polym, 26~J Ethnopharmacol, 27~Anal Sci, 28~J Am Oil Chem Soc, 29~Chromatographia, 30~Trac-Trend Anal Chem, 31~Cereal Chem, 32~J Aoac Int, 33~Chemometr Intell Lab, 34~Food Res Int, 35~Food Addit Contam, 36~Postharvest Biol Tec, 37~Eur Food Res Technol, 38~J Chemometr, 39~J Cereal Sci, 40~Phytother Res, 41~Anal Lett, 42~Food Hydrocolloid, 43~Trends Food Sci Tech, 44~Int J Biol Macromol, 45~Mass Spectrom Rev, 46~J Liq Chromatogr R T, 47~Lwt-Food Sci Technol, 48~Int J Food Sci Tech, 49~J Biochem Bioph Meth, 50~Bioelectrochemistry, 51~Z Naturforsch C, 52~Dry Technol, 53~Biochem Syst Ecol, 54~Food Qual Prefer, 55~Microchem J, 56~Fitoterapia, 57~Am J Enol Viticult, 58~Phytomedicine, 59~Food Technol-Chicago, 60~Vib Spectrosc, 61~Starch-Starke, 62~Eur J Lipid Sci Tech, 63~Flavour Frag J, 64~J Chromatogr Sci, 65~J Brazil Chem Soc, 66~Crit Rev Food Sci, 67~Nahrung, 68~Geostandard Newslett, 69~Phytochem Analysis, 70~Food Control, 71~J Essent Oil Res, 72~Biomed Chromatogr, 73~Int J Environ An Ch, 74~J Near Infrared Spec, 75~Accredit Qual Assur, 76~Ind Crop Prod, 77~Arch Pharm Res, 78~J Anal Chem+, 79~Chinese J Anal Chem, 80~Lc Gc N Am, 81~Cereal Food World, 82~Deut Lebensm-Rundsch, 83~Bunseki Kagaku, 84~Quim Nova, 85~Jpc-J Planar Chromat, 86~J Texture Stud, 87~Int J Food Sci Nutr, 88~Food Sci Technol Int, 89~Sci Aliment, 90~Pharm Biol, 91~J Am Soc Brew Chem, 92~J I Brewing, 93~J Sens Stud, 94~Am Lab, 95~Eur J Mass Spectrom, 96~Crit Rev Anal Chem, 97~Chem Anal-Warsaw, 98~J Food Quality, 99~J Food Biochem, 100~J Food Process Eng, 101~Atom Spectrosc, 102~Am J Chinese Med, 103~Spectrosc Spect Anal, 104~Food Agr Immunol, 105~Luminescence, 106~Ital J Food Sci, 107~Ann Chim-Rome, 108~Spectrosc Lett, 109~Plant Food Hum Nutr, 110~J Food Process Pres, 111~Spectroscopy, 112~J Food Sci Tech Mys, 113~J Food Hyg Soc Jpn, 114~J Hopkins Apl Tech D, 115~Food Bioprod Process, 116~J Jpn Soc Food Sci, 117~Food Rev Int, 118~Instrum Sci Technol, 119~J Food Drug Anal, 120~Grasas Aceites, 121~J Food Lipids, 122~Chem Listy, 123~Acta Aliment Hung, 124~Can J Anal Sci Spect, 125~Food Aust, 126~Genet Eng News, 127~Iran J Chem Chem Eng, 128~Bangladesh J Botany, 129~Cuban J Agr Sci\subsection*{{\large 15 Business \& Marketing (1.2\%)}\label{module15}}
\noindent 1~Acad Manage J, 2~Manage Sci, 3~J Appl Psychol, 4~Strategic Manage J, 5~Acad Manage Rev, 6~Admin Sci Quart, 7~Organ Sci, 8~J Marketing Res, 9~J Marketing, 10~Res Policy, 11~J Manage, 12~J Organ Behav, 13~Market Sci, 14~J Consum Res, 15~Mis Quart, 16~J Vocat Behav, 17~Hum Relat, 18~Pers Psychol, 19~J Acad Market Sci, 20~J Manage Stud, 21~J Int Bus Stud, 22~J Manage Inform Syst, 23~Inform Syst Res, 24~Organ Stud, 25~Inform Manage-Amster, 26~Calif Manage Rev, 27~J Bus Res, 28~Hum Resource Manage, 29~J Oper Manag, 30~J Occup Organ Psych, 31~Decis Support Syst, 32~J Retailing, 33~J Prod Innovat Manag, 34~Organization, 35~Leadership Quart, 36~Work Employ Soc, 37~Appl Psychol-Int Rev, 38~Int J Oper Prod Man, 39~Ind Market Manag, 40~J Bus Venturing, 41~Group Organ Manage, 42~Int J Res Mark, 43~Decision Sci, 44~J Advertising Res, 45~J Advertising, 46~Organ Dyn, 47~J Bus Ethics, 48~Ieee T Eng Manage, 49~Int J Select Assess, 50~Small Gr Res, 51~Hum Perform, 52~Psychol Market, 53~Int J Electron Comm, 54~J Public Policy Mark, 55~Career Dev Q, 56~Account Org Soc, 57~Int J Technol Manage, 58~Eur J Inform Syst, 59~RFd Manage, 60~J Organ Change Manag, 61~J Career Assessment, 62~Long Range Plann, 63~Technol Forecast Soc, 64~Internet Res, 65~Behav Inform Technol, 66~Res Technol Manage, 67~J World Bus, 68~J Bus Psychol, 69~New Tech Work Employ, 70~Technol Anal Strateg, 71~J Strategic Inf Syst, 72~J Inf Technol, 73~Int J Inform Manage, 74~Econ Ind Democracy, 75~Mil Psychol, 76~J Int Marketing, 77~J Manage Inquiry, 78~Manage Learn, 79~J Small Bus Manage, 80~Int J Serv Ind Manag, 81~Technovation, 82~Pers Rev, 83~Public Pers Manage, 84~Futures, 85~Inform Syst J, 86~Z Arb Organ, 87~J Career Dev, 88~Ind Manage Data Syst, 89~J Eng Technol Manage, 90~Brit J Guid Couns, 91~J Employment Couns, 92~J Sport Manage, 93~Adv Consum Res, 94~Inform Syst Manage, 95~J Consum Aff, 96~Serv Ind J, 97~Group Decis Negot, 98~J Comput Inform Syst, 99~Transport J, 100~Gruppendynamik Organ, 101~Adult Educ Quart\subsection*{{\large 16 Political Science (1.2\%)}\label{module16}}
\noindent 1~Am Polit Sci Rev, 2~Am J Polit Sci, 3~Int Organ, 4~World Polit, 5~Int Security, 6~J Polit, 7~J Democr, 8~Comp Polit Stud, 9~J Conflict Resolut, 10~Int Stud Quart, 11~Eur J Polit Res, 12~China Quart, 13~Comp Polit, 14~J Eur Public Policy, 15~J Peace Res, 16~J Common Mark Stud, 17~Brit J Polit Sci, 18~Annu Rev Polit Sci, 19~Polit Soc, 20~Public Admin Rev, 21~Polit Res Quart, 22~Public Opin Quart, 23~Elect Stud, 24~Legis Stud Quart, 25~Polit Psychol, 26~Party Polit, 27~Ps-Polit Sci Polit, 28~Public Admin, 29~J Theor Polit, 30~Third World Q, 31~China J, 32~Wash Quart, 33~Governance, 34~Rev Int Stud, 35~E Eur Polit Soc, 36~Int Aff, 37~Asian Surv, 38~Polit Stud-London, 39~Int Polit Sci Rev, 40~Int Interact, 41~Eur J Int Relat, 42~Soc Policy Admin, 43~Polit Sci Quart, 44~Parliament Aff, 45~Polit Theory, 46~Stud Comp Int Dev, 47~Secur Stud, 48~Polit Quart, 49~Mod China, 50~Armed Forces Soc, 51~Policy Polit, 52~Millennium-J Int St, 53~Europe-Asia Stud, 54~Communis Post-Commun, 55~Post-Sov Aff, 56~Hum Rights Quart, 57~J Eur Soc Policy, 58~Polit Behav, 59~J Soc Policy, 60~J Strategic Stud, 61~Policy Rev, 62~Issues Stud, 63~Lat Am Res Rev, 64~Scand Polit Stud, 65~Local Gov Stud, 66~Glob Gov, 67~J Lat Am Stud, 68~Judicature, 69~Pac Rev, 70~Admin Soc, 71~Defence Peace Econ, 72~Polit Vierteljahr, 73~Public Money Manage, 74~Gov Oppos, 75~Public Admin Develop, 76~Secur Dialogue, 77~Can J Polit Sci, 78~J Contemp Asia, 79~World Policy J, 80~Publius J Federalism, 81~Am Rev Public Adm, 82~Aust J Polit Sci, 83~Policy Stud J, 84~Alternatives, 85~Dados-Rev Cienc Soc, 86~Aust J Publ Admin, 87~Women Polit, 88~Aust J Int Aff, 89~Aust J Soc Issues, 90~Pac Aff, 91~Int Rev Adm Sci, 92~Can Public Admin, 93~Int J, 94~Polity, 95~Tidsskr Samfunnsfor, 96~Society, 97~J Baltic Stud, 98~Osteuropa, 99~Rev Etud Comp Est-O\subsection*{{\large 17 Fluid Mechanics (1.1\%)}\label{module17}}
\noindent 1~J Fluid Mech, 2~Phys Fluids, 3~J Comput Phys, 4~Int J Heat Mass Tran, 5~Aiaa J, 6~Siam J Sci Comput, 7~Siam J Numer Anal, 8~J Comput Appl Math, 9~Linear Algebra Appl, 10~Combust Flame, 11~Math Comput, 12~Numer Math, 13~Annu Rev Fluid Mech, 14~Siam J Appl Math, 15~J Heat Trans-T Asme, 16~Siam J Matrix Anal A, 17~Exp Fluids, 18~Siam Rev, 19~J Fluid Eng-T Asme, 20~Int J Multiphas Flow, 21~Energ Convers Manage, 22~Int J Numer Meth Fl, 23~Inverse Probl, 24~J Aircraft, 25~Int J Heat Fluid Fl, 26~Appl Numer Math, 27~Math Mod Meth Appl S, 28~J Fluid Struct, 29~Energy, 30~Prog Energ Combust, 31~Appl Therm Eng, 32~Adv Comput Math, 33~J Propul Power, 34~Renew Energ, 35~Int J Refrig, 36~Sol Energy, 37~Exp Therm Fluid Sci, 38~Combust Sci Technol, 39~Comput Fluids, 40~Build Environ, 41~Esaim-Math Model Num, 42~Bit, 43~Acm T Math Software, 44~Numer Heat Tr A-Appl, 45~Fire Safety J, 46~Ima J Numer Anal, 47~Int Commun Heat Mass, 48~J Turbomach, 49~Computing, 50~Int J Therm Sci, 51~J Spacecraft Rockets, 52~Eur J Mech B-Fluid, 53~Combust Theor Model, 54~Int J Energ Res, 55~Prog Aerosp Sci, 56~Energ Buildings, 57~Numer Algorithms, 58~J Eng Gas Turb Power, 59~Numer Linear Algebr, 60~Appl Energ, 61~Math Comput Simulat, 62~J Thermophys Heat Tr, 63~Numer Meth Part D E, 64~Heat Mass Transfer, 65~J Eng Math, 66~J Am Helicopter Soc, 67~Flow Turbul Combust, 68~Numer Heat Tr B-Fund, 69~Appl Math Model, 70~Atomization Spray, 71~Comput Sci Eng, 72~Fluid Dyn Res, 73~Energ Source, 74~Int J Comput Math, 75~J Electron Packaging, 76~HvacFr Res, 77~J Sol Energ-T Asme, 78~Theor Comp Fluid Dyn, 79~Fire Mater, 80~Heat Transfer Eng, 81~Ima J Appl Math, 82~J Enhanc Heat Transf, 83~Microscale Therm Eng, 84~Aerosp Sci Technol, 85~Transport Theor Stat, 86~Jsme Int J B-Fluid T, 87~Aeronaut J, 88~Shock Waves, 89~J Ship Res, 90~Int J Numer Method H, 91~J Comput Math, 92~Flow Meas Instrum, 93~Geophys Astro Fluid, 94~P I Mech Eng A-J Pow, 95~Int J Comput Fluid D, 96~Fire Technol, 97~J Porous Media, 98~Exp Heat Transfer, 99~Combust Explo Shock+, 100~Int J Turbo Jet Eng, 101~Russ J Numer Anal M, 102~J Fire Sci, 103~J Energ Resour-Asme, 104~J I Energy, 105~T Jpn Soc Aeronaut S, 106~P I Mech Eng G-J Aer, 107~Microgravity Sci Tec\subsection*{{\large 18 Medical Imaging (1.1\%)}\label{module18}}
\noindent 1~Radiology, 2~Int J Radiat Oncol, 3~Am J Roentgenol, 4~J Neurosurg, 5~Neurosurgery, 6~Magnet Reson Med, 7~Am J Neuroradiol, 8~J Nucl Med, 9~J Magn Reson Imaging, 10~Ieee T Med Imaging, 11~Eur Radiol, 12~Phys Med Biol, 13~Med Phys, 14~Radiother Oncol, 15~Radiographics, 16~Magn Reson Imaging, 17~J Vasc Interv Radiol, 18~J Comput Assist Tomo, 19~Neuroradiology, 20~Brit J Radiol, 21~J Neuro-Oncol, 22~Eur J Radiol, 23~Acta Neurochir, 24~Clin Radiol, 25~Invest Radiol, 26~Acad Radiol, 27~Nucl Med Biol, 28~J Ultras Med, 29~Surg Neurol, 30~Pediatr Radiol, 31~Acta Oncol, 32~Nmr Biomed, 33~Pediatr Neurosurg, 34~Radiol Clin N Am, 35~Skeletal Radiol, 36~Rofo-Fortschr Rontg, 37~Med Image Anal, 38~Child Nerv Syst, 39~Acta Radiol, 40~Cancer J, 41~Nucl Med Commun, 42~Abdom Imaging, 43~Clin Nucl Med, 44~Cardiovasc Inter Rad, 45~Brit J Neurosurg, 46~J Nucl Cardiol, 47~Semin Radiat Oncol, 48~J Clin Neurosci, 49~Strahlenther Onkol, 50~Semin Nucl Med, 51~J Cardiov Magn Reson, 52~J Clin Ultrasound, 53~Stereot Funct Neuros, 54~Neurosurg Clin N Am, 55~Ieee Eng Med Biol, 56~Neurol Med-Chir, 57~Int J Hyperther, 58~Minim Invas Neurosur, 59~Comput Med Imag Grap, 60~Radiologe, 61~Cancer Biother Radio, 62~Clin Oncol-Uk, 63~Clin Imag, 64~Neuroimag Clin N Am, 65~J Digit Imaging, 66~Clin Neuropathol, 67~J Thorac Imag, 68~Ann Nucl Med, 69~Semin Ultrasound Ct, 70~Neurosurg Rev, 71~Neurol Surg Tokyo, 72~J Radiol, 73~Neurochirurgie, 74~Nuklearmed-Nucl Med, 75~J Neuroradiology, 76~Interv Neuroradiol, 77~Semin Roentgenol, 78~Neurol India, 79~Ultraschall Med, 80~Can Assoc Radiol J, 81~Zbl Neurochir, 82~Neurosurg Quart, 83~Neurocirugia, 84~Riv Neuroradiol\subsection*{{\large 19 Material Engineering (1.1\%)}\label{module19}}
\noindent 1~Int J Solids Struct, 2~Comput Method Appl M, 3~Int J Numer Meth Eng, 4~J Sound Vib, 5~J Mech Phys Solids, 6~J Struct Eng-Asce, 7~Compos Sci Technol, 8~Cement Concrete Res, 9~P Roy Soc A-Math Phy, 10~Comput Struct, 11~J Eng Mech-Asce, 12~Eng Fract Mech, 13~Eng Struct, 14~J Appl Mech-T Asme, 15~Compos Struct, 16~Int J Fracture, 17~Aci Struct J, 18~Compos Part A-Appl S, 19~Smart Mater Struct, 20~Int J Mech Sci, 21~Int J Plasticity, 22~Earthq Eng Struct D, 23~Int J Eng Sci, 24~Int J Fatigue, 25~Mech Mater, 26~Comput Mech, 27~J Constr Steel Res, 28~Int J Nonlinear Mech, 29~J Compos Mater, 30~Fatigue Fract Eng M, 31~Cement Concrete Comp, 32~Aci Mater J, 33~Mater Struct, 34~Eur J Mech A-Solid, 35~Acta Mech, 36~Mech Syst Signal Pr, 37~Compos Part B-Eng, 38~J Wind Eng Ind Aerod, 39~J Vib Acoust, 40~J Intel Mat Syst Str, 41~J Eng Mater-T Asme, 42~Int J Impact Eng, 43~Int J Pres Ves Pip, 44~Eng Anal Bound Elem, 45~Finite Elem Anal Des, 46~Wave Motion, 47~Z Angew Math Mech, 48~Nonlinear Dynam, 49~Can J Civil Eng, 50~Thin Wall Struct, 51~Struct Multidiscip O, 52~J Elasticity, 53~J Mater Civil Eng, 54~Mag Concrete Res, 55~Commun Numer Meth En, 56~Exp Mech, 57~Struct Saf, 58~J Therm Stresses, 59~Constr Build Mater, 60~Arch Appl Mech, 61~Mech Res Commun, 62~Struct Eng Mech, 63~Probabilist Eng Mech, 64~Theor Appl Fract Mec, 65~P I Mech Eng C-J Mec, 66~Cmes-Comp Model Eng, 67~Int Appl Mech+, 68~J Strain Anal Eng, 69~J Reinf Plast Comp, 70~J Vib Control, 71~J Press Vess-T Asme, 72~Eng Comput-Germany, 73~Q J Mech Appl Math, 74~Adv Eng Softw, 75~Int J Vehicle Des, 76~Continuum Mech Therm, 77~Pci J, 78~Eng Computation, 79~J Test Eval, 80~P I Civil Eng-Str B, 81~Acta Mech Sinica, 82~P I Mech Eng D-J Aut, 83~Mech Compos Mater, 84~Math Mech Solids, 85~Jsme Int J A-Solid M, 86~Ksme Int J, 87~Appl Math Mech-Engl, 88~Meccanica, 89~Appl Compos Mater, 90~Shock Vib, 91~Adv Cem Res, 92~P I Mech Eng F-J Rai, 93~Mater Sci Res Int, 94~Adv Compos Lett, 95~Acta Mech Solida Sin, 96~Sampe J, 97~J Chin Inst Eng, 98~Int J Mater Prod Tec, 99~Cement Concrete Aggr, 100~Sadhana-Acad P Eng S, 101~Forsch Ingenieurwes, 102~Eng Fail Anal, 103~J Aerospace Eng, 104~Adv Compos Mater, 105~Defence Sci J, 106~Sci Eng Compos Mater, 107~Eng J Aisc\subsection*{{\large 20 Sociology (0.97\%)}\label{module20}}
\noindent 1~Am Sociol Rev, 2~Am J Sociol, 3~Annu Rev Sociol, 4~J Marriage Fam, 5~Soc Forces, 6~Demography, 7~Criminology, 8~Am Behav Sci, 9~Soc Probl, 10~Soc Psychol Quart, 11~Sociology, 12~Popul Dev Rev, 13~Soc Sci Quart, 14~Econ Soc, 15~Ethnic Racial Stud, 16~J Health Soc Behav, 17~Brit J Sociol, 18~Soc Sci Res, 19~Theor Cult Soc, 20~Theor Soc, 21~J Fam Issues, 22~J Sci Stud Relig, 23~Brit J Criminol, 24~Gender Soc, 25~Int Migr Rev, 26~Crime Delinquency, 27~Ann Am Acad Polit Ss, 28~J Quant Criminol, 29~J Res Crime Delinq, 30~Sociol Health Ill, 31~Work Occupation, 32~Pop Stud-J Demog, 33~Sociol Theor, 34~Soc Natur Resour, 35~Sociol Quart, 36~J Crim Just, 37~Sociol Rev, 38~Contemp Sociol, 39~Sociol Forum, 40~Sociol Relig, 41~Eur Sociol Rev, 42~Soc Networks, 43~Rural Sociol, 44~Disabil Soc, 45~Sociol Perspect, 46~Soc Polit, 47~Nonprof Volunt Sec Q, 48~Hum Stud, 49~J Contemp Ethnogr, 50~Inquiry, 51~Ageing Soc, 52~Symb Interact, 53~Hist Hum Sci, 54~Rev Relig Res, 55~Women Stud Int Forum, 56~J Comp Fam Stud, 57~Sociol Inq, 58~Eur J Popul, 59~Popul Res Policy Rev, 60~Kolner Z Soziol Soz, 61~Int Migr, 62~Deviant Behav, 63~J Law Soc, 64~Ration Soc, 65~Z Soziol, 66~Int Sociol, 67~Crime Law Social Ch, 68~Time Soc, 69~Childhood, 70~Policy Sci, 71~Acta Sociol, 72~Soc Legal Stud, 73~Arch Eur Sociol, 74~Population, 75~Int J Sociol Law, 76~Soc Anim, 77~J Black Stud, 78~Berl J Soziol, 79~Eur J Womens Stud, 80~Popul Environ, 81~Sociol Sport J, 82~Sci Soc Sante, 83~Aust Nz J Criminol, 84~Crim Law Rev, 85~Can Rev Soc Anthrop, 86~Can J Sociol, 87~Sociol Cas, 88~Anthrozoos, 89~Teach Sociol, 90~Sociol Spectrum, 91~Soz Welt, 92~Soc Sci J, 93~Desarrollo Econ, 94~Fed Probat, 95~Soc Compass, 96~J Psychohist\subsection*{{\large 21 Probability \& Statistics (0.87\%)}\label{module21}}
\noindent 1~J Am Stat Assoc, 2~Ann Stat, 3~J Roy Stat Soc B, 4~Biometrics, 5~Stat Med, 6~Biometrika, 7~J Stat Plan Infer, 8~Stoch Proc Appl, 9~Stat Probabil Lett, 10~Ann Probab, 11~Stat Sinica, 12~Ann Appl Probab, 13~Probab Theory Rel, 14~Bernoulli, 15~Stat Sci, 16~J Comput Graph Stat, 17~Scand J Stat, 18~Adv Appl Probab, 19~J Appl Probab, 20~J Multivariate Anal, 21~Comput Stat Data An, 22~Am Stat, 23~Technometrics, 24~Can J Stat, 25~Commun Stat-Theor M, 26~Queueing Syst, 27~J Roy Stat Soc C-App, 28~J Qual Technol, 29~Ann I Stat Math, 30~J Theor Probab, 31~Ann I H Poincare-Pr, 32~Sociol Method Res, 33~Stat Methods Med Res, 34~Stat Comput, 35~Theor Probab Appl+, 36~Int Stat Rev, 37~J Appl Stat, 38~J Roy Stat Soc A Sta, 39~Lifetime Data Anal, 40~Biometrical J, 41~Insur Math Econ, 42~Aust Nz J Stat, 43~Probab Eng Inform Sc, 44~Commun Stat-Simul C, 45~Metrika, 46~Stoch Anal Appl, 47~Brit J Math Stat Psy, 48~Environmetrics, 49~J Stat Comput Sim, 50~J Agr Biol Envir St, 51~Qual Reliab Eng Int, 52~Computation Stat, 53~Statistics, 54~Environ Ecol Stat, 55~Stat Neerl, 56~Stat Pap, 57~Qual Quant\subsection*{{\large 22 Astronomy \& Astrophysics (0.86\%)}\label{module22}}
\noindent 1~Astrophys J, 2~Astron Astrophys, 3~Mon Not R Astron Soc, 4~Astron J, 5~Astrophys J Suppl S, 6~Publ Astron Soc Pac, 7~Annu Rev Astron Astr, 8~Sol Phys, 9~Publ Astron Soc Jpn, 10~Astrophys Space Sci, 11~Iau Symp, 12~New Astron, 13~New Astron Rev, 14~Earth Moon Planets, 15~Acta Astronom, 16~Astron Nachr, 17~Astron Lett+, 18~Publ Astron Soc Aust, 19~Rev Mex Astron Astr, 20~Astron Rep+, 21~Exp Astron, 22~J Astrophys Astron, 23~Astron Geophys, 24~Chinese Astron Astr, 25~Observatory\subsection*{{\large 23 Gastroenterology (0.80\%)}\label{module23}}
\noindent 1~Gastroenterology, 2~Hepatology, 3~Am J Gastroenterol, 4~Gut, 5~J Hepatol, 6~Am J Physiol-Gastr L, 7~Gastrointest Endosc, 8~Aliment Pharm Therap, 9~Digest Dis Sci, 10~Liver Transplant, 11~Scand J Gastroentero, 12~Endoscopy, 13~J Pediatr Gastr Nutr, 14~Hepato-Gastroenterol, 15~J Gastroen Hepatol, 16~Eur J Gastroen Hepat, 17~Semin Liver Dis, 18~J Clin Gastroenterol, 19~J Viral Hepatitis, 20~Inflamm Bowel Dis, 21~J Gastroenterol, 22~Pancreas, 23~Digestion, 24~Gastroenterol Clin N, 25~Neurogastroent Motil, 26~Digest Liver Dis, 27~Helicobacter, 28~Int J Colorectal Dis, 29~Can J Gastroenterol, 30~Intervirology, 31~Gastroen Clin Biol, 32~Hepatol Res, 33~Best Pract Res Cl Ga, 34~Z Gastroenterol, 35~Dis Esophagus, 36~Metab Brain Dis, 37~Digest Dis, 38~Acta Gastro-Ent Belg, 39~Curr Opin Gastroen, 40~Rev Esp Enferm Dig\subsection*{{\large 24 Law (0.79\%)}\label{module24}}
\noindent 1~Yale Law J, 2~Harvard Law Rev, 3~Stanford Law Rev, 4~Columbia Law Rev, 5~Va Law Rev, 6~U Chicago Law Rev, 7~U Penn Law Rev, 8~Calif Law Rev, 9~Mich Law Rev, 10~New York U Law Rev, 11~Am J Int Law, 12~Tex Law Rev, 13~Vanderbilt Law Rev, 14~Northwest U Law Rev, 15~Georgetown Law J, 16~J Legal Stud, 17~Cornell Law Rev, 18~Ucla Law Rev, 19~Duke Law J, 20~Minn Law Rev, 21~Fordham Law Rev, 22~Law Soc Rev, 23~Iowa Law Rev, 24~Indiana Law J, 25~Notre Dame Law Rev, 26~South Calif Law Rev, 27~Boston U Law Rev, 28~U Illinois Law Rev, 29~Wisc Law Rev, 30~Bus Lawyer, 31~Harvard J Law Publ P, 32~George Wash Law Rev, 33~Law Social Inquiry, 34~Am Crim Law Rev, 35~Admin Law Rev, 36~U Cinci Law Rev, 37~Hastings Law J, 38~Harvard Int Law J, 39~Am J Comp Law, 40~Columbia J Trans Law, 41~Wash Law Rev, 42~J Crim Law Crim, 43~Buffalo Law Rev, 44~Antitrust Law J, 45~J Int Econ Law, 46~U Pitt Law Rev, 47~Harvard J Legis, 48~Am J Law Med, 49~J Copyright Soc Usa, 50~J Legal Educ, 51~Cornell Int Law J, 52~Ecol Law Quart, 53~Am Bus Law J, 54~U Pa J Int Econ Law, 55~Am Bankrupt Law J, 56~Common Mkt Law Rev, 57~Rutgers Law Rev, 58~Cathol U Law Rev, 59~Urban Lawyer, 60~Fam Law Quart, 61~Food Drug Law J, 62~Mil Law Rev, 63~Denver U Law Rev, 64~Law Philos, 65~Law Libr J, 66~Nat Resour J, 67~Columbia J Law Soc P, 68~J Legal Med, 69~Iic-Int Rev Intell P, 70~Secur Regul Law J, 71~J Marit Law Commer\subsection*{{\large 25 Chemical Engineering (0.77\%)}\label{module25}}
\noindent 1~Ind Eng Chem Res, 2~Chem Eng Sci, 3~J Catal, 4~Appl Catal A-Gen, 5~Catal Today, 6~J Membrane Sci, 7~Aiche J, 8~Appl Catal B-Environ, 9~Fuel, 10~Micropor Mesopor Mat, 11~Catal Lett, 12~Energ Fuel, 13~Powder Technol, 14~Thermochim Acta, 15~Fluid Phase Equilibr, 16~Desalination, 17~J Chem Eng Data, 18~Comput Chem Eng, 19~Chem Eng J, 20~Stud Surf Sci Catal, 21~Sep Sci Technol, 22~Top Catal, 23~Sep Purif Technol, 24~J Anal Appl Pyrol, 25~Fuel Process Technol, 26~Chem Eng Res Des, 27~Miner Eng, 28~Hydrometallurgy, 29~Biomass Bioenerg, 30~J Chem Thermodyn, 31~J Phys Chem Ref Data, 32~J Supercrit Fluid, 33~Int J Miner Process, 34~Can J Chem Eng, 35~Chem Eng Technol, 36~Int J Thermophys, 37~J Chem Eng Jpn, 38~J Petrol Sci Eng, 39~J Solution Chem, 40~Chem Eng Process, 41~Ultrason Sonochem, 42~J Loss Prevent Proc, 43~Solvent Extr Ion Exc, 44~Chem Eng Prog, 45~Spe Reserv Eval Eng, 46~React Kinet Catal L, 47~Chem-Ing-Tech, 48~Process Saf Environ, 49~J Can Petrol Technol, 50~Kinet Catal+, 51~Korean J Chem Eng, 52~Chem Eng Commun, 53~J Porous Mat, 54~Petrol Sci Technol, 55~Process Saf Prog, 56~Kagaku Kogaku Ronbun, 57~Oil Shale, 58~Gold Bull, 59~Chem Eng-New York, 60~Phys Chem Liq, 61~Miner Metall Proc, 62~J S Afr I Min Metall, 63~Indian J Chem Techn, 64~Chem Pap-Chem Zvesti, 65~Particul Sci Technol, 66~Adv Powder Technol, 67~Filtr Separat, 68~Rev Chim-Bucharest, 69~J Chin Inst Chem Eng, 70~Inz Chem Procesowa, 71~Chinese J Chem Eng, 72~Przem Chem, 73~T I Min Metall C, 74~Chem Biochem Eng Q, 75~Petrol Chem+\subsection*{{\large 26 Education (0.75\%)}\label{module26}}
\noindent 1~J Educ Psychol, 2~Am Educ Res J, 3~Phi Delta Kappan, 4~Educ Leadership, 5~Educ Psychol, 6~Read Res Quart, 7~J Res Sci Teach, 8~Sci Educ, 9~Educ Eval Policy An, 10~Teach Coll Rec, 11~Elem School J, 12~Int J Sci Educ, 13~Except Children, 14~Teach Teach Educ, 15~School Psychol Rev, 16~J School Psychol, 17~Cognition Instruct, 18~J Learn Sci, 19~J Teach Educ, 20~Contemp Educ Psychol, 21~J Learn Disabil-Us, 22~Sociol Educ, 23~Read Teach, 24~Harvard Educ Rev, 25~J Spec Educ, 26~Rem Spec Educ, 27~Brit J Sociol Educ, 28~J Adolesc Adult Lit, 29~Educ Psychol Rev, 30~Brit J Educ Psychol, 31~Learn Instr, 32~J Lit Res, 33~J Educ Res, 34~J Res Math Educ, 35~J Educ Policy, 36~Theor Pract, 37~Psychol Schools, 38~J Curriculum Stud, 39~School Psychol Quart, 40~Sch Eff Sch Improv, 41~Educ Urban Soc, 42~Urban Educ, 43~Early Child Res Q, 44~Top Early Child Spec, 45~J Emot Behav Disord, 46~High Educ, 47~Educ Policy, 48~Educ Admin Quart, 49~Anthropol Educ Quart, 50~Eur J Psychol Educ, 51~Comp Educ, 52~Oxford Rev Educ, 53~Res Teach Engl, 54~Brit J Educ Stud, 55~Educ Stud, 56~Curriculum Inq, 57~Instr Sci, 58~J Exp Educ, 59~Comput Educ, 60~Educ Res-Uk, 61~Learn Disability Q, 62~J Early Intervention, 63~Int J Educ Dev, 64~J Philos Educ, 65~School Psychol Int, 66~Z Padagog Psychol, 67~J Moral Educ, 68~Z Padagogik, 69~Am Biol Teach, 70~Stud High Educ, 71~Comp Educ Rev, 72~Young Children, 73~Gifted Child Quart, 74~Am Ann Deaf, 75~EtrFd-Educ Tech Res, 76~Brit J Educ Technol, 77~Interv Sch Clin, 78~J Psychoeduc Assess, 79~J Biol Educ, 80~Infant Young Child, 81~J Comput Assist Lear, 82~J Educ Gifted, 83~J Educ Psychol Cons, 84~Educ Train Dev Disab, 85~Educ Rev, 86~Res Pract Pers Sev D\subsection*{{\large 27 Telecommunication (0.68\%)}\label{module27}}
\noindent 1~Ieee T Inform Theory, 2~Ieee J Sel Area Comm, 3~P Ieee, 4~Ieee T Commun, 5~Ieee T Signal Proces, 6~Ieee Commun Mag, 7~Ieee Acm T Network, 8~Ieee Commun Lett, 9~Ieee Network, 10~Ieee T Veh Technol, 11~Comput Netw, 12~Signal Process, 13~Ieee T Aero Elec Sys, 14~Ieee Signal Proc Mag, 15~Ieee Spectrum, 16~Ieee Signal Proc Let, 17~Comput Commun, 18~Wirel Netw, 19~Perform Evaluation, 20~Eur T Telecommun, 21~Ieee T Consum Electr, 22~J High Speed Netw, 23~Bell Labs Tech J, 24~Iee P-Radar Son Nav, 25~Telecommun Syst, 26~Iee P-Commun, 27~Digit Signal Process, 28~Aeu-Int J Electron C, 29~Ieee T Broadcast, 30~Etri J, 31~Ann Telecommun, 32~Int J Commun Syst, 33~Ieee Aero El Sys Mag, 34~Comput Electr Eng, 35~Electron Comm Jpn 3, 36~Electron Comm Jpn 1, 37~Space Commun\subsection*{{\large 28 Orthopedics (0.68\%)}\label{module28}}
\noindent 1~Spine, 2~Clin Orthop Relat R, 3~J Bone Joint Surg Am, 4~Arch Phys Med Rehab, 5~J Biomech, 6~Am J Sport Med, 7~J Bone Joint Surg Br, 8~J Orthop Res, 9~Arthroscopy, 10~J Arthroplasty, 11~Clin Biomech, 12~J Biomech Eng-T Asme, 13~Acta Orthop Scand, 14~Eur Spine J, 15~J Shoulder Elb Surg, 16~Ann Biomed Eng, 17~J Pediatr Orthoped, 18~Foot Ankle Int, 19~Phys Ther, 20~Knee Surg Sport Tr A, 21~Gait Posture, 22~Injury, 23~Am J Phys Med Rehab, 24~J Orthop Trauma, 25~Brit J Sport Med, 26~Brain Injury, 27~J Hand Surg-Am, 28~Spinal Cord, 29~Orthopedics, 30~Arch Orthop Traum Su, 31~Clin Rehabil, 32~J Electromyogr Kines, 33~Orthop Clin N Am, 34~J Hand Surg-Brit Eur, 35~Scand J Med Sci Spor, 36~J Orthop Sport Phys, 37~J Head Trauma Rehab, 38~Med Biol Eng Comput, 39~Med Eng Phys, 40~Clin Sport Med, 41~Int Orthop, 42~J Rehabil Res Dev, 43~Biorheology, 44~P I Mech Eng H, 45~J Athl Training, 46~Orthopade, 47~Unfallchirurg, 48~J Am Podiat Med Assn, 49~Rehabil Psychol, 50~Hand Clin, 51~Z Orthop Grenzgeb, 52~Am J Occup Ther, 53~Knee, 54~Rev Chir Orthop, 55~J Manip Physiol Ther, 56~Neurorehabilitation, 57~Neuropsychol Rehabil, 58~Physician Sportsmed, 59~J Occup Rehabil, 60~J Appl Biomech, 61~Int J Rehabil Res, 62~Prosthet Orthot Int, 63~Biomed Tech, 64~Sports Med Arthrosc, 65~Oper Techn Sport Med, 66~J Sport Rehabil, 67~Exp Techniques, 68~J Neurosurg-Spine, 69~Phys Med Rehab Kuror, 70~Isokinet Exerc Sci, 71~Curr Orthopaed, 72~J Back Musculoskelet\subsection*{{\large 29 Control Theory (0.64\%)}\label{module29}}
\noindent 1~Ieee T Automat Contr, 2~Automatica, 3~Ieee T Neural Networ, 4~Fuzzy Set Syst, 5~Siam J Control Optim, 6~Ieee T Robotic Autom, 7~Syst Control Lett, 8~Ieee T Fuzzy Syst, 9~Int J Control, 10~Ieee T Syst Man Cy B, 11~Int J Robot Res, 12~J Guid Control Dynam, 13~Ieee T Contr Syst T, 14~J Mech Design, 15~Int J Robust Nonlin, 16~J Dyn Syst-T Asme, 17~Ieee Contr Syst Mag, 18~Mech Mach Theory, 19~Iee P-Contr Theor Ap, 20~Inform Sciences, 21~Control Eng Pract, 22~Ieee T Syst Man Cy A, 23~Robot Auton Syst, 24~Int J Intell Syst, 25~Auton Robot, 26~Celest Mech Dyn Astr, 27~Ieee-Asme T Mech, 28~Lect Notes Contr Inf, 29~Robotica, 30~Int J Approx Reason, 31~Int J Syst Sci, 32~Ieee Robot Autom Mag, 33~J Process Contr, 34~Acta Astronaut, 35~J Robotic Syst, 36~Int J Uncertain Fuzz, 37~Mechatronics, 38~Ieee T Syst Man Cy C, 39~Math Control Signal, 40~Vehicle Syst Dyn, 41~J Astronaut Sci, 42~J Franklin I, 43~Int J Adapt Control, 44~Neural Process Lett, 45~Adv Robotics, 46~Control Cybern, 47~Automat Rem Contr+, 48~Int J Gen Syst, 49~Jsme Int J C-Mech Sy, 50~Multidim Syst Sign P, 51~Kybernetika, 52~Discrete Event Dyn S, 53~J Intell Robot Syst, 54~Ind Robot, 55~J Intell Fuzzy Syst, 56~Assembly Autom, 57~Math Probl Eng, 58~Neural Comput Appl, 59~Isa T, 60~Dynam Cont Dis Ser A, 61~P I Mech Eng I-J Sys, 62~Optim Contr Appl Met, 63~T Can Soc Mech Eng\subsection*{{\large 30 Environmental Health (0.63\%)}\label{module30}}
\noindent 1~Environ Health Persp, 2~Toxicol Sci, 3~Drug Metab Dispos, 4~Toxicol Appl Pharm, 5~Toxicol Lett, 6~Toxicology, 7~Chem Res Toxicol, 8~Occup Environ Med, 9~Am J Ind Med, 10~Mutat Res-Gen Tox En, 11~J Occup Environ Med, 12~Food Chem Toxicol, 13~Scand J Work Env Hea, 14~Chem-Biol Interact, 15~Ergonomics, 16~Environ Res, 17~J Toxicol Env Health, 18~Risk Anal, 19~Int Arch Occ Env Hea, 20~J Occup Health, 21~J Expo Anal Env Epid, 22~Environ Mol Mutagen, 23~Arch Environ Health, 24~Arch Toxicol, 25~Xenobiotica, 26~Indoor Air, 27~Hum Exp Toxicol, 28~Appl Ergon, 29~Toxicol Pathol, 30~Int J Ind Ergonom, 31~Inhal Toxicol, 32~Mutagenesis, 33~Ann Occup Hyg, 34~Drug Metab Rev, 35~Work Stress, 36~Toxicol In Vitro, 37~Biol Trace Elem Res, 38~Reprod Toxicol, 39~Regul Toxicol Pharm, 40~Crit Rev Toxicol, 41~J Appl Toxicol, 42~Occup Med-Oxford, 43~Curr Drug Metab, 44~Safety Sci, 45~Biomarkers, 46~Ind Health, 47~J Biochem Mol Toxic, 48~Toxicol Ind Health, 49~Environment, 50~Environ Toxicol Phar, 51~J Safety Res, 52~Int J Hyg Envir Heal, 53~J Health Sci, 54~Atla-Altern Lab Anim, 55~Cell Biol Toxicol, 56~Polycycl Aromat Comp, 57~Ashrae J, 58~J Trace Elem Med Bio, 59~Exp Toxicol Pathol, 60~Int J Toxicol, 61~Drug Chem Toxicol, 62~Chem Ind-London, 63~Indoor Built Environ, 64~Eur J Drug Metab Ph, 65~Int J Environ Heal R, 66~Biomed Environ Sci, 67~Trace Elem Electroly, 68~Altex-Altern Tierexp, 69~Fluoride, 70~Med Probl Perform Ar, 71~Trav Humain, 72~J Toxicol-Cutan Ocul, 73~Acta Vet-Beograd\subsection*{{\large 31 Operations Research (0.59\%)}\label{module31}}
\noindent 1~Eur J Oper Res, 2~Transport Res Rec, 3~Oper Res, 4~Math Program, 5~Siam J Optimiz, 6~Int J Prod Res, 7~Math Oper Res, 8~J Oper Res Soc, 9~Int J Prod Econ, 10~Transport Res B-Meth, 11~Transport Res A-Pol, 12~Comput Oper Res, 13~J Optimiz Theory App, 14~Iie Trans, 15~Accident Anal Prev, 16~Ann Oper Res, 17~Transport Sci, 18~Oper Res Lett, 19~J Transp Eng-Asce, 20~Omega-Int J Manage S, 21~Transport Res C-Emer, 22~Interfaces, 23~Informs J Comput, 24~Comput Optim Appl, 25~Comput Ind Eng, 26~Nav Res Log, 27~J Prod Anal, 28~Networks, 29~Transportation, 30~J Global Optim, 31~Comput Ind, 32~Math Method Oper Res, 33~Appl Math Opt, 34~J Heuristics, 35~J Transp Econ Policy, 36~Transport Res E-Log, 37~Optim Method Softw, 38~Res Eng Des, 39~Prod Plan Control, 40~Robot Cim-Int Manuf, 41~J Intell Manuf, 42~Transport Rev, 43~Int J Comput Integ M, 44~Ite J, 45~Eng Optimiz, 46~Numer Func Anal Opt, 47~Concurrent Eng-Res A, 48~J Manuf Syst, 49~Set-Valued Anal, 50~Transport Res D-Tr E, 51~Infor, 52~Cybernet Syst, 53~Or Spectrum, 54~Int J Flex Manuf Sys, 55~J Adv Transport, 56~Rairo-Oper Res, 57~J Navigation, 58~Ai Edam, 59~J Eng Design, 60~J Oper Res Soc Jpn, 61~Asia Pac J Oper Res, 62~P I Civil Eng-Transp\subsection*{{\large 32 Ophthalmology (0.48\%)}\label{module32}}
\noindent 1~Invest Ophth Vis Sci, 2~Ophthalmology, 3~Am J Ophthalmol, 4~Arch Ophthalmol-Chic, 5~Brit J Ophthalmol, 6~J Cataract Refr Surg, 7~Exp Eye Res, 8~Cornea, 9~Graef Arch Clin Exp, 10~Surv Ophthalmol, 11~J Refract Surg, 12~Retina-J Ret Vit Dis, 13~Prog Retin Eye Res, 14~Eye, 15~Curr Eye Res, 16~Acta Ophthalmol Scan, 17~Optometry Vision Sci, 18~J Glaucoma, 19~Mol Vis, 20~Jpn J Ophthalmol, 21~J Aapos, 22~Clin Exp Ophthalmol, 23~Ophthalmologe, 24~Ophthal Physl Opt, 25~Eur J Ophthalmol, 26~Ophthal Plast Recons, 27~Klin Monatsbl Augenh, 28~Ophthalmologica, 29~J Ocul Pharmacol Th, 30~Ophthal Res, 31~J Pediat Ophth Strab, 32~Ocul Immunol Inflamm, 33~J Neuro-Ophthalmol, 34~J Fr Ophtalmol, 35~Can J Ophthalmol, 36~Neuro-Ophthalmology\subsection*{{\large 33 Crop Science (0.47\%)}\label{module33}}
\noindent 1~Theor Appl Genet, 2~Crop Sci, 3~Phytopathology, 4~Plant Dis, 5~Genome, 6~Euphytica, 7~Plant Cell Rep, 8~Weed Sci, 9~Hortscience, 10~Pest Manag Sci, 11~Crop Prot, 12~Annu Rev Phytopathol, 13~Mol Breeding, 14~Eur J Plant Pathol, 15~J Am Soc Hortic Sci, 16~Plant Pathol, 17~Plant Breeding, 18~Weed Technol, 19~Plant Cell Tiss Org, 20~J Nematol, 21~Physiol Mol Plant P, 22~Sci Hortic-Amsterdam, 23~Plant Growth Regul, 24~Ann Appl Biol, 25~Can J Plant Sci, 26~Nematology, 27~Can J Plant Pathol, 28~Weed Res, 29~J Phytopathol, 30~Pestic Biochem Phys, 31~Biol Plantarum, 32~Hereditas, 33~In Vitro Cell Dev-Pl, 34~Seed Sci Technol, 35~Genet Resour Crop Ev, 36~Seed Sci Res, 37~Exp Agr, 38~Z Pflanzenk Pflanzen, 39~Phytoparasitica, 40~J Jpn Soc Hortic Sci, 41~Vitis, 42~Breeding Sci, 43~Australas Plant Path, 44~Genet Mol Biol, 45~Am J Potato Res, 46~Maydica, 47~Nematropica, 48~Bot Bull Acad Sinica, 49~Caryologia, 50~Cereal Res Commun, 51~New Zeal J Crop Hort, 52~Phytoprotection, 53~Trop Grasslands, 54~Soil Crop Sci Soc Fl, 55~Ber Landwirtsch, 56~Acta Biol Cracov Bot, 57~J Plant Biochem Biot, 58~Trop Agr, 59~Pakistan J Bot, 60~Jarq-Jpn Agr Res Q, 61~Listy Cukrov\subsection*{{\large 34 Geography (0.47\%)}\label{module34}}
\noindent 1~Urban Stud, 2~Environ Plann A, 3~Int J Urban Regional, 4~Prog Hum Geog, 5~Reg Stud, 6~Ann Assoc Am Geogr, 7~Environ Plann D, 8~Antipode, 9~Polit Geogr, 10~T I Brit Geogr, 11~Geoforum, 12~Hous Policy Debate, 13~Int J Geogr Inf Sci, 14~Urban Aff Rev, 15~Prof Geogr, 16~Econ Geogr, 17~J Hist Geogr, 18~Area, 19~J Urban Aff, 20~J Am Plann Assoc, 21~Housing Stud, 22~J Rural Stud, 23~Sociol Ruralis, 24~Environ Plann B, 25~Urban Geogr, 26~Int Regional Sci Rev, 27~Pap Reg Sci, 28~Environ Plann C, 29~Rev Int Polit Econ, 30~Cities, 31~Ann Regional Sci, 32~Geogr Anal, 33~Growth Change, 34~Tijdschr Econ Soc Ge, 35~Eur Urban Reg Stud, 36~J Plan Educ Res, 37~Econ Dev Q, 38~Int Soc Sci J, 39~Environ Urban, 40~Can Geogr-Geogr Can, 41~Geogr Rev, 42~Habitat Int, 43~Land Use Policy, 44~Mt Res Dev, 45~Geogr J, 46~Aust Geogr, 47~Singapore J Trop Geo, 48~J Geogr Higher Educ, 49~Appl Geogr, 50~J Archit Plan Res, 51~J Urban Technol, 52~J Urban Plan D-Asce, 53~Geography, 54~Geogr Z, 55~Cartogr J, 56~Interdiscipl Sci Rev\subsection*{{\large 35 Anthropology (0.45\%)}\label{module35}}
\noindent 1~Am J Phys Anthropol, 2~Public Culture, 3~Curr Anthropol, 4~Am Ethnol, 5~Am Anthropol, 6~J Hum Evol, 7~J Archaeol Sci, 8~Cult Anthropol, 9~Annu Rev Anthropol, 10~Signs, 11~J S Afr Stud, 12~Am J Primatol, 13~Comp Stud Soc Hist, 14~Feminist Stud, 15~J Roy Anthropol Inst, 16~Int J Primatol, 17~Evol Anthropol, 18~Afr Affairs, 19~J Asian Stud, 20~Am Antiquity, 21~Hum Biol, 22~Africa, 23~Contemp Pacific, 24~Am J Hum Biol, 25~J Mod Afr Stud, 26~J Anthropol Archaeol, 27~Folia Primatol, 28~Hum Organ, 29~Oceania, 30~Primates, 31~Ann Hum Biol, 32~Glq-J Lesbian Gay St, 33~Hum Ecol, 34~Mod Asian Stud, 35~J Polynesian Soc, 36~J Afr Hist, 37~Crit Anthropol, 38~Archaeometry, 39~Identities-Glob Stud, 40~Lat Am Perspect, 41~Anthropol Quart, 42~J Peasant Stud, 43~Int J Osteoarchaeol, 44~Ethnohistory, 45~J Hist Sociol, 46~Cult Stud, 47~J Mat Cult, 48~Ethnology, 49~J Anthropol Res, 50~Bijdr Taal-Land-V, 51~Ann Anat, 52~Anthropologie, 53~Homme, 54~Frontiers, 55~Plains Anthropol, 56~Anthropol Sci, 57~Contrib Indian Soc, 58~J Gender Stud, 59~Collegium Antropol, 60~Anthropos, 61~Homo, 62~Arctic Anthropol\subsection*{{\large 36 Veterinary (0.45\%)}\label{module36}}
\noindent 1~Vet Rec, 2~Javma-J Am Vet Med A, 3~Vet Microbiol, 4~Am J Vet Res, 5~Vet Immunol Immunop, 6~Equine Vet J, 7~J Vet Intern Med, 8~Prev Vet Med, 9~J Vet Diagn Invest, 10~Avian Dis, 11~Vet Pathol, 12~J Vet Med Sci, 13~J Wildlife Dis, 14~Vet Surg, 15~J Small Anim Pract, 16~J Comp Pathol, 17~Avian Pathol, 18~Res Vet Sci, 19~Vet Clin N Am-Small, 20~Rev Sci Tech Oie, 21~Aust Vet J, 22~Vet Res, 23~Vet J, 24~Vet Radiol Ultrasoun, 25~J Vet Med B, 26~Can Vet J, 27~Comp Cont Educ Pract, 28~Berl Munch Tierarztl, 29~Vet Clin N Am-Equine, 30~Can J Vet Res, 31~J Vet Pharmacol Ther, 32~Vet Quart, 33~Acta Vet Scand, 34~Vet Dermatol, 35~Vet Hum Toxicol, 36~J Vet Med A, 37~J Zoo Wildlife Med, 38~Vet Clin N Am-Food A, 39~Deut Tierarztl Woch, 40~Vet Res Commun, 41~Vet Comp Orthopaed, 42~Schweiz Arch Tierh, 43~Tierarztl Prax, 44~New Zeal Vet J, 45~Equine Vet Educ, 46~Acta Vet Hung, 47~Clin Tech Small An P, 48~Wien Tierarztl Monat, 49~Prakt Tierarzt, 50~J Equine Vet Sci, 51~J Avian Med Surg, 52~Vet Med-Czech, 53~Comp Immunol Microb, 54~J Vet Med Educ, 55~Acta Vet Brno, 56~Tierarztl Umschau, 57~Semin Avian Exot Pet, 58~Vet Clin Path, 59~Vet Med-Us, 60~Kleintierpraxis, 61~Z Jagdwiss, 62~In Practice, 63~Pferdeheilkunde, 64~Rev Med Vet-Toulouse, 65~Med Weter, 66~Point Vet, 67~Prat Med Chir Anim, 68~Aust Vet Pract, 69~Vlaams Diergen Tijds, 70~Tijdschr Diergeneesk, 71~J Camel Pract Res, 72~B Vet I Pulawy, 73~Arch Med Vet, 74~Pesquisa Vet Brasil, 75~Irish Vet J, 76~Magy Allatorvosok, 77~Ann Med Vet\subsection*{{\large 37 Computer Imaging (0.43\%)}\label{module37}}
\noindent 1~Ieee T Pattern Anal, 2~Ieee T Image Process, 3~Pattern Recogn, 4~Int J Comput Vision, 5~Ieee T Circ Syst Vid, 6~Comput Vis Image Und, 7~Pattern Recogn Lett, 8~Image Vision Comput, 9~Comput Aided Design, 10~Acm T Graphic, 11~Ieee Comput Graph, 12~Ieee T Vis Comput Gr, 13~Comput Aided Geom D, 14~Comput Graph Forum, 15~Comput Graph-Uk, 16~Ieee Multimedia, 17~J Vis Commun Image R, 18~Signal Process-Image, 19~Presence-Teleop Virt, 20~J Math Imaging Vis, 21~Int J Pattern Recogn, 22~Visual Comput, 23~Multimedia Syst, 24~J Electron Imaging, 25~Pattern Anal Appl, 26~Iee P-Vis Image Sign, 27~Multimed Tools Appl, 28~Mach Vision Appl, 29~Int J Imag Syst Tech, 30~Real-Time Imaging, 31~Imaging Sci J\subsection*{{\large 38 Agriculture (0.40\%)}\label{module38}}
\noindent 1~J Dairy Sci, 2~J Anim Sci, 3~Theriogenology, 4~Poultry Sci, 5~Meat Sci, 6~Livest Prod Sci, 7~Int Dairy J, 8~Anim Genet, 9~Appl Anim Behav Sci, 10~Anim Reprod Sci, 11~Aust J Agr Res, 12~Anim Feed Sci Tech, 13~Anim Sci, 14~Brit Poultry Sci, 15~Aust J Exp Agr, 16~Domest Anim Endocrin, 17~Small Ruminant Res, 18~J Dairy Res, 19~Genet Sel Evol, 20~Can J Anim Sci, 21~Lait, 22~Reprod Fert Develop, 23~Asian Austral J Anim, 24~Milchwissenschaft, 25~Grass Forage Sci, 26~Reprod Domest Anim, 27~J Anim Breed Genet, 28~Anim Welfare, 29~Reprod Nutr Dev, 30~World Poultry Sci J, 31~Aust J Dairy Technol, 32~Rev Bras Zootecn, 33~J Anim Physiol An N, 34~Acta Agr Scand A-An, 35~Indian J Anim Sci, 36~Pesqui Agropecu Bras, 37~Arch Anim Nutr, 38~Arch Tierzucht, 39~Turk J Vet Anim Sci, 40~Indian Vet J, 41~J Anim Feed Sci, 42~Arch Geflugelkd, 43~Trop Anim Health Pro, 44~Zuchtungskunde, 45~Int J Dairy Technol, 46~Czech J Anim Sci, 47~S Afr J Anim Sci, 48~J Appl Anim Res, 49~Fleischwirtschaft, 50~Prod Anim, 51~J Muscle Foods, 52~Irish J Agr Food Res, 53~Outlook Agr, 54~Arq Bras Med Vet Zoo, 55~Wool Tech Sheep Bree, 56~Rev Cient-Fac Cien V\subsection*{{\large 39 Parasitology (0.37\%)}\label{module39}}
\noindent 1~Am J Trop Med Hyg, 2~Int J Parasitol, 3~J Parasitol, 4~Vet Parasitol, 5~T Roy Soc Trop Med H, 6~Trop Med Int Health, 7~Parasitology, 8~Trends Parasitol, 9~Dis Aquat Organ, 10~J Med Entomol, 11~J Eukaryot Microbiol, 12~Parasitol Res, 13~Acta Trop, 14~Mem I Oswaldo Cruz, 15~Ann Trop Med Parasit, 16~Exp Parasitol, 17~J Fish Dis, 18~Parasite Immunol, 19~Med Vet Entomol, 20~Fish Shellfish Immun, 21~Exp Appl Acarol, 22~J Am Mosquito Contr, 23~Protist, 24~Adv Parasit, 25~B Eur Assoc Fish Pat, 26~J Aquat Anim Health, 27~Eur J Protistol, 28~Folia Parasit, 29~J Helminthol, 30~Syst Parasitol, 31~Parasite, 32~Fish Pathol, 33~J Vector Ecol, 34~Acta Protozool, 35~Acta Parasitol, 36~Onderstepoort J Vet, 37~Helminthologia, 38~Arch Sci\subsection*{{\large 40 Dentistry (0.37\%)}\label{module40}}
\noindent 1~J Dent Res, 2~J Periodontol, 3~Oral Surg Oral Med O, 4~J Clin Periodontol, 5~Int J Oral Max Impl, 6~Clin Oral Implan Res, 7~J Oral Maxil Surg, 8~J Prosthet Dent, 9~Dent Mater, 10~J Am Dent Assoc, 11~Am J Orthod Dentofac, 12~J Oral Rehabil, 13~J Dent, 14~Arch Oral Biol, 15~J Endodont, 16~Oper Dent, 17~Int J Prosthodont, 18~Am J Dent, 19~Int J Oral Max Surg, 20~Eur J Oral Sci, 21~Int Endod J, 22~J Periodontal Res, 23~Brit Dent J, 24~Caries Res, 25~Periodontol 2PPP, 26~Community Dent Oral, 27~Crit Rev Oral Biol M, 28~Quintessence Int, 29~J Oral Pathol Med, 30~Angle Orthod, 31~Oral Microbiol Immun, 32~Int J Periodont Rest, 33~Eur J Orthodont, 34~Acta Odontol Scand, 35~Brit J Oral Max Surg, 36~J Cranio Maxill Surg, 37~Oral Dis, 38~J Orofac Pain, 39~J Public Health Dent, 40~Dentomaxillofac Rad, 41~Swed Dent J, 42~Cranio, 43~Aust Dent J\subsection*{{\large 41 Dermatology (0.34\%)}\label{module41}}
\noindent 1~J Am Acad Dermatol, 2~Brit J Dermatol, 3~Plast Reconstr Surg, 4~Arch Dermatol, 5~Ann Plas Surg, 6~Dermatol Surg, 7~Brit J Plast Surg, 8~Dermatology, 9~Laser Surg Med, 10~Int J Dermatol, 11~Clin Exp Dermatol, 12~Contact Dermatitis, 13~Acta Derm-Venereol, 14~Wound Repair Regen, 15~Burns, 16~Am J Dermatopath, 17~Cleft Palate-Cran J, 18~J Cutan Pathol, 19~Eur J Dermatol, 20~J Eur Acad Dermatol, 21~J Craniofac Surg, 22~Cutis, 23~Pediatr Dermatol, 24~J Reconstr Microsurg, 25~J Burn Care Rehabil, 26~Microsurg, 27~Dermatol Clin, 28~Clin Plast Surg, 29~Ann Dermatol Vener, 30~Clin Dermatol, 31~Photodermatol Photo, 32~Aesthet Plast Surg, 33~Hautarzt, 34~Scand J Plast Recons, 35~Semin Cutan Med Surg, 36~J Clin Laser Med Sur, 37~Laser Med Sci, 38~Wounds\subsection*{{\large 42 Urology (0.32\%)}\label{module42}}
\noindent 1~J Urology, 2~Urology, 3~Bju Int, 4~Eur Urol, 5~J Endourol, 6~Int J Impot Res, 7~Neurourol Urodynam, 8~Urol Clin N Am, 9~Int Urogynecol J Pel, 10~Scand J Urol Nephrol, 11~World J Urol, 12~J Sex Marital Ther, 13~Int J Urol, 14~Urol Res, 15~Urol Int, 16~Prostate Cancer P D, 17~Andrologia, 18~Arch Andrology, 19~Prog Urol, 20~Urologe A, 21~Asian J Androl, 22~Ann Urol, 23~Aktuel Urol\subsection*{{\large 43 Rheumatology (0.32\%)}\label{module43}}
\noindent 1~Arth RheumOar C Res, 2~J Rheumatol, 3~Ann Rheum Dis, 4~Rheumatology, 5~Osteoarthr Cartilage, 6~Curr Opin Rheumatol, 7~Lupus, 8~Clin Exp Rheumatol, 9~Rheum Dis Clin N Am, 10~Scand J Rheumatol, 11~Semin Arthritis Rheu, 12~Clin Rheumatol, 13~Rheumatol Int, 14~Best Pract Res Cl Rh, 15~Joint Bone Spine, 16~Z Rheumatol, 17~J Musculoskelet Pain, 18~Jcr-J Clin Rheumatol, 19~Int J Tissue React, 20~Aktuel Rheumatol\subsection*{{\large 44 Applied Acoustics (0.30\%)}\label{module44}}
\noindent 1~J Acoust Soc Am, 2~J Speech Lang Hear R, 3~Hearing Res, 4~Ultrasound Med Biol, 5~Ieee T Ultrason Ferr, 6~Ieee T Speech Audi P, 7~Ultrasonics, 8~Ieee J Oceanic Eng, 9~Speech Commun, 10~Appl Psycholinguist, 11~Ear Hearing, 12~NdtFe Int, 13~Audiol Neuro-Otol, 14~Jaro-J Assoc Res Oto, 15~Int J Lang Comm Dis, 16~J Phonetics, 17~J Comput Acoust, 18~Appl Acoust, 19~Lang Speech Hear Ser, 20~J Commun Disord, 21~Clin Linguist Phonet, 22~Comput Speech Lang, 23~Mater Eval, 24~Insight, 25~J Audio Eng Soc, 26~Phonetica, 27~J Fluency Disord, 28~Acoust Phys+, 29~Ultrasonic Imaging, 30~Int J Aviat Psychol, 31~Top Lang Disord, 32~Phys Medica, 33~Mar Technol Soc J, 34~Noise Control Eng J, 35~Volta Rev, 36~J Nondestruct Eval\subsection*{{\large 45 Pharmacology (0.30\%)}\label{module45}}
\noindent 1~Biomaterials, 2~J Control Release, 3~Int J Pharm, 4~Pharm Res, 5~Adv Drug Deliver Rev, 6~J Pharm Sci-Us, 7~J Pharm Pharmacol, 8~Tissue Eng, 9~Eur J Pharm Sci, 10~J Mater Sci-Mater M, 11~Eur J Pharm Biopharm, 12~J Biomat Sci-Polym E, 13~Annu Rev Biomed Eng, 14~Drug Dev Ind Pharm, 15~Pharmazie, 16~J Microencapsul, 17~J Drug Target, 18~Pharm Dev Technol, 19~Biopharm Drug Dispos, 20~Crit Rev Ther Drug, 21~Bio-Med Mater Eng, 22~Drug Deliv, 23~Artif Cell Blood Sub, 24~J Biomater Appl, 25~J Bioact Compat Pol, 26~Pda J Pharm Sci Tech, 27~Pharm Ind, 28~J Drug Deliv Sci Tec\subsection*{{\large 46 Pathology (0.27\%)}\label{module46}}
\noindent 1~Am J Surg Pathol, 2~Hum Pathol, 3~Modern Pathol, 4~Am J Clin Pathol, 5~J Clin Pathol, 6~Arch Pathol Lab Med, 7~Histopathology, 8~Virchows Arch, 9~Pathol Int, 10~Acta Cytol, 11~Diagn Cytopathol, 12~Cancer Cytopathol, 13~Int J Gynecol Pathol, 14~Pathol Res Pract, 15~Pediatr Devel Pathol, 16~Pathology, 17~Appl Immunohisto M M, 18~Semin Diagn Pathol, 19~Diagn Mol Pathol, 20~Adv Anat Pathol, 21~Anal Quant Cytol, 22~Ultrastruct Pathol, 23~Endocr Pathol, 24~Int J Surg Pathol, 25~Cytopathology, 26~Pathologe, 27~Ann Pathol, 28~J Histotechnol\subsection*{{\large 47 History \& Philosophy of Science (0.19\%)}\label{module47}}
\noindent 1~Philos Sci, 2~Soc Stud Sci, 3~Stud Hist Philos Sci, 4~Am Math Mon, 5~Arch Hist Exact Sci, 6~Isis, 7~Brit J Philos Sci, 8~Biol Philos, 9~Hist Sci, 10~Sci Context, 11~Hist Math, 12~Osiris, 13~Am Sci, 14~Synthese, 15~Notes Rec Roy Soc, 16~Math Intell, 17~Brit J Hist Sci, 18~Technol Cult, 19~Hist Psychiatr, 20~Ann Sci, 21~Med Hist, 22~Mind Mach, 23~B Hist Med, 24~J Hist Biol, 25~Philos Soc Sci, 26~Soc Hist Med, 27~J Hist Behav Sci, 28~Minerva, 29~J Hist Med All Sci, 30~Hist Stud Phys Biol, 31~J Hist Sexuality, 32~Hist Phil Life Sci\subsection*{{\large 48 Otolaryngology (0.19\%)}\label{module48}}
\noindent 1~Laryngoscope, 2~Otolaryng Head Neck, 3~Arch Otolaryngol, 4~Ann Oto Rhinol Laryn, 5~Head Neck-J Sci Spec, 6~Acta Oto-Laryngol, 7~J Laryngol Otol, 8~Int J Pediatr Otorhi, 9~Otol Neurotol, 10~Otolaryng Clin N Am, 11~Eur Arch Oto-Rhino-L, 12~Am J Rhinol, 13~Clin Otolaryngol All, 14~J Voice, 15~Orl J Oto-Rhino-Lary, 16~Am J Otolaryng, 17~Dysphagia, 18~J Otolaryngol, 19~Folia Phoniatr Logo, 20~Hno\subsection*{{\large 49 Electromagnetic Engineering (0.18\%)}\label{module49}}
\noindent 1~Ieee T Microw Theory, 2~Ieee T Antenn Propag, 3~Microw Opt Techn Let, 4~Ieee Microw Wirel Co, 5~Ieee Antennas Propag, 6~Ieee T Adv Packaging, 7~Ieee T Electromagn C, 8~Bioelectromagnetics, 9~Int J Rf Microw C E, 10~Iee P-Microw Anten P, 11~J Electromagnet Wave, 12~Int J Infrared Milli, 13~Microwave J, 14~Electromagnetics, 15~Int J Numer Model El, 16~Electr Eng, 17~Izv Vuz Radioelektr+\subsection*{{\large 50 Information Science (0.17\%)}\label{module50}}
\noindent 1~J Am Soc Inf Sci Tec, 2~Inform Process Manag, 3~Scientometrics, 4~Coll Res Libr, 5~J Doc, 6~J Inf Sci, 7~J Acad Libr, 8~Libr Trends, 9~Libr Inform Sci Res, 10~Gov Inform Q, 11~Learn Publ, 12~Ref User Serv Q, 13~Libri, 14~Libr Quart, 15~Aslib Proc, 16~Inform Technol Libr, 17~Libr Resour Tech Ser, 18~J Libr Inf Sci, 19~Electron Libr, 20~J Gov Inform, 21~Knowl Organ, 22~Z Bibl Bibl, 23~Interlend Doc Supply\subsection*{{\large 51 Circuits (0.17\%)}\label{module51}}
\noindent 1~Ieee J Solid-St Circ, 2~Ieee T Circuits-I, 3~Ieee T Circuits-Ii, 4~Ieee T Comput Aid D, 5~Ieee T Vlsi Syst, 6~Ieee Des Test Comput, 7~Int J Electron, 8~J Vlsi Sig Proc Syst, 9~Int J Circ Theor App, 10~Integration, 11~Analog Integr Circ S, 12~J Electron Test, 13~Circ Syst Signal Pr, 14~Frequenz, 15~J Circuit Syst Comp, 16~Electron World\subsection*{{\large 52 Media \& Communication (0.16\%)}\label{module52}}
\noindent 1~J Commun, 2~Journalism Mass Comm, 3~Polit Commun, 4~Commun Res, 5~Hum Commun Res, 6~Commun Monogr, 7~J Broadcast Electron, 8~Sci Technol Hum Val, 9~Harv Int J PressOpol, 10~Commun Theor, 11~Media Cult Soc, 12~Writ Commun, 13~Inform Soc, 14~J Lang Soc Psychol, 15~Public Underst Sci, 16~Telecommun Policy, 17~Eur J Commun, 18~Int J Public Opin R, 19~Q J Speech, 20~Sci Commun, 21~Tech Commun, 22~J Bus Tech Commun, 23~J Media Econ, 24~Public Relat Rev\subsection*{{\large 53 Power Systems (0.16\%)}\label{module53}}
\noindent 1~Ieee T Power Syst, 2~Ieee T Ind Electron, 3~Ieee T Ind Appl, 4~Ieee T Power Electr, 5~Ieee T Power Deliver, 6~Ieee T Energy Conver, 7~Ieee T Dielect El In, 8~Iee P-Elect Pow Appl, 9~Iee P-Gener Transm D, 10~Int J Elec Power, 11~J Electrostat, 12~Electr Pow Syst Res, 13~Ieee T Educ, 14~Eur T Electr Power, 15~Kunstst-Plast Eur, 16~Ieee Electr Insul M, 17~Int J Elec Eng Educ, 18~Electr Eng Jpn\subsection*{{\large 54 Tribology (0.16\%)}\label{module54}}
\noindent 1~J Mater Process Tech, 2~Wear, 3~Int J Mach Tool Manu, 4~J Manuf Sci E-T Asme, 5~Cirp Ann-Manuf Techn, 6~J Tribol-T Asme, 7~Tribol Int, 8~Int J Adv Manuf Tech, 9~Tribol T, 10~Tribol Lett, 11~P I Mech Eng B-J Eng, 12~Mater Design, 13~P I Mech Eng J-J Eng, 14~Mach Sci Technol, 15~J Jpn Soc Tribologis, 16~Tribol Lubr Technol\subsection*{{\large 55 History (0.15\%)}\label{module55}}
\noindent 1~J Econ Hist, 2~Am Hist Rev, 3~J Am Hist, 4~Econ Hist Rev, 5~Slavic Rev, 6~Environ Hist, 7~Explor Econ Hist, 8~J Mod Hist, 9~Past Present, 10~Continuity Change, 11~Bus Hist, 12~J Urban Hist, 13~J Fam Hist, 14~Labor Hist, 15~Int Rev Soc Hist, 16~J Soc Hist, 17~Soc Sci Hist, 18~J Interdiscipl Hist, 19~Agr Hist, 20~Endeavour, 21~E Eur Quart, 22~Crit Rev, 23~J Sport Hist\subsection*{{\large 56 Geotechnology (0.14\%)}\label{module56}}
\noindent 1~J Geotech Geoenviron, 2~Geotechnique, 3~Can Geotech J, 4~Int J Numer Anal Met, 5~Int J Rock Mech Min, 6~Eng Geol, 7~Soil Dyn Earthq Eng, 8~Geotech Test J, 9~P I Civil Eng-Geotec, 10~Geotext Geomembranes, 11~Geosynth Int, 12~Comput Geotech, 13~Tunn Undergr Sp Tech, 14~Rock Mech Rock Eng, 15~Q J Eng Geol Hydroge, 16~P I Civil Eng-Civ En\subsection*{{\large 57 Wood Products (0.12\%)}\label{module57}}
\noindent 1~Holzforschung, 2~J Pulp Pap Sci, 3~Nord Pulp Pap Res J, 4~Forest Prod J, 5~Pap Puu-Pap Tim, 6~Appita J, 7~Wood Fiber Sci, 8~J Wood Sci, 9~Wood Sci Technol, 10~Cellulose, 11~Holz Roh Werkst, 12~J Wood Chem Technol, 13~Pulp Pap-Canada, 14~Iawa J, 15~Mokuzai Gakkaishi, 16~Wochenbl Papierfabr, 17~Cell Chem Technol, 18~Restaurator\subsection*{{\large 58 Radiation (0.10\%)}\label{module58}}
\noindent 1~Radiat Res, 2~Int J Radiat Biol, 3~Appl Radiat Isotopes, 4~Radiat Prot Dosim, 5~J Radioanal Nucl Ch, 6~Health Phys, 7~J Environ Radioactiv, 8~Radiochim Acta, 9~Radiat Environ Bioph, 10~J Radiat Res, 11~P I Mech Eng E-J Pro, 12~Kerntechnik, 13~Nucl Energ-J Br Nucl\subsection*{{\large 59 Linguistics (0.093\%)}\label{module59}}
\noindent 1~J Pragmatics, 2~Discourse Soc, 3~Lang Soc, 4~Mod Lang J, 5~Tesol Quart, 6~Appl Linguist, 7~Res Lang Soc Interac, 8~Aphasiology, 9~Lang Learn, 10~Lang Commun, 11~Can Mod Lang Rev, 12~Foreign Lang Ann, 13~Am Speech\subsection*{{\large 60 Social Work (0.079\%)}\label{module60}}
\noindent 1~Soc Work, 2~Fam Soc-J Contemp H, 3~Child Youth Serv Rev, 4~Soc Serv Rev, 5~Res Social Work Prac, 6~Soc Work Res, 7~Brit J Soc Work, 8~J Soc Work Educ, 9~Soc Work Health Care, 10~Health Soc Work, 11~Juvenile Fam Court J, 12~Affilia J Wom Soc Wo, 13~Int Soc Work, 14~Smith Coll Stud Soc, 15~Clin Soc Work J, 16~Admin Soc Work, 17~J Soc Serv Res\subsection*{{\large 61 Psychoanalysis (0.077\%)}\label{module61}}
\noindent 1~J Am Psychoanal Ass, 2~Int J Psychoanal, 3~Psychoanal Quart, 4~Psychoanal Inq, 5~Psychoanal Psychol, 6~Psyche-Z Psychoanal, 7~Contemp Psychoanal, 8~Forum Psychoanal, 9~J Anal Psychol\subsection*{{\large 62 Middle Eastern Studies (0.073\%)}\label{module62}}
\noindent 1~Int J Middle E Stud, 2~New Left Rev, 3~Middle Eastern Stud, 4~Middle East J, 5~Mon Rev, 6~Sci Soc, 7~Race Class\subsection*{{\large 63 Forensic Science (0.067\%)}\label{module63}}
\noindent 1~Forensic Sci Int, 2~J Forensic Sci, 3~Int J Legal Med, 4~J Anal Toxicol, 5~Am J Foren Med Path, 6~Med Sci Law, 7~Sci Justice, 8~Kriminalistik\subsection*{{\large 64 Transfusion (0.066\%)}\label{module64}}
\noindent 1~Transfusion, 2~Vox Sang, 3~Syst Dynam Rev, 4~Transfusion Med, 5~Syst Res Behav Sci, 6~Transfus Med Rev, 7~Syst Pract Act Res, 8~Biologicals, 9~Kybernetes, 10~J Clin Apheresis, 11~Transfus Clin Biol\subsection*{{\large 65 Mycology (0.055\%)}\label{module65}}
\noindent 1~Mycol Res, 2~Mycologia, 3~Mycotaxon, 4~Lichenologist, 5~Sydowia, 6~Persoonia, 7~Cryptogamie Mycol, 8~Mikol Fitopatol\subsection*{{\large 66 Nuclear Energy (0.055\%)}\label{module66}}
\noindent 1~Nucl Eng Des, 2~Ann Nucl Energy, 3~Nucl Technol, 4~J Nucl Sci Technol, 5~Nucl Sci Eng, 6~Prog Nucl Energ, 7~J Atom Energ Soc Jpn, 8~Atw-Int J Nucl Power\subsection*{{\large 67 offshore Engineering (0.036\%)}\label{module67}}
\noindent 1~Coast Eng, 2~Ocean Eng, 3~J Waterw Port C-Asce, 4~Appl Ocean Res, 5~J Offshore Mech Arct, 6~Mar Technol Sname N, 7~Nav Eng J\subsection*{{\large 68 Environmental Ethics (0.034\%)}\label{module68}}
\noindent 1~Ethics, 2~Soc Philos Policy, 3~Environ Ethics, 4~Environ Value, 5~J Agr Environ Ethic\subsection*{{\large 69 Entomology (0.033\%)}\label{module69}}
\noindent 1~P Entomol Soc Wash, 2~Entomol News, 3~T Am Entomol Soc, 4~J New York Entomol S, 5~Pan-Pac Entomol, 6~Orient Insects\subsection*{{\large 70 Higher Education (0.028\%)}\label{module70}}
\noindent 1~J Coll Student Dev, 2~Res High Educ, 3~J High Educ, 4~Rev High Educ\subsection*{{\large 71 Refineries (0.024\%)}\label{module71}}
\noindent 1~Oil Gas J, 2~Hydrocarb Process, 3~Neft Khoz\subsection*{{\large 72 Reliability Engineering (0.023\%)}\label{module72}}
\noindent 1~Reliab Eng Syst Safe, 2~Ieee T Reliab, 3~P Rel Maint S\subsection*{{\large 73 Other (0.023\%)}\label{module73}}
\noindent 1~J Leisure Res, 2~Leisure Sci, 3~J Am Leather Chem As, 4~J Soc Leath Tech Ch, 5~J Sci Ind Res India\subsection*{{\large 74 Civil Engineering (0.022\%)}\label{module74}}
\noindent 1~J Comput Civil Eng, 2~J Constr Eng M Asce, 3~Build Res Inf, 4~J Surv Eng-Asce, 5~J Prof Iss Eng Ed Pr\subsection*{{\large 75 Lab Veterinary (0.017\%)}\label{module75}}
\noindent 1~Lab Anim-Uk, 2~Comparative Med, 3~Contemp Top Lab Anim, 4~Lab Animal, 5~Scand J Lab Anim Sci\subsection*{{\large 76 Music (0.017\%)}\label{module76}}
\noindent 1~Music Percept, 2~J New Music Res, 3~Comput Music J\subsection*{{\large 77 Tourism (0.017\%)}\label{module77}}
\noindent 1~Ann Tourism Res, 2~Tourism Manage\subsection*{{\large 78 Textiles (0.015\%)}\label{module78}}
\noindent 1~Text Res J, 2~Sen-I Gakkaishi, 3~Indian J Fibre Text, 4~Tekstil\subsection*{{\large 79 Creativity Research (0.014\%)}\label{module79}}
\noindent 1~Creativity Res J, 2~J Creative Behav\subsection*{{\large 80 Travel Sociology (0.012\%)}\label{module80}}
\noindent 1~Sociol Trav, 2~Rev Fr Sociol, 3~Mouvement Soc\subsection*{{\large 81 Medical Informatics (0.011\%)}\label{module81}}
\noindent 1~Comput Meth Prog Bio, 2~Comput Biol Med, 3~Meas Control-Uk\subsection*{{\large 82 Leprosy (0.011\%)}\label{module82}}
\noindent 1~Leprosy Rev, 2~Int J Leprosy\subsection*{{\large 83 Sociology (Russian) (0.011\%)}\label{module83}}
\noindent 1~Psikhol Zh, 2~Vop Psikhol+, 3~Sotsiol Issled+\subsection*{{\large 84 Cryobiology (0.010\%)}\label{module84}}
\noindent 1~Cryobiology, 2~Cryoletters\subsection*{{\large 85 Death Studies (0.0095\%)}\label{module85}}
\noindent 1~Death Stud, 2~Omega-J Death Dying\subsection*{{\large 86 Rehabilitation Counseling (0.0074\%)}\label{module86}}
\noindent 1~Rehabil Couns Bull, 2~J Rehabil\subsection*{{\large 87 Steel (0.0072\%)}\label{module87}}
\noindent 1~Steel Res Int, 2~Stahl Eisen\subsection*{{\large 88 Futurist (0.0030\%)}\label{module88}}
\noindent 1~Futurist\section*{{\Large Fields of social science}\label{socialscience}}
\subsection*{{\large 1 Economics            (19\%)}\label{moduleSS1}}
\noindent 1~Am Econ Rev, 2~Q J Econ, 3~J Financ, 4~Econometrica, 5~J Polit Econ, 6~J Financ Econ, 7~Econ J, 8~J Econ Perspect, 9~J Econometrics, 10~J Monetary Econ, 11~Rev Econ Stat, 12~Rev Econ Stud, 13~J Econ Lit, 14~Eur Econ Rev, 15~J Int Econ, 16~Rev Financ Stud, 17~J Econ Theory, 18~J Public Econ, 19~World Dev, 20~Rand J Econ, 21~J Health Econ, 22~J Labor Econ, 23~J Hum Resour, 24~J Dev Econ, 25~Int Econ Rev, 26~Econ Lett, 27~J Bus Econ Stat, 28~J Econ Behav Organ, 29~J Urban Econ, 30~J Financ Quant Anal, 31~Ind Labor Relat Rev, 32~Game Econ Behav, 33~J Law Econ Organ, 34~Natl Tax J, 35~J Bus, 36~J Money Credit Bank, 37~J Law Econ, 38~Economet Theor, 39~J Econ Growth, 40~Am J Agr Econ, 41~J Dev Stud, 42~World Bank Econ Rev, 43~J Bank Financ, 44~J Appl Econom, 45~J Environ Econ Manag, 46~Public Choice, 47~J Account Econ, 48~Ecol Econ, 49~Int J Ind Organ, 50~Econ Theor, 51~Can J Econ, 52~Appl Econ, 53~Brookings Pap Eco Ac, 54~J Risk Uncertainty, 55~J Accounting Res, 56~Health Econ, 57~J Econ Dyn Control, 58~Econ Inq, 59~Ind Relat, 60~Oxford B Econ Stat, 61~J Ind Econ, 62~Econ Educ Rev, 63~South Econ J, 64~J Policy Anal Manag, 65~Reg Sci Urban Econ, 66~J Econ Manage Strat, 67~J Int Money Financ, 68~Soc Choice Welfare, 69~Land Econ, 70~Oxford Econ Pap, 71~Economica, 72~Energ Policy, 73~Econ Dev Cult Change, 74~Account Rev, 75~J Comp Econ, 76~Math Financ, 77~Scand J Econ, 78~Cambridge J Econ, 79~Mon Labor Rev, 80~Oxford Rev Econ Pol, 81~Math Soc Sci, 82~World Econ, 83~Financ Manage, 84~J Popul Econ, 85~J Math Econ, 86~Environ Resour Econ, 87~Int J Game Theory, 88~Energy J, 89~J Financ Intermed, 90~Econ Transit, 91~Int Tax Public Finan, 92~Rev Ind Organ, 93~J Prod Anal, 94~Small Bus Econ, 95~Energ Econ, 96~Theor Decis, 97~Resour Energy Econ, 98~Agr Econ, 99~J Econ Psychol, 100~J Labor Res, 101~Real Estate Econ, 102~Kyklos, 103~Macroecon Dyn, 104~Int J Forecasting, 105~Scot J Polit Econ, 106~J Forecasting, 107~J Econ Educ, 108~Int Rev Law Econ, 109~J Real Estate Financ, 110~Appl Econ Lett, 111~J Afr Econ, 112~J Econ, 113~J Regul Econ, 114~Eur Rev Agric Econ, 115~J Portfolio Manage, 116~J Futures Markets, 117~J Jpn Int Econ, 118~J Inst Theor Econ, 119~J Risk Insur, 120~J Hous Econ, 121~Food Policy, 122~J Econ Issues, 123~Can Public Pol, 124~China Econ Rev, 125~J Agr Econ, 126~Int Labour Rev, 127~Econ Philos, 128~J Policy Model, 129~J Macroecon, 130~Jpn World Econ, 131~Contemp Econ Policy, 132~Am J Econ Sociol, 133~Relat Ind-Ind Relat, 134~J Roy Stat Soc A Sta, 135~J Evol Econ, 136~Econ Model, 137~Econ Rec, 138~S Afr J Econ, 139~Auditing-J Pract Th, 140~Insur Math Econ, 141~Int J Manpower, 142~B Indones Econ Stud, 143~J Post Keynesian Ec, 144~Economist-Netherland, 145~Aust J Agr Resour Ec, 146~Eastern Eur Econ, 147~Polit Ekon, 148~Jahrb Natl Stat, 149~Dev Econ, 150~Betrieb Forsch Prax, 151~Resour Policy, 152~Open Econ Rev, 153~Ekon Cas, 154~Can J Dev Stud, 155~Trimest Econ\subsection*{{\large 2 Psychology           (18\%)}\label{moduleSS2}}
\noindent 1~J Pers Soc Psychol, 2~Child Dev, 3~Am Psychol, 4~Psychol Bull, 5~Psychol Sci, 6~Pers Soc Psychol B, 7~Psychol Rev, 8~Dev Psychol, 9~J Exp Psychol Learn, 10~Trends Cogn Sci, 11~Cognition, 12~Annu Rev Psychol, 13~J Exp Psychol Human, 14~Neuropsychologia, 15~Mem Cognition, 16~J Cognitive Neurosci, 17~Psychon B Rev, 18~Pers Indiv Differ, 19~J Mem Lang, 20~Behav Brain Sci, 21~J Exp Psychol Gen, 22~Percept Psychophys, 23~J Exp Soc Psychol, 24~Psychophysiology, 25~Psychol Aging, 26~J Pers, 27~Cognitive Psychol, 28~Organ Behav Hum Dec, 29~Brain Lang, 30~Cognition Emotion, 31~J Soc Issues, 32~Q J Exp Psychol-A, 33~J Exp Child Psychol, 34~Curr Dir Psychol Sci, 35~J Appl Soc Psychol, 36~Eur J Soc Psychol, 37~Pers Soc Psychol Rev, 38~J Fam Psychol, 39~Psychol Rep, 40~Psychol Inq, 41~Sex Roles, 42~Neuropsychology, 43~Brain Cognition, 44~J Cross Cult Psychol, 45~Acta Psychol, 46~Perception, 47~Cognitive Sci, 48~Lang Cognitive Proc, 49~J Res Pers, 50~Evol Hum Behav, 51~Appl Cognitive Psych, 52~Biol Psychol, 53~Cogn Neuropsychol, 54~Soc Dev, 55~Vis Cogn, 56~Memory, 57~Eur J Personality, 58~Behav Res Meth Ins C, 59~Int J Behav Dev, 60~Intelligence, 61~J Adolescence, 62~Psychol Res-Psych Fo, 63~Cognitive Dev, 64~J Youth Adolescence, 65~Int J Psychophysiol, 66~Brit J Soc Psychol, 67~J Clin Exp Neuropsyc, 68~J Math Psychol, 69~Percept Motor Skill, 70~Soc Cognition, 71~J Psycholinguist Res, 72~Pers Relationship, 73~Brit J Dev Psychol, 74~J Adolescent Res, 75~J Soc Psychol, 76~Psychol Women Quart, 77~Dev Neuropsychol, 78~Brit J Psychol, 79~J Exp Psychol-Appl, 80~Infant Behav Dev, 81~J Soc Clin Psychol, 82~Adolescence, 83~J Soc Pers Relat, 84~J Res Adolescence, 85~Merrill Palmer Quart, 86~J Behav Decis Making, 87~Inf Mental Hlth J, 88~J Child Lang, 89~J Early Adolescence, 90~J Motor Behav, 91~Dev Rev, 92~Conscious Cogn, 93~Eur J Cogn Psychol, 94~Basic Appl Soc Psych, 95~Aggressive Behav, 96~Can J Exp Psychol, 97~J Psychol, 98~Int J Psychol, 99~J Appl Dev Psychol, 100~Motiv Emotion, 101~Soc Indic Res, 102~Arch Clin Neuropsych, 103~Scand J Psychol, 104~Int J Intercult Rel, 105~Am J Psychol, 106~J Consciousness Stud, 107~Discourse Process, 108~Hum Nature-Int Bios, 109~Theor Psychol, 110~Hum Dev, 111~Risk Anal, 112~Hum Movement Sci, 113~J Genet Psychol, 114~Mind Lang, 115~J Nonverbal Behav, 116~J Neurolinguist, 117~Ethos, 118~Anxiety Stress Copin, 119~Aging Neuropsychol C, 120~Spatial Vision, 121~Soc Behav Personal, 122~Lang Speech, 123~Ann Psychol, 124~Can J Behav Sci, 125~New Ideas Psychol, 126~Eur J Psychol Assess, 127~J Gen Psychol, 128~Music Percept, 129~J Theor Soc Behav, 130~Teach Psychol, 131~Ecol Psychol, 132~Fem Psychol, 133~Genet Soc Gen Psych, 134~J Community Appl Soc, 135~J Psychophysiol, 136~Psychol Belg, 137~Cult Psychol, 138~Psychologist, 139~J Psychol Theol, 140~J Adult Dev, 141~Cah Psychol Cogn, 142~Curr Psychol, 143~Psicothema, 144~Soc Sci Inform, 145~Philos Psychol, 146~Aust J Psychol, 147~Zygon, 148~Swiss J Psychol, 149~Ethics Behav, 150~Z Entwickl Padagogis, 151~Comput Linguist, 152~Humor, 153~Psychol Rundsch, 154~Z Sozialpsychol, 155~Psychologia, 156~J Relig Health, 157~J Mind Behav, 158~Z Psychol, 159~J Humanist Psychol, 160~J Parapsychol, 161~J Constr Psychol, 162~Cesk Psychol, 163~Stud Psychol, 164~New Zeal J Psychol, 165~Rev Lat Am Psicol\subsection*{{\large 3 Psychiatry           (16\%)}\label{moduleSS3}}
\noindent 1~Am J Psychiat, 2~Arch Gen Psychiat, 3~J Consult Clin Psych, 4~Brit J Psychiat, 5~J Am Acad Child Psy, 6~J Clin Psychiat, 7~Psychol Med, 8~J Abnorm Psychol, 9~Schizophr Res, 10~J Child Psychol Psyc, 11~Addiction, 12~Psychiat Serv, 13~Acta Psychiat Scand, 14~Behav Res Ther, 15~Dev Psychopathol, 16~J Affect Disorders, 17~Health Psychol, 18~Psychosom Med, 19~J Stud Alcohol, 20~Schizophrenia Bull, 21~J Nerv Ment Dis, 22~J Psychosom Res, 23~Int J Eat Disorder, 24~J Abnorm Child Psych, 25~Psychiat Res, 26~Clin Psychol Rev, 27~Addict Behav, 28~Psychol Assessment, 29~Soc Psych Psych Epid, 30~Clin Psychol-Sci Pr, 31~J Subst Abuse Treat, 32~J Trauma Stress, 33~Int J Geriatr Psych, 34~J Clin Psychol, 35~Psychol Addict Behav, 36~Am J Geriat Psychiat, 37~Aust Nz J Psychiat, 38~Child Abuse Neglect, 39~Compr Psychiat, 40~Ann Behav Med, 41~J Interpers Violence, 42~J Autism Dev Disord, 43~Cognitive Ther Res, 44~Psychosomatics, 45~Gen Hosp Psychiat, 46~Psychother Psychosom, 47~Prof Psychol-Res Pr, 48~Subst Use Misuse, 49~Psychiat Clin N Am, 50~J Psychiat Res, 51~Behav Genet, 52~Psycho-Oncol, 53~Depress Anxiety, 54~Am J Drug Alcohol Ab, 55~J Pers Assess, 56~Exp Clin Psychopharm, 57~Behav Ther, 58~Psychol Health, 59~J Pers Disord, 60~J Anxiety Disord, 61~Eur Psychiat, 62~J Am Coll Health, 63~Am J Orthopsychiat, 64~Suicide Life-Threat, 65~Brit J Clin Psychol, 66~J Pediatr Psychol, 67~Crim Justice Behav, 68~J Behav Med, 69~Child Adol Psych Cl, 70~Community Ment Hlt J, 71~Psychiatry, 72~J Dev Behav Pediatr, 73~Eur Child Adoles Psy, 74~J Drug Issues, 75~J Psychoactive Drugs, 76~Behav Sci Law, 77~Aggress Violent Beh, 78~Am J Addiction, 79~Assessment, 80~Int Psychogeriatr, 81~Harvard Rev Psychiat, 82~Psychother Res, 83~J Fam Violence, 84~Int J Soc Psychiatr, 85~Alcohol Res Health, 86~Eur Eat Disord Rev, 87~Curr Opin Psychiatr, 88~J Psychiatr Neurosci, 89~J Psychopathol Behav, 90~Psychopathology, 91~Int J Law Psychiat, 92~Drug Alcohol Rev, 93~Int Rev Psychiatr, 94~Int J Psychiat Med, 95~Brit J Health Psych, 96~Psychiatr Rehabil J, 97~Psychiat Ann, 98~J Am Acad Psychiatry, 99~J Addict Dis, 100~Int J Behav Med, 101~Clin Psychol Psychot, 102~Int J Offender Ther, 103~Psychiat Quart, 104~Nord J Psychiat, 105~Psychother Psych Med, 106~Psychiat Prax, 107~J Child Adoles Subst, 108~J Behav Ther Exp Psy, 109~J Psychosoc Oncol, 110~Eur Addict Res, 111~Child Psychiat Hum D, 112~J Psychosom Obst Gyn, 113~B Menninger Clin, 114~J Drug Educ, 115~J Ect, 116~Diagnostica, 117~Adm Policy Ment Hlth, 118~Hist Psychiatr, 119~Psychotherapeut, 120~Z Psychosom Med Psyc, 121~Behav Change, 122~Behav Med, 123~Aust Psychol, 124~J Clin Psychol Med S, 125~Israel J Psychiat, 126~Arch Psychiat Nurs, 127~Child Fam Behav Ther, 128~Drug-Educ Prev Polic, 129~Int J Group Psychoth, 130~Z Kl Psych Psychoth, 131~Can Psychol, 132~Appl Psychophys Biof, 133~Z Kinder Jug-Psych, 134~Verhaltenstherapie, 135~Child Health Care, 136~Salud Ment, 137~Eur J Psychiat, 138~Prax Kinderpsychol K, 139~Gruppenpsychother Gr, 140~Gesundheitswesen, 141~Acad Psychiatr, 142~Med Sci Law, 143~Art Psychother, 144~J Music Ther, 145~Indian J Soc Work, 146~Salud Publica Mexico, 147~Z Klin Psych Psychia, 148~Evol Psychiatr, 149~Nord Psykol, 150~Kriminalistik\subsection*{{\large 4 Healthcare           (7.2\%)}\label{moduleSS4}}
\noindent 1~Am J Public Health, 2~Soc Sci Med, 3~Health Affair, 4~Med Care, 5~J Am Geriatr Soc, 6~Health Serv Res, 7~Gerontologist, 8~J Gerontol B-Psychol, 9~J Adv Nurs, 10~J Adolescent Health, 11~J Gerontol A-Biol, 12~Tob Control, 13~J Health Soc Behav, 14~Med Care Res Rev, 15~Am J Commun Psychol, 16~Public Health Rep, 17~Annu Rev Publ Health, 18~Milbank Q, 19~Sociol Health Ill, 20~J Health Polit Polic, 21~Health Educ Res, 22~Future Child, 23~Patient Educ Couns, 24~Health Educ Behav, 25~J Community Psychol, 26~Am J Manag Care, 27~Aids Care, 28~J School Health, 29~Am J Health Promot, 30~Nurs Res, 31~Res Nurs Health, 32~Int J Health Serv, 33~Health Policy, 34~Inquiry-J Health Car, 35~Aids Educ Prev, 36~Health Policy Plann, 37~Aust Nz J Publ Heal, 38~Res Aging, 39~Qual Health Res, 40~Aging Ment Health, 41~J Clin Nurs, 42~J Nurs Admin, 43~Health Care Manage R, 44~J Am Med Inform Assn, 45~J Aging Health, 46~Stud Family Plann, 47~Women Health, 48~Health Place, 49~Generations, 50~Int J Aging Hum Dev, 51~J Nurs Scholarship, 52~J Rural Health, 53~Public Health Nurs, 54~Am J Health Behav, 55~Eval Program Plann, 56~Health Soc Care Comm, 57~Int J Nurs Stud, 58~Cancer Nurs, 59~Evaluation Rev, 60~Med Anthropol Q, 61~Ageing Soc, 62~Hispanic J Behav Sci, 63~Eur J Public Health, 64~Int Fam Plan Perspec, 65~Western J Nurs Res, 66~J Biosoc Sci, 67~Adv Nurs Sci, 68~J Health Care Poor U, 69~Am J Nurs, 70~J Nurs Educ, 71~Scand J Public Healt, 72~J Aging Stud, 73~Nurs Outlook, 74~Can J Public Health, 75~Am J Occup Ther, 76~J Public Health Med, 77~Int J Qual Health C, 78~Birth-Iss Perinat C, 79~Aids Patient Care St, 80~Nurs Educ Today, 81~J Healthc Manag, 82~Cult Med Psychiat, 83~J Health Commun, 84~Health Promot Int, 85~Scand J Caring Sci, 86~Geriatr Nurs, 87~Can J Aging, 88~Nurs Ethics, 89~J Appl Gerontol, 90~J Prof Nurs, 91~J Commun Health, 92~Appl Nurs Res, 93~Public Health, 94~Child Care Hlth Dev, 95~Health Commun, 96~J Palliative Care, 97~Women Health Iss, 98~Nurs Sci Quart, 99~J Nurs Care Qual, 100~J Women Aging, 101~Midwifery, 102~J Elder Abuse Negl, 103~Nurs Clin N Am, 104~J Aging Phys Activ, 105~Int J Health Plan M, 106~Educ Gerontol, 107~Fam Community Health, 108~J Rehabil Res Dev, 109~Eval Health Prof, 110~Am J Eval, 111~Sci Soc Sante, 112~J Public Health Pol, 113~J Midwifery Wom Heal, 114~Rev Saude Publ, 115~Z Gerontol Geriatr, 116~Top Geriatr Rehabil, 117~J Perinat Neonat Nur\subsection*{{\large 5 Political Science    (4.9\%)}\label{moduleSS5}}
\noindent 1~Am Polit Sci Rev, 2~Am J Polit Sci, 3~Int Organ, 4~World Polit, 5~J Polit, 6~Int Security, 7~J Conflict Resolut, 8~Comp Polit Stud, 9~J Democr, 10~Int Stud Quart, 11~Eur J Polit Res, 12~Comp Polit, 13~J Eur Public Policy, 14~J Peace Res, 15~Annu Rev Polit Sci, 16~J Common Mark Stud, 17~Brit J Polit Sci, 18~Polit Soc, 19~Public Opin Quart, 20~Polit Res Quart, 21~Polit Psychol, 22~Ps-Polit Sci Polit, 23~Legis Stud Quart, 24~Elect Stud, 25~J Theor Polit, 26~Third World Q, 27~Party Polit, 28~Wash Quart, 29~Governance, 30~Polit Stud-London, 31~Int Aff, 32~Rev Int Stud, 33~E Eur Polit Soc, 34~Int Polit Sci Rev, 35~Int Interact, 36~Polit Sci Quart, 37~Polit Theory, 38~Stud Comp Int Dev, 39~Eur J Int Relat, 40~Secur Stud, 41~Armed Forces Soc, 42~Hum Rights Quart, 43~Europe-Asia Stud, 44~Millennium-J Int St, 45~Post-Sov Aff, 46~Polit Behav, 47~Communis Post-Commun, 48~Lat Am Res Rev, 49~J Lat Am Stud, 50~Scand Polit Stud, 51~J Strategic Stud, 52~Judicature, 53~Glob Gov, 54~Policy Rev, 55~Polit Vierteljahr, 56~Defence Peace Econ, 57~Can J Polit Sci, 58~Publius J Federalism, 59~Gov Oppos, 60~Women Polit, 61~World Policy J, 62~Alternatives, 63~Dados-Rev Cienc Soc, 64~Aust J Polit Sci, 65~Polity, 66~Int J, 67~Aust J Int Aff, 68~Society, 69~Tidsskr Samfunnsfor, 70~Secur Dialogue, 71~J Baltic Stud, 72~Osteuropa, 73~Rev Etud Comp Est-O\subsection*{{\large 6 Sociology (Behavioral)  (4.4\%)}\label{moduleSS6}}
\noindent 1~Am Sociol Rev, 2~Am J Sociol, 3~J Marriage Fam, 4~Annu Rev Sociol, 5~Soc Forces, 6~Demography, 7~Criminology, 8~Am Behav Sci, 9~Soc Probl, 10~Popul Dev Rev, 11~Soc Psychol Quart, 12~Soc Sci Quart, 13~Ethnic Racial Stud, 14~Soc Sci Res, 15~J Fam Issues, 16~Theor Soc, 17~J Sci Stud Relig, 18~Law Soc Rev, 19~Gender Soc, 20~Pop Stud-J Demog, 21~J Quant Criminol, 22~Ann Am Acad Polit Ss, 23~Int Migr Rev, 24~J Res Crime Delinq, 25~Crime Delinquency, 26~Work Occupation, 27~Sociol Method Res, 28~Sociol Theor, 29~Sociol Quart, 30~Contemp Sociol, 31~Sociol Forum, 32~J Crim Just, 33~Eur Sociol Rev, 34~Sociol Relig, 35~Rural Sociol, 36~Sociol Perspect, 37~Soc Networks, 38~Youth Soc, 39~Eur J Popul, 40~Symb Interact, 41~J Comp Fam Stud, 42~Popul Res Policy Rev, 43~J Contemp Ethnogr, 44~Rev Relig Res, 45~Ration Soc, 46~Sociol Inq, 47~Deviant Behav, 48~Int Migr, 49~Kolner Z Soziol Soz, 50~Int Sociol, 51~Population, 52~Acta Sociol, 53~Z Soziol, 54~J Black Stud, 55~Qual Quant, 56~Can Rev Soc Anthrop, 57~Arch Eur Sociol, 58~Popul Environ, 59~Soc Sci J, 60~Can J Sociol, 61~Berl J Soziol, 62~Juvenile Fam Court J, 63~Soz Welt, 64~Sociol Spectrum, 65~Desarrollo Econ, 66~Soc Compass, 67~Soc Anim, 68~J Psychohist, 69~Man India\subsection*{{\large 7 Management           (4.1\%)}\label{moduleSS7}}
\noindent 1~J Appl Psychol, 2~Acad Manage J, 3~Acad Manage Rev, 4~Strategic Manage J, 5~Manage Sci, 6~Admin Sci Quart, 7~Organ Sci, 8~Res Policy, 9~J Organ Behav, 10~J Manage, 11~J Vocat Behav, 12~Pers Psychol, 13~Hum Relat, 14~J Manage Stud, 15~J Int Bus Stud, 16~Organ Stud, 17~Calif Manage Rev, 18~J Occup Organ Psych, 19~Hum Resource Manage, 20~Organization, 21~Leadership Quart, 22~Appl Psychol-Int Rev, 23~Group Organ Manage, 24~J Prod Innovat Manag, 25~Hum Perform, 26~J Bus Venturing, 27~Int J Select Assess, 28~Work Stress, 29~J Bus Ethics, 30~Small Gr Res, 31~Organ Dyn, 32~Ieee T Eng Manage, 33~Account Org Soc, 34~Int J Technol Manage, 35~J Bus Psychol, 36~J Organ Change Manag, 37~RFd Manage, 38~Long Range Plann, 39~J World Bus, 40~Int J Oper Prod Man, 41~Res Technol Manage, 42~Mil Psychol, 43~J Manage Inquiry, 44~Manage Learn, 45~Public Pers Manage, 46~J Small Bus Manage, 47~Technol Forecast Soc, 48~Time Soc, 49~Pers Rev, 50~Futures, 51~J Employment Couns, 52~Z Arb Organ, 53~Negotiation J, 54~Trav Humain, 55~Adult Educ Quart, 56~Gruppendynamik Organ\subsection*{{\large 8 Law                  (3.9\%)}\label{moduleSS8}}
\noindent 1~Yale Law J, 2~Harvard Law Rev, 3~Stanford Law Rev, 4~Columbia Law Rev, 5~Va Law Rev, 6~Calif Law Rev, 7~U Chicago Law Rev, 8~U Penn Law Rev, 9~Mich Law Rev, 10~New York U Law Rev, 11~Am J Int Law, 12~Tex Law Rev, 13~Georgetown Law J, 14~Vanderbilt Law Rev, 15~Northwest U Law Rev, 16~J Legal Stud, 17~Cornell Law Rev, 18~Ucla Law Rev, 19~Duke Law J, 20~Minn Law Rev, 21~Fordham Law Rev, 22~Iowa Law Rev, 23~Indiana Law J, 24~Notre Dame Law Rev, 25~South Calif Law Rev, 26~Boston U Law Rev, 27~U Illinois Law Rev, 28~Wisc Law Rev, 29~Bus Lawyer, 30~Harvard J Law Publ P, 31~George Wash Law Rev, 32~Admin Law Rev, 33~Law Social Inquiry, 34~Am Crim Law Rev, 35~U Cinci Law Rev, 36~Am J Law Med, 37~Hastings Law J, 38~Harvard Int Law J, 39~Am J Comp Law, 40~Columbia J Trans Law, 41~J Crim Law Crim, 42~Wash Law Rev, 43~Buffalo Law Rev, 44~Antitrust Law J, 45~Harvard J Legis, 46~U Pitt Law Rev, 47~Ecol Law Quart, 48~J Copyright Soc Usa, 49~J Int Econ Law, 50~Cornell Int Law J, 51~J Legal Educ, 52~Am Bus Law J, 53~U Pa J Int Econ Law, 54~Common Mkt Law Rev, 55~Rutgers Law Rev, 56~Cathol U Law Rev, 57~Fam Law Quart, 58~Food Drug Law J, 59~Am Bankrupt Law J, 60~Urban Lawyer, 61~Law Philos, 62~Denver U Law Rev, 63~Law Libr J, 64~Mil Law Rev, 65~J Legal Med, 66~Nat Resour J, 67~Columbia J Law Soc P, 68~Iic-Int Rev Intell P, 69~Issues Law Med, 70~Secur Regul Law J, 71~J Marit Law Commer\subsection*{{\large 9 Education            (2.7\%)}\label{moduleSS9}}
\noindent 1~J Educ Psychol, 2~Am Educ Res J, 3~Educ Psychol, 4~Phi Delta Kappan, 5~Educ Leadership, 6~Read Res Quart, 7~J Res Sci Teach, 8~Sci Educ, 9~Educ Eval Policy An, 10~Teach Coll Rec, 11~Elem School J, 12~Contemp Educ Psychol, 13~Cognition Instruct, 14~Int J Sci Educ, 15~J Learn Sci, 16~Sociol Educ, 17~Teach Teach Educ, 18~J Teach Educ, 19~Brit J Educ Psychol, 20~Read Teach, 21~Harvard Educ Rev, 22~Educ Psychol Rev, 23~Learn Instr, 24~J Educ Res, 25~J Adolesc Adult Lit, 26~Theor Pract, 27~J Res Math Educ, 28~J Lit Res, 29~Early Child Res Q, 30~Sch Eff Sch Improv, 31~J Curriculum Stud, 32~Urban Educ, 33~Educ Urban Soc, 34~Eur J Psychol Educ, 35~Educ Admin Quart, 36~High Educ, 37~Anthropol Educ Quart, 38~Educ Policy, 39~Res Teach Engl, 40~J Exp Educ, 41~Instr Sci, 42~Educ Res-Uk, 43~Curriculum Inq, 44~Comput Educ, 45~Stud High Educ, 46~Young Children, 47~EtrFd-Educ Tech Res, 48~Gifted Child Quart, 49~J Comput Assist Lear, 50~Brit J Educ Technol, 51~Interv Sch Clin, 52~J Educ Gifted\subsection*{{\large 10 Geography            (2.5\%)}\label{moduleSS10}}
\noindent 1~Urban Stud, 2~Environ Plann A, 3~Prog Hum Geog, 4~Int J Urban Regional, 5~Reg Stud, 6~Ann Assoc Am Geogr, 7~Environ Plann D, 8~Antipode, 9~T I Brit Geogr, 10~Polit Geogr, 11~Geoforum, 12~Hous Policy Debate, 13~Prof Geogr, 14~Urban Aff Rev, 15~J Hist Geogr, 16~Econ Geogr, 17~Area, 18~J Urban Aff, 19~J Am Plann Assoc, 20~Sociol Ruralis, 21~J Rural Stud, 22~Environ Plann B, 23~Housing Stud, 24~Urban Geogr, 25~Int J Geogr Inf Sci, 26~J Regional Sci, 27~Environ Plann C, 28~Pap Reg Sci, 29~Int Regional Sci Rev, 30~Rev Int Polit Econ, 31~Cities, 32~Geogr Anal, 33~Growth Change, 34~Ann Regional Sci, 35~Tijdschr Econ Soc Ge, 36~Eur Urban Reg Stud, 37~Landscape Urban Plan, 38~Econ Dev Q, 39~J Plan Educ Res, 40~Int Soc Sci J, 41~Can Geogr-Geogr Can, 42~Geogr Rev, 43~Land Use Policy, 44~Habitat Int, 45~Geogr J, 46~Environ Urban, 47~Singapore J Trop Geo, 48~Aust Geogr, 49~Geography, 50~Appl Geogr, 51~J Geogr Higher Educ, 52~Cartogr J, 53~J Archit Plan Res, 54~J Urban Technol, 55~Interdiscipl Sci Rev, 56~Geogr Z, 57~J Urban Plan D-Asce\subsection*{{\large 11 Physical Anthropology (1.5\%)}\label{moduleSS11}}
\noindent 1~Am J Phys Anthropol, 2~J Hum Evol, 3~Curr Anthropol, 4~J Archaeol Sci, 5~Am Anthropol, 6~Evol Anthropol, 7~Annu Rev Anthropol, 8~Am Antiquity, 9~J Anthropol Archaeol, 10~Int J Osteoarchaeol, 11~Am J Hum Biol, 12~Hum Organ, 13~Anthropologie, 14~Hum Ecol, 15~J Anthropol Res, 16~Scientist, 17~Plains Anthropol, 18~Arctic Anthropol, 19~Homo, 20~Collegium Antropol\subsection*{{\large 12 Cultural Anthropology (1.2\%)}\label{moduleSS12}}
\noindent 1~Public Culture, 2~Am Ethnol, 3~Cult Anthropol, 4~Signs, 5~J S Afr Stud, 6~Comp Stud Soc Hist, 7~Ids Bull-I Dev Stud, 8~J Roy Anthropol Inst, 9~Dev Change, 10~J Asian Stud, 11~Afr Affairs, 12~Feminist Stud, 13~Africa, 14~J Mod Afr Stud, 15~Glq-J Lesbian Gay St, 16~J Afr Hist, 17~Mod Asian Stud, 18~Identities-Glob Stud, 19~Lat Am Perspect, 20~Ethnohistory, 21~Anthropol Quart, 22~J Hist Sociol, 23~Ethnology, 24~Cult Stud, 25~Crit Anthropol, 26~J Peasant Stud, 27~Bijdr Taal-Land-V, 28~Disasters, 29~Homme, 30~Anthropos, 31~J Gender Stud, 32~Contrib Indian Soc, 33~Frontiers\subsection*{{\large 13 Marketing            (1.0\%)}\label{moduleSS13}}
\noindent 1~J Marketing Res, 2~J Marketing, 3~Market Sci, 4~J Consum Res, 5~J Acad Market Sci, 6~J Bus Res, 7~J Retailing, 8~J Advertising Res, 9~Int J Res Mark, 10~J Advertising, 11~Ind Market Manag, 12~Psychol Market, 13~J Public Policy Mark, 14~Decision Sci, 15~J Int Marketing, 16~Int J Serv Ind Manag, 17~Adv Consum Res, 18~J Consum Aff, 19~Serv Ind J\subsection*{{\large 14 Information Science  (0.99\%)}\label{moduleSS14}}
\noindent 1~J Am Soc Inf Sci Tec, 2~Scientometrics, 3~Coll Res Libr, 4~J Doc, 5~Inform Process Manag, 6~J Inf Sci, 7~Libr Trends, 8~J Acad Libr, 9~Libr Inform Sci Res, 10~Libr Quart, 11~Aslib Proc, 12~Learn Publ, 13~Ref User Serv Q, 14~Gov Inform Q, 15~Libri, 16~Inform Technol Libr, 17~Libr Resour Tech Ser, 18~J Libr Inf Sci, 19~Knowl Organ, 20~J Gov Inform, 21~Electron Libr, 22~Interlend Doc Supply, 23~Z Bibl Bibl\subsection*{{\large 15 Philosophy of Science (0.98\%)}\label{moduleSS15}}
\noindent 1~Stud Hist Philos Sci, 2~Philos Sci, 3~Soc Stud Sci, 4~Arch Hist Exact Sci, 5~Brit J Philos Sci, 6~Sci Context, 7~Biol Philos, 8~Osiris, 9~Hist Sci, 10~Isis, 11~Hist Math, 12~Synthese, 13~Brit J Hist Sci, 14~B Hist Med, 15~Med Hist, 16~Soc Hist Med, 17~Ann Sci, 18~Technol Cult, 19~J Hist Biol, 20~J Hist Behav Sci, 21~Philos Soc Sci, 22~J Hist Med All Sci, 23~Minerva, 24~Hist Stud Phys Biol, 25~Hist Phil Life Sci\subsection*{{\large 16 Sociology (Institutional) (0.95\%)}\label{moduleSS16}}
\noindent 1~Sociology, 2~Econ Soc, 3~Brit J Sociol, 4~Theor Cult Soc, 5~Work Employ Soc, 6~Brit J Sociol Educ, 7~Sociol Rev, 8~Brit J Ind Relat, 9~J Educ Policy, 10~Hist Hum Sci, 11~Women Stud Int Forum, 12~Oxford Rev Educ, 13~Soc Res, 14~Comp Educ, 15~Educ Stud, 16~Brit J Educ Stud, 17~New Tech Work Employ, 18~Z Padagog Psychol, 19~Int J Educ Dev, 20~Econ Ind Democracy, 21~Childhood, 22~Comp Educ Rev, 23~J Moral Educ, 24~Eur J Ind Relat, 25~Z Padagogik, 26~J Philos Educ, 27~Teach Sociol, 28~Psychol Erz Unterr, 29~Educ Rev, 30~Eur J Womens Stud\subsection*{{\large 17 Communication        (0.80\%)}\label{moduleSS17}}
\noindent 1~J Commun, 2~Journalism Mass Comm, 3~Commun Res, 4~Polit Commun, 5~Hum Commun Res, 6~Commun Monogr, 7~J Broadcast Electron, 8~Sci Technol Hum Val, 9~Commun Theor, 10~Media Cult Soc, 11~Harv Int J PressOpol, 12~J Lang Soc Psychol, 13~Public Underst Sci, 14~Inform Soc, 15~Writ Commun, 16~Eur J Commun, 17~Telecommun Policy, 18~Int J Public Opin R, 19~Q J Speech, 20~Sci Commun, 21~Tech Commun, 22~J Media Econ, 23~J Bus Tech Commun, 24~Public Relat Rev\subsection*{{\large 18 Educational Assessment (0.70\%)}\label{moduleSS18}}
\noindent 1~Psychol Methods, 2~Psychometrika, 3~Educ Psychol Meas, 4~Appl Psych Meas, 5~Multivar Behav Res, 6~J Educ Meas, 7~J Educ Behav Stat, 8~Brit J Math Stat Psy, 9~Appl Meas Educ, 10~J Classif\subsection*{{\large 19 Educational Psychology (0.66\%)}\label{moduleSS19}}
\noindent 1~J School Psychol, 2~School Psychol Rev, 3~Except Children, 4~J Learn Disabil-Us, 5~J Spec Educ, 6~Rem Spec Educ, 7~Psychol Schools, 8~School Psychol Quart, 9~J Emot Behav Disord, 10~Top Early Child Spec, 11~J Early Intervention, 12~School Psychol Int, 13~Learn Disability Q, 14~Am Ann Deaf, 15~J Psychoeduc Assess, 16~Infant Young Child, 17~J Educ Psychol Cons, 18~Volta Rev\subsection*{{\large 20 Human-Computer Interface (0.54\%)}\label{moduleSS20}}
\noindent 1~Mis Quart, 2~Inform Syst Res, 3~J Manage Inform Syst, 4~Comput Hum Behav, 5~Inform Manage-Amster, 6~Int J Hum-Comput St, 7~Cyberpsychol Behav, 8~Omega-Int J Manage S, 9~Behav Inform Technol, 10~Soc Sci Comput Rev, 11~Int J Electron Comm, 12~Int J Inform Manage, 13~J Inf Technol, 14~Interact Comput, 15~Int J Hum-Comput Int, 16~Inform Syst J, 17~Group Decis Negot\subsection*{{\large 21 Applied Linguistics  (0.51\%)}\label{moduleSS21}}
\noindent 1~J Pragmatics, 2~Discourse Soc, 3~Lang Soc, 4~Res Lang Soc Interac, 5~Tesol Quart, 6~Mod Lang J, 7~Appl Linguist, 8~Language, 9~Lang Learn, 10~Lang Commun, 11~Linguistics, 12~Can Mod Lang Rev, 13~Lingua, 14~Am Speech, 15~Foreign Lang Ann\subsection*{{\large 22 Experimental Psychology (0.49\%)}\label{moduleSS22}}
\noindent 1~J Exp Anal Behav, 2~J Exp Psychol Anim B, 3~Physiol Behav, 4~Behav Process, 5~J Comp Psychol, 6~Q J Exp Psychol-B, 7~Psychol Rec, 8~Learn Motiv, 9~Behav Analyst, 10~Neurobiol Learn Mem, 11~Integr Phys Beh Sci, 12~Jpn Psychol Res\subsection*{{\large 23 History              (0.46\%)}\label{moduleSS23}}
\noindent 1~Am Hist Rev, 2~J Am Hist, 3~Environ Hist, 4~Slavic Rev, 5~J Mod Hist, 6~Past Present, 7~Continuity Change, 8~J Fam Hist, 9~Labor Hist, 10~Int Rev Soc Hist, 11~J Soc Hist, 12~J Urban Hist, 13~J Hist Sexuality, 14~E Eur Quart, 15~Crit Rev\subsection*{{\large 24 Social Work          (0.40\%)}\label{moduleSS24}}
\noindent 1~Soc Work, 2~Child Youth Serv Rev, 3~Fam Soc-J Contemp H, 4~Soc Serv Rev, 5~Res Social Work Prac, 6~Soc Work Res, 7~Soc Work Health Care, 8~Brit J Soc Work, 9~Health Soc Work, 10~J Soc Work Educ, 11~Affilia J Wom Soc Wo, 12~J Soc Serv Res, 13~Smith Coll Stud Soc, 14~Int Soc Work, 15~Clin Soc Work J, 16~Admin Soc Work\subsection*{{\large 25 Speech And Hearing    (0.40\%)}\label{moduleSS25}}
\noindent 1~J Speech Lang Hear R, 2~Appl Psycholinguist, 3~Int J Lang Comm Dis, 4~J Phonetics, 5~Lang Speech Hear Ser, 6~Clin Linguist Phonet, 7~J Commun Disord, 8~Phonetica, 9~J Fluency Disord, 10~Top Lang Disord, 11~Folia Phoniatr Logo\subsection*{{\large 26 Disabilities          (0.39\%)}\label{moduleSS26}}
\noindent 1~Am J Ment Retard, 2~J Intell Disabil Res, 3~J Appl Behav Anal, 4~Behav Modif, 5~Disabil Soc, 6~Res Dev Disabil, 7~Ment Retard, 8~J Appl Res Intellect, 9~J Dev Phys Disabil, 10~Res Pract Pers Sev D\subsection*{{\large 27 Transportation       (0.39\%)}\label{moduleSS27}}
\noindent 1~Transport Res A-Pol, 2~J Oper Res Soc, 3~Transport Res B-Meth, 4~Interfaces, 5~Transport Sci, 6~Transportation, 7~J Transp Econ Policy, 8~Transport Res E-Log, 9~Syst Dynam Rev, 10~Syst Res Behav Sci, 11~Transport Res D-Tr E, 12~Transport Rev, 13~Syst Pract Act Res, 14~Transport J\subsection*{{\large 28 Psychoanalysis       (0.38\%)}\label{moduleSS28}}
\noindent 1~J Am Psychoanal Ass, 2~Int J Psychoanal, 3~Psychoanal Quart, 4~Psychoanal Inq, 5~Psychoanal Psychol, 6~Psyche-Z Psychoanal, 7~Contemp Psychoanal, 8~Forum Psychoanal, 9~J Anal Psychol\subsection*{{\large 29 Guidance Counseling  (0.36\%)}\label{moduleSS29}}
\noindent 1~J Couns Psychol, 2~Couns Psychol, 3~J Couns Dev, 4~J Career Assessment, 5~Meas Eval Couns Dev, 6~Career Dev Q, 7~J Multicult Couns D, 8~J Career Dev, 9~Brit J Guid Couns, 10~Fed Probat, 11~Women Ther\subsection*{{\large 30 Middle Eastern Studies (0.35\%)}\label{moduleSS30}}
\noindent 1~Int J Middle E Stud, 2~New Left Rev, 3~Middle Eastern Stud, 4~Middle East J, 5~Mon Rev, 6~Sci Soc, 7~Race Class\subsection*{{\large 31 East Asian Studies    (0.35\%)}\label{moduleSS31}}
\noindent 1~China Quart, 2~China J, 3~Asian Surv, 4~Mod China, 5~Issues Stud, 6~Pac Rev, 7~J Contemp Asia, 8~Pac Aff\subsection*{{\large 32 Ergonomics           (0.34\%)}\label{moduleSS32}}
\noindent 1~Ergonomics, 2~Accident Anal Prev, 3~Hum Factors, 4~Appl Ergon, 5~Int J Ind Ergonom, 6~J Safety Res, 7~Int J Aviat Psychol, 8~J Occup Rehabil\subsection*{{\large 33 Medical Ethics        (0.33\%)}\label{moduleSS33}}
\noindent 1~Hastings Cent Rep, 2~J Med Ethics, 3~J Law Med Ethics, 4~Bioethics, 5~Kennedy Inst Ethic J, 6~J Clin Ethic, 7~J Med Philos, 8~Health Care Anal, 9~Camb Q Healthc Ethic, 10~Sci Eng Ethics, 11~Theor Med Bioeth\subsection*{{\large 34 Public Administration  (0.26\%)}\label{moduleSS34}}
\noindent 1~Public Admin Rev, 2~Soc Natur Resour, 3~Nonprof Volunt Sec Q, 4~Admin Soc, 5~Policy Sci, 6~Policy Stud J, 7~Am Rev Public Adm, 8~Int Rev Adm Sci, 9~Can Public Admin\subsection*{{\large 35 Ethics               (0.26\%)}\label{moduleSS35}}
\noindent 1~Ethics, 2~Inquiry, 3~Soc Philos Policy, 4~Environ Ethics, 5~Hum Stud, 6~Environ Value, 7~J Agr Environ Ethic\subsection*{{\large 36 Economic History     (0.26\%)}\label{moduleSS36}}
\noindent 1~J Econ Hist, 2~Econ Hist Rev, 3~Explor Econ Hist, 4~Bus Hist, 5~Soc Sci Hist, 6~J Interdiscipl Hist, 7~Agr Hist\subsection*{{\large 37 Sport Psychology     (0.25\%)}\label{moduleSS37}}
\noindent 1~Res Q Exercise Sport, 2~J Sport Exercise Psy, 3~Quest, 4~J Appl Sport Psychol, 5~J Teach Phys Educ, 6~Sport Psychol, 7~Int J Sport Psychol, 8~Sociol Sport J, 9~J Sport Hist\subsection*{{\large 38 Public Affairs        (0.22\%)}\label{moduleSS38}}
\noindent 1~Public Admin, 2~Parliament Aff, 3~Polit Quart, 4~Local Gov Stud, 5~Public Money Manage, 6~Public Admin Develop, 7~Aust J Publ Admin\subsection*{{\large 39 Social Policy         (0.19\%)}\label{moduleSS39}}
\noindent 1~Soc Policy Admin, 2~Policy Polit, 3~J Eur Soc Policy, 4~J Soc Policy, 5~Soc Polit, 6~Aust J Soc Issues\subsection*{{\large 40 Family Relations      (0.18\%)}\label{moduleSS40}}
\noindent 1~Fam Relat, 2~Fam Process, 3~J Marital Fam Ther, 4~Am J Fam Ther, 5~J Fam Ther, 6~Contemp Fam Ther\subsection*{{\large 41 Law And Behavior      (0.17\%)}\label{moduleSS41}}
\noindent 1~Law Human Behav, 2~Psychol Public Pol L, 3~Psychol Crime Law\subsection*{{\large 42 Criminology           (0.16\%)}\label{moduleSS42}}
\noindent 1~Brit J Criminol, 2~J Law Soc, 3~Crime Law Social Ch, 4~Soc Legal Stud, 5~Int J Sociol Law, 6~Aust Nz J Criminol, 7~Crim Law Rev\subsection*{{\large 43 Sexuality             (0.16\%)}\label{moduleSS43}}
\noindent 1~Arch Sex Behav, 2~J Sex Res, 3~J Sex Marital Ther, 4~J Homosexual\subsection*{{\large 44 Higher Education      (0.14\%)}\label{moduleSS44}}
\noindent 1~J Coll Student Dev, 2~Res High Educ, 3~J High Educ, 4~Rev High Educ\subsection*{{\large 45 Leisure Studies (0.14\%)}\label{moduleSS45}}
\noindent 1~Environ Behav, 2~J Environ Psychol, 3~J Leisure Res, 4~Leisure Sci\subsection*{{\large 46 Neurorehabilitation   (0.14\%)}\label{moduleSS46}}
\noindent 1~J Head Trauma Rehab, 2~Rehabil Psychol, 3~Neurorehabilitation, 4~Rehabil Couns Bull, 5~J Rehabil, 6~Int J Rehabil Res\subsection*{{\large 47 Pacific Studies       (0.10\%)}\label{moduleSS47}}
\noindent 1~Contemp Pacific, 2~Oceania, 3~J Polynesian Soc, 4~J Mat Cult\subsection*{{\large 48 Tourism               (0.093\%)}\label{moduleSS48}}
\noindent 1~Ann Tourism Res, 2~Tourism Manage\subsection*{{\large 49 Creativity            (0.073\%)}\label{moduleSS49}}
\noindent 1~Creativity Res J, 2~J Creative Behav\subsection*{{\large 50 Death And Dying       (0.061\%)}\label{moduleSS50}}
\noindent 1~Death Stud, 2~Omega-J Death Dying\subsection*{{\large 51 Sociology (French)    (0.059\%)}\label{moduleSS51}}
\noindent 1~Psikhol Zh, 2~Vop Psikhol+, 3~Sociol Cas, 4~Sotsiol Issled+\subsection*{{\large 52 Sociology (Eastern Europe)  (0.054\%)}\label{moduleSS52}}
\noindent 1~Sociol Trav, 2~Rev Fr Sociol, 3~Mouvement Soc\subsection*{{\large 53 Maritime Law          (0.048\%)}\label{moduleSS53}}
\noindent 1~Mar Policy, 2~Ocean Dev Int Law\subsection*{{\large 54 Hypnosis             (0.028\%)}\label{moduleSS54}}
\noindent 1~Int J Clin Exp Hyp, 2~Am J Clin Hypn
\end{document}